\documentclass{article}%
\usepackage{amsmath}
\usepackage{amssymb}
\usepackage{geometry}%
\usepackage{amsfonts}%
\usepackage{graphicx}
\providecommand{\U}[1]{\protect \rule{.1in}{.1in}}

\newcommand{\mathnf}{\mathsf}

\begin{document}

\title{Sufficientarian Grading Rules and Rankings: Characterizations and Implementation}
\author{Marcello Basili\thanks{University of Siena, Department of Economics and
Statistics, marcello.basili@unisi.it}
\and Ernesto Savaglio \thanks{University of Chieti-Pescara, Department of Business
Economics, \& GRASS, ernesto@unich.it}
\and Stefano Vannucci \thanks{University of Siena, Department of Economics and
Statistics, stefano.vannucci@unisi.it}}
\date{\today }
\maketitle

\begin{abstract}
Sufficientarian grading rules are defined using a finite family of
sufficientarian judgements on individual capability assignments as embodied in
a sufficientarian binary grading function (BGF). Both sufficientarian grading
rules and the sufficientarian total preorders on capability-type assignments
they induce are characterized. Moreover, several further total preorders based
upon sufficiency-gap information provided by a sufficientarian grading rule
are explicitly defined and some of them are also characterized. It is also
shown that there exists a class of inclusive, unanimity-respecting and
suitably strategy-proof protocols (including simple majority when the number
of agents is odd) which can be deployed in order to select one specific
sufficientarian grading rule.



\end{abstract}

\section{\textbf{Introduction}}

In the last few decades a considerable amount of work has been devoted to
\textit{sufficientarianism}, the class of distribution rules establishing that
`\textit{everyone should have enough' }to live a properly accomplished and
socially respected life, or `enjoy sufficient freedom' as someone might
perhaps like to express that very notion. Thus, any such sufficientarian rule
also provides in a most straightforward way both \textit{ratings }as based on
\textit{benchmarks }(e.g. sets of goals or targets) and the resulting
\textit{rankings }(e.g. preorders, including of course \textit{total}
preorders). Such ratings and rankings are meant to enable, respectively,
`\textit{intrinsic}' assessments of \textit{individual} assignments and of the
resulting \textit{overall }assignments of the relevant affordances to agents,
and \textit{comparative} assessments of such overall assignments. Both of them
are to be used in certain social situations of interest including possibly as
criteria for general assessments of \textit{social progress} at large.

As a matter of fact, in the extant literature sufficientarian views and rules
and the underlying principles and motivations are discussed and scrutinized at
length from different perspectives. But more often than not advocates of
sufficientarianism insist on the need to rely on \textit{absolute} as opposed
to \textit{comparative} judgements: a theoretical stance that suggests
precisely a view of the relevant rankings as a derivative notion of previously
established \textit{ratings, }as suggested above \footnote{The
distinction/opposition between ratings and rankings, especially when it comes
to aggregation problems, has by now a long history which goes back at least as
far as Huntington (1938) and has been recently revitalized, mostly as a result
of the work of Balinski and Laraki (2007, 2011, 2014).}. Yet, while
characterizations of several versions of sufficientarian rules are also
available, such characterizations are typically focussed on sufficientarian
\textit{rankings }as opposed to ratings (more details on those two somewhat
contrasting approaches to sufficientarianism will be presented below in
Section 2).

The general aim of the present work is to provide a quite comprehensive study
of sufficientarian principles starting on the contrary from
\textit{sufficientarian rating rules}, and relying on them in order to
introduce sufficientarian ranking rules as a \textit{derivative} notion of the
former. Incidentally, such an approach may also help bridging the two strands
of literature mentioned above.

Indeed, coming back to the core sufficientarian principle requiring that
`every one should have enough' and parsing that statement accurately, it seems
to be quite clear that the basic components of any sufficientarian rule are
ultimately a \textit{common} language and framework enabling the expression of
\textit{judgements} involving a population of \textit{agents} (the set of
relevant `individuals'), and providing for each one of them a `yes/no' answer
to the following question: `does this particular agent \textit{have enough?'
.}

Obviously, any such answer requires in turn, as an indispensable input,
answers to \textit{two} further underlying questions, namely:

$(I)$ `Enough of\emph{\ what}?', first and foremost, and then $(II)$ `What is
\emph{enough}\textit{?}'.

Concerning question (I), it is broadly speaking \textit{affordances
-}resulting in \textit{access to achievements }of some sorts- that are to be
considered here when defining \textit{assignments} to agents. But then, again,
what kinds of affordances/achievements precisely? Several distinct proposals
have been advanced in the literature, including \textit{welfare levels},
\textit{income levels,} \textit{consumption bundles of perfectly divisible
private goods}, \textit{capabilities} as subsets of a suitably defined space
of \textit{functionings, }and occasionally proposals to adjoin \textit{burdens
}to the family of relevant affordances.

The present work relies heavily on \emph{a specific definition of the relevant
affordance/achievement space as the (finite) set }$X$\emph{\ of all possible
capability-types (i.e.,\ combinations of levels of a finite family of
affordances/achievements, each one of them being represented by some finite
set of linearly ordered levels}\textbf{\ }at which it can be possibly made
available\emph{)}. We also rely on the working assumption that \emph{all the
affordances/achievements that are represented in }$X$\emph{, and their levels,
are both observable and verifiable characteristics.}\textit{\ }

While such a finiteness assumption is indisputably quite strong, it is
arguably the case that it is also very much consistent with any approach to
sufficientarian rules which is seriously concerned with their possible
practical applications. Anyway, \emph{it should be emphasized at the outset
that such a formulation of the affordance/achievement space plays a pivotal
role in this paper}\textit{.}

Concerning question (II), establishing what is `enough' clearly amounts to
defining some \textit{threshold} or \textit{threshold system }on the relevant
affordance/achievement space as provided by the answer to question (I). Once
such a threshold system has been properly specified, a \textit{benchmark} is
available to form and express the required \textit{judgments} and related
binary $1/0$ ratings of individual assignments of affordances/achievements. As
a result, we have precisely a basic sufficientarian \textit{benchmark-based
rating }rule that consists in a certain type of function that takes individual
affordance/achievement assignments to agents as inputs and returns a list of
$1/0$ ratings, one for each \textit{individual} assignment, as output. Indeed,
such a basic sufficientarian \textit{rating }rule provides immediately a most
natural \textit{simple }$1/0$ \textit{rating of assignments} themselves by
attributing rate $1$ precisely to those assignments which obtain rate $1$ for
\textit{each one }of their components, i.e., individual assignments to agents.
A supplementary, refined \textit{sufficiency-count rating }rule is also
immediately obtained by attaching to each affordance/achievement assignment
the rational number given by the \textit{ratio} between the number of agents
with a $1$-rated individual assignment and the number of agents of the entire
population under consideration. Of course, both a simple sufficientarian
ranking and a sufficiency-count ranking, respectively, can be immediately
defined relying on the aforementioned pair of sufficientarian ratings. It
should also be noticed, however, that there are not only critics but also
advocates of a sufficientarian view who, being committed to using
sufficientarian rankings as a guide to redistributive policies, reject or
regard anyway as highly disputable or simply implausible any use of
sufficiency-count ratings and/or their induced rankings in the latter
capacity, that they clearly see as a crucial one (see, e.g., Casal (2007) and
Huseby (2020), respectively).

More precisely, for any population $[n]:=\left \{  1,...,n\right \}  $ of $n$
agents, once an explicitly defined affordance/achievement space $\mathbf{A}$
endowed with some structure is in place, such a description of basic
sufficientarian rating rules as a particular subclass of functions
$f:\mathbf{A}^{n}\longrightarrow \left \{  0,1\right \}  ^{n}$ makes it possible
(at least in principle) to characterize them through properties that rely on
the very structure of $\mathbf{A}$, with \textit{no mention whatsoever of
thresholds}. And that is indeed the case when one picks our finite
capability-type space $\mathbf{X}$ as the relevant space, considers binary
rating functions $f:\mathbf{X}^{n}\longrightarrow \left \{  0,1\right \}  ^{n}$
on that space (which we denote here, following Balinski and Laraki (2011) as
\textit{binary grading functions (BGFs)), }defines a \textit{threshold system}
of $\mathbf{X}$ as a set of capability-types or vectors of $\mathbf{X}$ which
are \textit{pairwise incomparable} w.r.t. its component-wise order (i.e., form
an \textit{antichain }of that order), and relies on threshold systems thus
defined to define in turn (binary) \emph{sufficientarian grading
rules}\textit{\ }(or, equivalently, \emph{basic sufficientarian rating rules}:
henceforth, we shall use those two terms as synonyms)\textit{\ }as that
particular subclass of \textit{BGFs }which satisfy the following

\emph{(Sufficientarian BGF property)} there exists a \emph{threshold system}

$\  \  \  \  \  \  \  \  \  \  \  \  \  \  \  \  \  \  \  \  \  \  \  \  \  \  \  \  \mathcal{X}^{\ast
}:=\left \{  \mathbf{x}_{1}^{\ast},...,\mathbf{x}_{k}^{\ast}\right \}
\subseteq \mathbf{X}$ \  \  \  \  \  \  \  \  \  \  \  \  \  \  \  \  \  \  \  \  \  \  \  \  \  \  \  \  \ 

such that, for any capability-type component $\mathbf{x}_{i}$ of any
capability-type assignment $\mathbf{x}$ in $\mathbf{X}^{n}$, $f$ attaches
grade/rate $1$ to $\mathbf{x}_{i}$ if and only if $\mathbf{x}_{j}^{\ast
}\leqslant \mathbf{x}_{i}$ for some threshold $\mathbf{x}_{j}^{\ast}$ of the
given threshold system (namely, if and only if there exists \textit{at least
one} capability-type $\mathbf{x}_{j}^{\ast}$ of the given threshold system
such that $\mathbf{x}_{i}$ either exceeds or is equal to $\mathbf{x}_{j}%
^{\ast}$).

Specifically, one of the key results of the present work establishes that a
BGF $f$ $:\mathbf{X}^{n}\longrightarrow \left \{  0,1\right \}  ^{n}$ is a
(binary) \emph{sufficientarian grading rule} if and only if it satisfies three
quite natural and mutually independent properties namely:

\begin{itemize}
\item \emph{Isotony} (higher capability-levels are \textit{never} conducive to
lower grades/rates),

\item \emph{Separability} (for any possible assignment of capability-types the
grade/rate of any individual capability-type is not affected by other
individual capability-types of the same assignment, or to put it otherwise the
capabilities to be considered are indeed \textit{individual }capabilities),

\item \emph{Symmetry}\textit{\ (}for any possible assignment of
capability-types, the grade/rate of any individual capability-type is
independent of the agent it is assigned to, i.e., the threshold system is a
\textit{universal} one).
\end{itemize}

Moreover, characterizations of both the two-indifference-class or
\textit{simple sufficientarian total preorder} and the
\textit{sufficiency-count total preorder} induced by a basic sufficientarian
rating rule are also provided. Thus, precisely as claimed above, we do obtain
in fact \emph{characterizations of basic sufficientarian rating rules (and of
the simple and sufficiency-count sufficientarian ranking rules they induce)
that dispense entirely with any single property referring to thresholds either
explicitly or implicitly}\textbf{. }It should be remarked that this feature of
such characterizations puts them apart from almost any previous
characterization of sufficientarian (ranking) rules the authors are aware of
(see, e.g., Alcantud, Mariotti and Veneziani (2022), Bossert, Cato and Kamaga
(2022, 2023), Adler, Bossert, Cato and Kamaga (2025, 2026), Nakada and
Sakamoto (2024)) with \textit{a single partial exception} concerning the
\textit{special} case of \textit{sufficiency-count rankings }as characterized
in Chambers and Ye (2024), and its further restriction to the subclass of
\textit{limitarian }sufficiency-count rankings as characterized in Ferreira
and Savva (2025): more on this point is to follow in Section 2 below).

It is our contention that sufficientarian rating rules and (rating-based)
ranking rules thus characterized offer a sound \emph{minimal common core} for
any sort of \emph{sufficientarian stance}\textit{. }Arguably, a
sufficientarian ranking (or rating) rule is meant to establish whether the
actual assignment of affordances/achievements to agents \textit{does }satisfy
the appropriate sufficientarian benchmark, and if that is \textit{not} the
case it simply signals that \textit{some} remedial action or policy should be
considered and implemented. Such remedial policies should be carefully
designed according to distributive criteria that \textit{may} possibly, but
\textit{need not,} rely in turn on sufficientarian rather than, say, some sort
(or mixture) of \ (generalized) egalitarian and/or utilitarian criteria.

However, such an understanding is definitely \emph{not} the prevailing
attitude among proponents of sufficientarian principles (more details on that
and on what follows in Section 2 below). As a matter of fact, a large part of
the proponents of a sufficientarian view subscribe to the so-called
\emph{Negative Thesis} (`no redistribution needed among agents whose
individual assignments are located \textit{either above or on} the
sufficiency-threshold'). And virtually \textit{all of them} regard as an
essential part of a sufficientarian view acceptance of the so-called
\emph{Positive Thesis} (`remedial policies must \textit{prioritize}
improvements for agents whose individual assignments are located
\textit{below} the sufficiency-threshold'). Furthermore, \textit{many of them
}also insist on redistributive policies that \textit{the more} prioritize
improvements for the agents whose individual assignments are located
\textit{below} the sufficiency-threshold, \textit{the farther} their
individual assignments are located from that threshold. In other words, such
authors advocate a \textit{strong }version of sufficientarianism, invoking
some sort of \textit{sufficientarian rankings as pivotal criteria for the
required remedial policies} in order to ensure that the latter aim at some
kind of \textit{insufficiency minimization} (see, e.g., Huseby (2020), Timmer
(2022)). But then, such a strong version of sufficientarianism requires by
definition sufficientarian \textit{rankings} that are \textit{both} (much)
more refined than the two-class simple sufficientarian ranking \textit{and}
arguably \textit{different }from the \textit{sufficiency-count ranking, }which
as mentioned above is openly rejected as a sound guidance for redistributive
policies by many authors, including some advocates of sufficientarian
principles.\textit{\ }In particular, insufficiency minimization apparently
requires definition of a suitable metric over the space $\mathbf{A}^{n}$ of
affordance/achievement assignments (a definition which incidentally, and apart
from any other consideration, sufficiency-count strictly speaking does
\textit{not }provide\footnote{To see this, observe that for any assignment
$\mathbf{a=(a}_{1},...,\mathbf{a}_{n})$ in $\mathbf{A}^{n}$ and any nontrivial
permutation $\sigma$ of $\left \{  1,...,n\right \}  $, the resulting permuted
assignment {}$\mathbf{a}_{\sigma}:=\mathbf{(a}_{\sigma(1)},...,\mathbf{a}%
_{\sigma(n)})$ \ is such that $|\left \{  i\in \left \{  1,...,n\right \}
:f_{i}(\mathbf{a)=}1\right \}  |=|\left \{  i\in \left \{  1,...,n\right \}
:f_{i}(\mathbf{a}_{\sigma}\mathbf{)}=1\right \}  |$ yet, by construction,
$\mathbf{a\neq a}_{\sigma}$. Thus, the function $\delta^{SC}$ on
$\mathbf{A}^{n}\times \mathbf{A}^{n}$ defined by the rule $\delta
^{SC}(\mathbf{a,b):=|(}|\left \{  i\in \left \{  1,...,n\right \}  :f_{i}%
(\mathbf{a})=1\right \}  |-|\left \{  i\in \left \{  1,...,n\right \}
:f_{i}(\mathbf{b})=1\right \}  \mathbf{|}$ does not satisfy the `Identity of
Indiscernibles' condition of metrics (the other conditions to be satisfied by
metrics being of course non-negativity, identity recognition, symmetry, and
triangular inequality which are indeed satisfied by $\delta^{SC}$. It follows
that $\delta^{SC}$ is \textit{not }a metric (but, rather, just a
\textit{pseudometric}) on $\mathbf{A}^{n}$.}). And, again, a promising and
natural way to obtain such a sound `sufficientarian' metric is to rely on the
\textit{metric }structure of the affordance/achievement space $\mathbf{A}$
itself (if any such metric is available) and on its \textit{threshold system}
(once it has been specified). The present paper addresses that issue precisely
in that manner, proceeding by \textit{two steps: }

\emph{Step 1}: we rely on the natural, `intrinsic' metric structure of finite
capability-type space $\mathbf{X}$ as endowed with a fixed threshold system
$\mathcal{X}^{\ast}:=\left \{  \mathbf{x}_{1}^{\ast},...,\mathbf{x}_{k}^{\ast
}\right \}  $ in order to compute the distance of any capability-type from that
fixed threshold system. Such a metric arises immediately from the very
structure of $\mathbf{X}$ which is by definition a \textit{finite }cartesian
product of \textit{finite} linearly ordered sets. Indeed, $\mathbf{X}$ itself
can be endowed with the partial order $\leqslant$ induced component-wise by
its linear orders, and it can be easily checked that for every pair of
capability-types $\mathbf{x,y}$ of $\mathbf{X}$ both their least upper-bound
or \textit{join} $\mathbf{x\vee y}$ and their greatest lower-bound or
\textit{meet} $\mathbf{x\wedge y}$ are well-defined. Therefore, $(\mathbf{X}%
,\leqslant)$ is also a lattice $(\mathbf{X},\vee,\wedge)$ such that
$\mathbf{x\leqslant y}$ holds if and only if $\mathbf{x\vee y=y}$ or
equivalently $\mathbf{x\wedge y=x}$. Moreover, $(\mathbf{X},\leqslant)$ is
\textit{bounded} by construction, i.e., it is endowed with both a
\textit{maximum} and a \textit{minimum, }and as a product of linear orders it
is also \textit{distributive, }i.e., $\mathbf{x\vee(y\wedge z)=(x\vee
y)\wedge(x\vee z)}$, or equivalently $\mathbf{x\wedge(y\vee z)=(x\wedge
y)\vee(x\wedge z)}$ for all $\mathbf{x,y,z}$ $\in$ $\mathbf{X.} $\textit{\ }
But then, it can be shown that: (a) for any $\mathbf{x,y}$ of $\mathbf{X}$ the
\textit{length} $l(\left[  \mathbf{x\wedge y,x\vee y}\right]  ) $ of
\textit{interval} $\left[  \mathbf{x\wedge y,x\vee y}\right]  :=\left \{
\mathbf{z\in X}\text{: }\mathbf{x\wedge y\leqslant z\leqslant x\vee
y}\right \}  $, namely $k-1$ where $k$ is the size of any maximal chain (or
linearly ordered subset) included in that interval, is indeed a
\textit{uniquely defined} non-negative integer number, and (b) the function
$d:\mathbf{X\times X\longrightarrow}\mathbb{Z}_{+}$, defined by the rule
$d(\mathbf{x,y):=}l(\left[  \mathbf{x\wedge y,x\vee y}\right]  )$ for any
$\mathbf{x,y}$ $\in \mathbf{X}$, is in fact a well-defined \textit{metric }(see
also Section 3.3, and Barbut and Monjardet (1970), for the relevant details).
It follows that, relying on the arbitrarily fixed threshold system (or
antichain) $\mathcal{X}^{\ast}$ of $\mathbf{X}$, it is possible to attach a
distance $d^{\ast}(\mathbf{x},\mathcal{X}^{\ast}\mathbf{)}$ from threshold
system $\mathcal{X}^{\ast}$ to any capability-type $\mathbf{x}$ in
$\mathbf{X,}$ defined as the \textit{minimum} $d-$distance of $\mathbf{x}$
from a capability-type $\mathbf{x}_{j}^{\ast}$ of $\mathcal{X}^{\ast}$.

\emph{Step 2}: Moving now to an \textit{entire} capability-type assignment
$(\mathbf{x}_{1},...,\mathbf{x}_{n})$ in $\mathbf{X}^{n}$ we can uniquely
attach to that assignment the non-negative integer vector $(d^{\ast
}(\mathbf{x}_{1},\mathcal{X}^{\ast}\mathbf{),...,}d^{\ast}(\mathbf{x}%
_{n},\mathcal{X}^{\ast}\mathbf{)}$ of the respective extended $d$-distances of
its individual capability-types from threshold system $\mathcal{X}^{\ast}$. Of
course, we are interested precisely in the distance of each capability-type
assignment $(\mathbf{x}_{1},...,\mathbf{x}_{n})$ from threshold system
$\mathcal{X}^{\ast}$ and in the ranking (i.e., total preorder) over
$\mathbf{X}^{n}$ induced by such distances. To compute the latter distances of
capability-type \textit{assignments} from threshold system $\mathcal{X}^{\ast
}$we start precisely from the vector $(d^{\ast}(\mathbf{x}_{1},\mathcal{X}%
^{\ast}\mathbf{),...,}d^{\ast}(\mathbf{x}_{n},\mathcal{X}^{\ast})\mathbf{)}$
of (extended) distances from $\mathcal{X}^{\ast}$ of the individual
capability-types of such assignments, and proceed to an aggregation of the
components of that distance-vector in order to obtain a \textit{single
`summary' distance} of the entire assignment from $\mathcal{X}^{\ast}$ to be
\textit{minimized.} Such an aggregation can be made in several ways: we focus
on taking \textit{the sum }(or equivalently \textit{the average}) or
\textit{the lexicographic maximum} (or \textit{leximax})\textit{\ }of the
distances of individual assignments from $\mathcal{X}^{\ast}$, respectively,
and provide a characterization of the two \textit{total preorders} they induce
over $\mathbf{X}^{n}$.

Thus the present work also provides \emph{a characterization of two
sufficientarian ranking rules (defined through basic sufficientarian rating
rules) that might be deployed as insufficiency-minimization prioritizing
criteria for policy formulation or assessment.}

And finally, a further crucial issue that is rarely raised but lurks behind
\textit{any} serious attempt to advocate \textit{any} version of \textit{a
sufficientarian stance} is also addressed here: namely, \emph{selection of the
threshold system }itself. Indeed, the basic sufficientarian rating rules (and
the sufficientarian ranking rules based upon them characterized in the present
paper, or for that matter in other works) amount in fact to an entire
\textit{family of rules \ which is parameterized by the class of \ threshold
systems} as defined above. But then, it follows that in actual practice any
conceivable attempt to adopt and implement a sufficientarian rule requires
first and foremost \ the \textit{identification} and \textit{selection }of a
\textit{single, specific }universal threshold system. And, arguably, it also
follows that a full-fledged formulation of any version of a sufficientarian
stance should include an explicit discussion and presentation of some
well-behaved \textit{protocol }to be adopted by the relevant
\textit{deliberative bodies }in order to select the required specific
threshold system. Yet, to the best of the authors' knowledge, the extant
literature on sufficientarianism is remarkably elusive when it comes to the
issue of threshold identification, evoking sometimes the pivotal role of an
`impartial observer or spectator', or invoking some general criteria to be
used in actual practice to make sure the threshold is properly adapted to
specific features of the population of agents under consideration, and only
occasionally suggesting the opportunity of a collective choice of the
threshold by means of `fair' democratic procedures (see, e.g., Crisp (2003),
Hassoun (2021), Timmer (2022), respectively).

The present paper addresses that open issue from a plain mechanism design
perspective, \emph{establishing the existence of inclusive and
unanimity-respecting opinion aggregation rules (including inclusive quorum
systems and the simple majority rule) that are also strategy-proof on a large
and `natural' domain of single-peaked preferences over threshold systems. Each
one of such opinion aggregation rules can work as the key component of a
protocol to be used in order to select one specific sufficientarian grading
rule}\footnote{Or perhaps more than one, if and when required (more on that
topic in Section 2).}\textbf{\ }\emph{by choosing its characteristic threshold
system }\footnote{That result is obtained as a joint corollary to previous
results in Savaglio, Vannucci (2019) and Vannucci (2019) as combined with a
classic theorem due to Dilworth (1960) establishing that the set of antichains
of a finite partially ordered set is a distributive lattice under a very
natural order.}\textbf{. }

Summing up, the main contributions of the present paper can be described by
the following four points.

\begin{itemize}
\item \emph{(i) Definition of the affordance/achievement space as a
capability-type space given by a finite product of finite linearly ordered
sets;}

\item \emph{(ii) Characterizations of (binary) sufficientarian grading rules
and of the simple and sufficiency-count sufficientarian ranking rules they
induce, with no explicit or implicit reference to thresholds};

\item \emph{(iii)} \emph{Characterizations of two sufficiency-gap ranking
rules (the min-average and min-leximax sufficiency-gap rules, whose
definitions actually rely on sufficientarian grading rules), to be possibly
deployed as insufficiency-minimization criteria in the design and assessment
of remedial policies};

\item \emph{(iv) A mechanism-design-theoretic possibility result, establishing
the existence of anonymous, inclusive, unanimity-respecting selection
protocols for threshold systems that also enjoy a quite robust
strategy-proofness property, and can be effectively used in order to select a
specific sufficientarian rating and/or ranking rule thanks to the one-to-one
correspondence between threshold systems and sufficientarian grading rules.}
\end{itemize}

The relevance and significance of those four points will be further clarified
in the next section by discussing their relationships to the previous
literature on sufficientarianism.

The rest of the paper is organized as follows. Section 2 provides an extensive
yet selective review of the literature on sufficientarianism whose main aim is
to help the reader to locate and appreciate the marginal contribution of the
present work, and its underlying motivation and structure. Section 3 first
introduces the basic notation and definitions of the model, and characterizes
(binary) sufficientarian grading rules (or basic sufficientarian rating
rules), defined as a subclass of binary grading rules over capability-type
assignments. Then, some sufficientarian rankings are introduced and
characterized, including the sufficiency-count ranking, the min-average
sufficiency-gap and the min-leximax sufficiency-gap ranking. Section 4
addresses the issue concerning the definition and existence of well-behaved
protocols of threshold-selection. Section 5 offers some concluding remarks and
suggests a few possible extensions of binary grading functions to richer
capability-type spaces as a topic for future research. All the main proofs are
collected in an Appendix.

\bigskip

\section{Related literature}

Remarkably, recent contributions on sufficientarian principles and rules come
from the perspectives of quite distinct subdisciplines including social choice
theory, normative economics, political philosophy and social ethics (see,
e.g., Crisp (2003), Roemer (2004), Benbaji (2005, 2006), Huseby (2010, 2019,
2020), Axelsen and\ Nielsen (2015), Hirose (2016), Nielsen (2019, 2019b),
Timmer (2021, 2022), Alcantud, Mariotti and Veneziani (2022), Bossert, Cato
and\ Kamaga (2022, 2023), Chambers and Ye (2024), Nakada and Sakamoto (2024),
Harting (2024), Adler, Bossert, Cato and Kamaga (2025)).

Accordingly, several distinct perspectives and understandings on the aim and
scope of sufficientarian principles are proposed and subscribed to by
different authors. Some of them regard sufficientarian principles as a
comprehensive theory of distributive justice (or even, more generally, of
social ethics) to be contrasted with rival approaches such as, say,
(generalized) egalitarianism, (generalized) utilitarianism, or
prioritarianism, i.e., a variety of generalized utilitarianism which confers
some priority to the worse-off: see, e.g., Crisp (2003), Benbaji (2005, 2006),
Huseby (2010, 2020), Axelsen and Nielsen (2015), Nielsen and Axelsen (2017),
Nielsen (2019, 2019b), Herlitz (2019), Timmer (2021, 2022), and Harting (2024)
for a somewhat less demanding `hybrid' stance advocating the combination of
sufficientarian and `relational egalitarian' distributive criteria. As a
result, such contributions share two key features: (i) they typically take for
granted that sufficientarian rules are to be used and validated \textit{both
}as a general and convenient benchmarking device \textit{and} as the basic
guidance to remedial (re)distributive policies whenever actual
affordance/achievement assignments fail to satisfy the appropriate
sufficientarian benchmarks, and (ii) the discussions and articulations of
sufficientarian principles they propose are significantly shaped\ by the
intent to defend sufficientarianism against the criticisms advanced by
supporters of alternative principles of distributive justice, including
egalitarianism, prioritarianism and other versions of generalized
utilitarianism, or mixtures of sufficientarianism and other distributive
principles (see, e.g., Arneson (2005), Casal (2007) and Cohen (2011), Parfit
(1997), Brown (2005), Shields (2012) and Knight (2022), respectively).
Moreover, some authors that are also prepared to consider sufficientarianism
as a general principle of social and/or population ethics propose generalized
characterizations of sufficientarian rankings in a \textit{variable
population} setting (see, e.g., Bossert, Cato and Kamaga (2022, 2023), and
Hirose (2016) for an early suggestion in that vein).

Generally speaking, the body of relevant literature is by now considerable if
not vast, and the range of both issues of contention and analitical methods
deployed is also considerable. Clearly, this is not the place for a
comprehensive review or discussion of all of those issues. Moreover, the
present paper is only concerned with a detailed analysis and characterization
of sufficientarian rules as rating and ranking criteria for
capability-assignments without any underlying assumption about their status as
a (comprehensive or partial) theory of distributive justice, or even as a
prominent tool in guiding formation, selection and assessment of
redistributive policies.

Therefore, we shall mostly focus on those contributions that, either raising
or reacting to specific challenges to sufficientarian principles, have been
shaping some widely held understandings on the possibly critical blind spots
or open issues of any sufficientarian view, to which the present article is
meant to bring some clarification, or contribute a solution.

The most immediate source of inspiration for the literature on sufficientarian
principles comes apparently from Frankfurt (1987, 1997a, 1997b, 2015) who
advances what he calls `the doctrine of sufficiency' as an alternative to the
notion that `economic equality' should be treated as a \textit{basic}
normative distributive principle. His argument relies heavily on the view
(previously alluded to in the Introduction), that any \textit{basic} normative
principle should make use of \textit{absolute} judgements as opposed to
\textit{comparative} ones. Moreover, Frankfurt maintains that such a
sufficientarian alternative lends support to policies focussing on welfare
enhancement for those who \textit{have not enough}, including of course
poverty abatement. Such policies \textit{may} also, \textit{but }at least in
principle \textit{need not, }include economic inequality control and
mitigation. Notice that Frankfurt's indictment only concerns economic equality
as a \textit{fundamental} principle of social ethics, but is consistent with
advocacy of inequality abatement policies as an \textit{instrument} to pursue
other goals, including possibly intertemporal allocative \textit{efficiency}
to the extent that economic inequality is regarded as a \textit{negative
externality} (as explicitly suggested, e.g., by Stiglitz (2012), or Nyborg
St\o stad and Cowell (2024)). More recently, sufficientarianism has also been
advocated as an `indispensable' standpoint to cope with the ethical issues
concerning the proper treatment of the very badly off, whether or not due to
their own choices\footnote{It should be remarked that even `\textit{luck
egalitarianism' , }as defined and advocated by Cohen (2011) and others,
disavows egalitarian redress against disadvantages that are to be classified
as a result of deliberate choices of an agent, as opposed to just `bad luck'.}
(see Herlitz (2019)). Furthermore, it is also worth mentioning that a
sufficientarian stance of some sort is arguably gaining credit as a promising
forceful or even ultimately indispensable framework to address the serious
unemployment problems to be possibly expected in the near future, as a result
of the massive diffusion of new information technologies (including AI and its
burgeoning applications).

In order to gain and take advantage of a broader perspective on
sufficientarian views, it should be noticed, however, that there are at least
\textit{two major streams} of earlier contributions pointing to distribution
rules which embody some version of sufficientarian principles. One of them is
the advocacy of a \textit{universal basic income (UBI) } which can be traced
back to Russell (1918), and has been revived in several versions and under
several labels in the last few decades (see, e.g., Van Parijs
(1995)\footnote{It should be emphasized that Van Parijs's own favored
principle of distributive justice in order to achieve `real freedom for all'
can be described as `UBI maximization subject to undominated diversity' (see
Section 3 for a short presentation of the latter notion, than can be regarded
as a generalized egalitarian principle). Thus, strictly speaking Van Parijs
does not advocate a proper sufficientarian principle, but rather a
sufficientarian rule of sorts subject to a generalized egalitarian constraint.
It should be noticed, however, that the latter constraint might also be
construed as a clause requiring that everyone is granted `enough real
freedom'.}\textit{, }Widerquist (2024) among many others). The other one
originates from the famous distribution rule `\textit{from each according to
their ability, to each according to their needs}' due to Marx (1875), who
envisages that distribution rule as the one that would/should prevail within
what he calls `the realm of freedom' (namely, the advanced or mature
`communist' stage of a `socialist' society, that is indeed supposed to achieve
`real freedom' for all).

It is undoubtedly the case that the foregoing distinct strands of
\textit{`sufficientarian' }principles exhibit some remarkable mutual
differences. For instance, it is quite clear that `having enough' in
Frankfurt's sense does imply enjoying a `non-poor' status. But it is not
entirely clear whether or not the reverse is also the case, though Frankfurt
seems to be willing to distinguish between `having enough' and `being
non-poor'. By contrast, it is arguably quite clear that Marx's `realm of
freedom' is expected to admit (and in fact largely rely upon) his `communist'
distribution rule precisely because it is supposed to afford a considerable
degree of diffuse affluence, due to the implicit assumption of massively
improved technological basis and productive capabilities\footnote{Which does
not necessarily mean the achievement of some utopian `state of abundance',
however. An extra-bonus of our model based on finite capability-type spaces is
that it makes crystal clear that achievement of full sufficientarian
capability-type assignments does not amount to, or require, a state of
`abundance' in the proper sense of `lack of scarcity'. That hypothetical, and
arguably utopian, state of affairs would presumably make the full achievement
of sufficientarian goals very easy or even trivial, but it is \textit{not at
all} implied or required by the latter.}. In other terms, the `needs' that are
mentioned in Marx's rule are \textit{not }just the needs that must be
satisfied in order to escape the `poor' status:\ being `non-poor' does not
imply `having enough' or full-fledged `need-satisfaction' in Marx's own sense
(that is in fact typically connected to achievement of conditions enabling a
`flourishing of human personality'\footnote{Notice, however, that Marx does
\textit{not} single out any particular set of activities or behaviours as
either `typycally human' or intrinsically superior to others. Hence, Marx's
distribution rule for the `realm of freedom' does \textit{not} imply or rely
on a `\textit{perfectionist}' ethical stance, and the same arguably holds for
most current formulations of sufficientarianism.}). On the other hand, and
contrary to what is sometimes wrongly taken for granted, Marx's `communist'
distribution rule does \textit{not} require \textit{at all }unconditional
`economic equality'\textit{, }precisely as Frankfurt's `doctrine of
sufficiency'. Furthermore, both such Marxian distribution rule and Frankfurt's
sufficiency standard are meant to provide first and foremost
\textit{benchmarks }to support (binary) \textit{ratings }\footnote{It is worth
recalling that, in Marx's own view, implementation of his `communist'
distribution rule works as the underlying benchmark of an overarching, grand
binary rating of social progress in human societies. Indeed, Marx claims that
widespread adoption of such a distribution rule is the hallmark of nothing
less than `the end of human prehistory' and, accordingly, the
\textit{beginning }of human history (properly so said), \textit{not} its
\textit{end}.}\textit{\ }of social states, which in turn \textit{may but need
not be used }to define \textit{rankings} for guidance and assessment
of\textit{\ }remedial policies. In fact, Marx (1875) simply ignores the whole
issue of remedial policies while Frankfurt (1987) does take into consideration
remedial policies to deal with social states that \textit{fail} to satisfy the
\textit{sufficiency benchmark} and \textit{rankings} or criteria to shape such
policies, but firmly distinguishes such rankings from the sufficiency
benchmark itself.

Coming to \textit{UBI advocacy}, its key distinctive features are precisely
its definition as a certain fixed amount of units of a convenient medium of
exchange and payment (e.g., money), and its universality or unconditional
nature. Its cash--like unidimensional denomination is clearly at variance with
the more nuanced and possibly multidimensional character of the capabilities
everyone should have \textit{enough }access to according to Frankfurt's and
Marx's distribution rules \footnote{It should be mentioned however that,
apparently, Frankfurt does not rule out use of money-denominated thresholds
(see e.g. Frankfurt (1987), p. 37).}. And its unconditional nature would also
be possibly inconsistent with any interpretation of Marx's rule which insists
that its first component (`\textit{from each according to their ability')
}amounts in fact to a conditionality clause (an interpretation that is in our
view scarcely compelling, but at the same time not unconceivable\footnote{But
see the last section of this paper for more observations on that point, and on
the way `burdens' might be taken into account through a straightforward
extension of our capability-type space.}). Concerning the relationship of UBI
proposals to poverty and sufficiency thresholds or systems of thresholds, it
is arguably the case that its nature depends again on the specific version of
UBI one has in mind. Under some of its versions, UBI amounts to a
\textit{sufficiency guarantee} that goes possibly \textit{far beyond }the
minimum level required to escape poverty (see e.g. Van Parijs (1995)). But
there are versions of UBI proposals that insist to keep it low enough to
preserve adequately strong work incentives (perhaps even \textit{close }to the
relevant \textit{minimum no-poverty threshold, }as it\textit{\ is }arguably
the case for the original proposal advanced by Russell (1918)), and even
further minimalist proposals that also allow for an UBI whose level is located
\textit{below the poverty threshold} (see e.g. Widerquist (2024) for a
comprehensive discussion of several proposed versions of UBI).

As mentioned above, the recent literature on sufficientarianism itself is by
now quite extensive and focussed on a broad spectrum of issues ranging from
interpretations of its core-notions to formulation of an appropriate
axiomatization, with its underlying framework. To begin with, there any many
different views concerning the appropriate definition of the underlying
affordance/achievement space. Some authors assume it consists of
unidimensional \textit{welfare levels} under various interpretations including
possibly as lifetime well-being indicators (e.g., Alcantud, Mariotti and
Veneziani (2022), Bossert, Cato and Kamaga (2022, 2023), Hirose (2016), Huseby
(2020)), or even just \textit{income levels} (e.g., Frankfurt (1987),
Widerquist (2010)). Others insist on a \textit{multidimensional}
representation of the relevant space and propose to focus either on
\textit{capabilities} as subsets of a suitably defined space of
functionings\footnote{Along the lines of Sen (1985,1997). Notice that
capability-spaces are inherently multidimensional, and functionings (as
opposed to welfare levels or utilities) are typically meant to be
\textit{observable and verifiable.}} (e.g., Axelsen and Nielsen (2015),
Nielsen and Axelsen (2017)) or on \textit{prospects }consisting of
\textit{state-contingent welfare levels} (see Adler, Bossert, Cato and Kamaga
(2025, 2026)). Thus, both in the unidimensional and the multidimensional case
there are both proposals of affordance/achievement spaces consisting of
variables that are in principle observable and verifiable (money units,
capabilities) and proposals of affordance/achievement spaces that on the
contrary include variables whose values are \textit{private information }of
the agents and typically can only be accessed to by an appropriate
\textit{elicitation }protocol (welfare levels, prospects\textit{).}

A remarkable exception is provided by Chambers and Ye (2024) who work with
three sorts of affordance/achievement spaces and their respective
sufficiency-sets, namely (i) a set $A$ of \textit{indexes of unspecified
appropriate characteristics }with a \textit{sufficiency-set }$S$ consisting in
an arbitrary subset of $A$; (ii) a standard multidimensional commodity space
$A^{\prime}$ endowed with a \textit{preorder} $\preccurlyeq$ (i.e., a
reflexive and transitive binary relation), with a sufficiency-set $S^{\prime}$
given by an \textit{upward closed }subset of $A^{\prime}$ (i.e., a
\textit{preorder filter }of $(A^{\prime},\preccurlyeq)$) and (iii) a partially
ordered space $A^{\prime \prime}$ with a \textit{partial order} $\leqslant$
(i.e., a reflexive, transitive and antisymmetric binary relation) that is also
a \textit{meet-semilattice }(i.e., such that the greatest lower bound is
well-defined for any pair of elements of $A^{\prime \prime}$), with a
sufficiency-set $S^{\prime \prime}$ given by an upward closed subset of
$A^{\prime \prime}$ which is also meet-closed (i.e., a \textit{latticial }order
filter of $(A^{\prime \prime},\leqslant)$). It is worth noticing that all of
those three specifications of the affordance/space proposed in Chambers and
Ye's contribution formally qualify as \textit{generalizations }of the space of
capacity-types introduced and deployed in the present paper.

\emph{Yet, none of such sufficiency-sets proposed in Chambers and Ye (2024) do
qualify as a sound generalization of the sufficiency-sets induced by the
sufficientarian threshold systems we define and deploy in the present paper,
whose key property here is that they are in a one-to-one correspondence with
basic sufficientarian ratings and the sufficientarian rankings they
induce.}\textbf{\ }

That is so because a threshold system of a finite capability-type space
consists of the \textit{minimal elements} of an upward closed subset, i.e., of
an \textit{order filter} of the (ordered) set of capability-types: thus, the
sufficiency-set attached to an arbitrary threshold system is an order filter.
However, order filters cannot be defined in an unstructured set such as $A$.
Thus, generally speaking, a sufficiency-set $S$ of $A$ is \textit{not} an
order filter, its minimal elements are not defined, and $S$ has no threshold
system whatsoever attached to itself. Preorder filters can be defined in
$(A^{\prime},\preccurlyeq)$ and a sufficiency-set $S^{\prime}$ is indeed a
preorder filter\footnote{A preorder filter is defined in the obvious way as a
subset which is upward closed with respect to the given preorder (see e.g.
Savaglio and Vannucci (2007) for the introduction of preorder filters in a
study of opportunity inequality).}. However, without some further discreteness
condition on $A^{\prime}$\footnote{Such as No Bounded Infinite Chain or the
Descending Chain Condition (i.e., no infinite descending chain). Without any
such discreteness conditions $(A^{\prime},\preccurlyeq)$ may well be in fact a
collection of open half lines in $\mathbb{R}_{+}^{m}$ endowed with the
restriction of the `natural' partial order $\leq$ of $\mathbb{R}_{+}^{m}$ to
that very collection. In that case any subcollection of the foregoing
collection would be an order filter of $(A^{\prime},\preccurlyeq)$ with no
minimal points, hence with no threshold or threshold system attached to it.},
$S^{\prime}$ may well have \textit{no threshold or threshold system }of its
own at all. And finally, the order filters of semilattice $(A^{\prime \prime
},\leqslant)$ are by construction \textit{principal, }i.e. they have
invariably exactly \textit{one }minimal element, hence they only admit a
\textit{special} and \textit{trivial} sort of threshold system consisting of a
\textit{single }threshold\footnote{Thus, it is not surprising that, in order
to characterize sufficiency-count rankings on very general spaces without
relying on sufficientarian binary grading rules, Chambers and Ye (2024) have
to add a further axiom to monotonicity or isotony, separability and symmetry.
Yet, at the same time, the supplementary axiom employed, labeled
\textit{`}Sufficientarian Judgement\textit{', }is not specifically related to
thresholds. That it so because it only requires that if any uniform assignment
is worsened by a change of the individual assignment of a single agent then
any further change of the individual assignment of another single agent will
result in an assignment which is never a strict improvement of the previous
one (hence a fortiori it cannot compensate for the first change). Clearly,
that is in fact the behaviour of a preorder which relies on a target subset
$S$ and ranks assignments according to the sizes of the sets of agents whose
individual assignments belong to $S$. Nothing is said about $S$. For instance,
it can be induced by a threshold (sufficientarian case), by a cap (limitarian
case), or by a threshold and a cap (limitarian-sufficientarian case).
Arguably, `Sufficientarian Judgment' is in fact a misnomer for such a
condition. `Target Set Inclusion-Count Judgment' would make perhaps a more
accurate descriptive label for it.}.

Finally, it should also be recalled here that, as previously mentioned in the
Introduction, there are also some suggestions to the effect that the relevant
affordance space should include both achievements \textit{and burdens. }Those
suggestions come from both advocates and critics of sufficientarian views and
rules such as Nielsen (2019) and Knight (2022), respectively, who also concur
in regarding a proper treatment of burdens as a major challenge for
sufficientarian views. A proposal concerning a possible way to extend our
capability-type spaces in order to accommodate burdens will be discussed in
the final section of this paper.

As it should be expected, a significant amount of discussion and controversy
in the literature concerns the sufficiency-threshold itself as considered from
several perspectives: its role and meaning, the precise formulation and
characterization of the sufficientarian rules that are required to embody the
sufficiency-threshold, the role and aim of such sufficientarian rules and, to
a lesser extent, how sufficiency-thresholds could and/or should be determined.

Concerning the role of the sufficiency-threshold, its ostensible effect is to
obtain a bipartition of the underlying affordance/achievement space into two
blocks consisting of `sufficient' and `insufficient' states, respectively. And
that bipartition is precisely what is needed in order to state a sound
representation of the main content of a full-fledged sufficientarian view in
terms of what are widely regarded as its \textit{two basic theses}. Indeed, as
previously observed in the Introduction, most critics and many advocates
apparently concur on the description of sufficientarianism as the view that
there is a `sufficiency threshold' in the relevant space which satisfies the
following two conditions: (a) (`\textit{Positive Thesis}') since everyone
should have enough, priority must be given to those who stay below the
threshold and (b) (`\textit{Negative Thesis}') distributive considerations
concerning those who stay on or above the threshold are essentially irrelevant
(see Arneson (2005), Brown (2005), Casal (2007), Axelsen and Nielsen (2015),
Nielsen (2019, 2019b), Knight (2022), and also Crisp (2003) where the two
theses are conflated into a single `\textit{principle of compassion}'). But
then, critics insist that the existence of a threshold having such a twofold
property is most counterintuitive and implausible: hence it is not clear if
and how such a threshold can be successfully defined \textit{at all}. Anyway,
all of them maintain that specific `priority' criteria -possibly egalitarian
or generalized utilitarian- should replace, or at least be adjoined to,
`sufficiency' benchmarks in order to guide policies aimed at improving the
access to achievements of those agents whose assignment is located
\textit{below} the threshold (see e.g. Arneson (2005) and Casal (2007), or
Brown (2005) and Knight (2022), respectively).

The answers to that criticism on the part of authors who advocate some version
of a sufficientarian stance clearly reflect somewhat different meanings
attached to sufficiency-thresholds, and are accordingly remarkably varied.
Nielsen (2019b) suggests that the sufficiency-threshold should ensure that
\textit{all} `reasons of justice' are `\textit{completely} sated' once it is
reached, so that in a sense the former is to be placed so `high' in the
relevant affordance/capability space that the `Negative Thesis' is in fact
satisfied, but only \textit{trivially} so\footnote{Notice that under such an
interpretation the sufficiency-threshold also qualifies as an extreme,
trivialized version of a \textit{limitarian}-threshold or \textit{cap
}(namely, a `threshold' such that any individual assignment that is located
above it should be avoided because it is `\textit{too much}': see, e.g.,
Ferreira and Savva (2025) for a study and characterization of
\textit{limitarian-sufficientarian }ranking rules).}. On the contrary, some
authors rather propose a considerable relaxation of the `Negative Thesis'. In
particular,\ Shields (2012) suggests that the sufficiency-threshold also marks
a discontinuous \textit{shift} that merely \textit{weakens} (without
cancelling) the reasons to further benefit those agents who have reached it.
Benbaji (2005, 2006) proposes a \textit{multiplicity} of ordered thresholds
such that the `Positive Thesis' applies to each one of them, with a priority
that is the higher the lower their location in the affordance/achievement
space: satisfaction of the `Negative Thesis', however, is reinterpreted as
indifference between the individual assignments located \textit{between} any
two thresholds (see also Casal (2007), for a similar approach which relies on
just \textit{two} ordered thresholds). In a somewhat similar vein, Timmer
(2022) also allows for \textit{several} (ordered) thresholds, claiming that
`priority' criteria favoring improvements of affordance/achievement bundles
located below \textit{some} thresholds (with a further prioritization of
\textit{lower }thresholds amongst the latter) are to be regarded as an
essential component of sufficientarianism. Nakada and Sakamoto (2024)
introduce and characterize \textit{multi-threshold generalized sufficientarian
rankings} where several ordered thresholds are represented by suitably defined
`discontinuity points' of the ranking. Moreover, Huseby suggests (in Huseby
(2010) and in Huseby (2020), respectively) \textit{two distinct versions} of a
sufficientarian view which \textit{both} rely on \textit{two}
(ordered)\ thresholds. In the first version (Huseby (2010) the upper or
maximal sufficiency-threshold represents a welfare-level of
`\textit{subjective contentment'} to which both the `Positive Thesis' and the
`Negative Thesis' do apply, while the lower or minimal sufficiency-threshold
specifies a subsistence welfare level that corresponds to satisfaction of
`\textit{basic human needs'} and is only meant to signal possible priority in
favor of those agents whose individual assignments are located below the
maximal threshold (thus, the lower threshold works as a sort of poverty
threshold, but neither the `Positive Thesis' nor the `Negative Thesis' are,
strictly speaking, implied here). In the second version (Huseby (2020)) the
`Positive Thesis' only refers to the lower sufficiency-threshold, while the
`Negative Thesis' only refers to the upper sufficiency-threshold. Furthermore,
while invoking `prospect utilitarianism' (a certain type of utilitarian
ranking relying on non-expected utility functions) as an improvement upon
sufficientarian rankings, Cato and Chung (2026) argue that the best possible
version of a sufficientarian ranking should rely on the sufficiency threshold
and an additional upper threshold: The basic reason they advance to favor
their own proposal is that the `prospect-utilitarian' ranking is
\textit{continuous}, while the two-threshold sufficientarian ranking they
consider has a \textit{discontinuity} point at the sufficiency-threshold. It
is worth noticing here that the finite capability-type space framework used in
the present paper makes any such \textit{continuity}-argument simply
\textit{irrelevant} \footnote{That is so because in any finite or more
generally discrete setting, the natural topology is the discrete one, which
forces \textit{any} function to be continuous. Cato and Chung do acknowledge
explicitly this point (see Cato and Chung (2026), note 28, p.366). However,
they also apparently claim that such a consideration does only apply to
sufficiency-count rankings, and not to other `non-headcount' sufficientarian
rankings (which include of course our sufficiency-gap rankings). But,
generally speaking, the latter claim is simply wrong, since under the discrete
topology any binary relation (hence in particular any preorder) is also
continuous.}.

Arguably, such a proliferation of `thresholds' within a sufficientarian
framework and stance is scarcely surprising if not just unavoidable, since as
mentioned above several of its proponents regard the sufficiency-threshold as
the counterpart of a considerably `high' standard of living: a view which in
turn, at a minimum, invites a comparison with a supplementary and obviously
\textit{distinct} poverty-threshold or no-poverty-benchmark. It should also be
noticed, incidentally, that a sufficientarian grading rule as defined in the
present paper can be immediately deployed in poverty analysis, simply
reinterpreting its threshold system as a \textit{basic-needs\ }or
\textit{no-poverty-threshold}\footnote{As a matter of fact, other possible
applications of binary grading functions can also be envisaged, including the
analysis of \textit{exploitation}, and \textit{expertise.} Some more
observations on that point are made in the final section of the present
paper.}. But then, one might perhaps even suggest that a key difference
between the deployment of sufficientarian grading rules in sufficientarian and
poverty analysis, respectively, is precisely the fact that while the latter
only requires a \textit{single} sufficientarian binary grading rule, the
former actually requires \textit{at least two of them}, and possibly more. For
instance, one might also consider a further auxiliary \textit{upper} threshold
located above the sufficiency-threshold, marking its upper contour as the
region of the capability-type space that is a possible prioritary target of
certain strongly progressive taxation policies (thus providing a sort of
`limitarian' variety of the sufficientarian stance).

Concerning the actual formulation and characterization of the relevant
sufficientarian rules and criteria, the bulk of the literature is most recent
(see Alcantud, Mariotti and Veneziani (2022), Bossert, Cato and Kamaga (2022,
2023), Chambers and Ye (2024), Nakada and Sakamoto (2024), Ferreira and Savva
(2025), Adler, Bossert, Cato and Kamaga (2025, 2026), with Roemer (2004) and
Hirose (2016) as early precursors).

As it happens, and somewhat surprisingly, the sufficientarian rules or
criteria that \textit{all }of those contributions focus on, and proceed to
characterize, are sufficientarian\textit{\ rankings }and in particular, with
the single exception of Chambers and Ye (2024), \textit{social welfare
orderings }(i.e., essentially, single-profile social welfare functions in the
Bergson-Samuelson tradition:\ see, e.g., Gevers (1979), Roberts (1980a,1980b),
d'Aspremont (1985), Moulin (1988), and some earlier related works such as Sen
(1977), Hammond (1976, 1979), d'Aspremont and Gevers (1977), Deschamps and
Gevers (1978)). And, as previously mentioned, almost all of those
characterizations (with the single exception of Chambers and Ye (2024) as
previously discussed above) include an explicit reference to an exogenously
given threshold. Thus, such contributions tend to disregard the widely shared
view that, as mentioned above, a sufficientarian stance should rely first and
foremost on sufficientarian \textit{ratings} of affordance/achievement
assignments, and only derivatively on the rankings induced by those ratings.
As previously mentioned in the Introduction, \emph{one of the main aims of the
present paper is precisely, in that respect, to redress the state of things by
taking into consideration sufficientarian rating rules. We do so by
introducing (binary) sufficientarian grading rules, and characterizing them
and the sufficientarian rankings they induce without any reference whatsoever
to thresholds.}

Furthermore, while endorsement of the sufficientarian `Positive Thesis' as
defined above invites remedial policies whenever some agents fail to reach the
sufficiency-threshold, it is in fact unclear whether and to what extent
sufficientarian rankings of some sort are required to be used in order to
design and/or assess such policies. That is so especially because the
sufficientarian rankings that have been characterized (see, e.g., Alcantud,
Mariotti and Veneziani (2022), Chambers and Ye (2024)) are indeed versions of
the sufficiency-count ranking, whose possible use as a pivotal policy
criterion is bound to be controversial (see, e.g., Huseby (2020), among
others, for an explicit, flat rejection of such an use of the
sufficiency-count ranking). Thus, \emph{the explicit introduction and
characterization of sufficiency-gap rankings in the present paper fills indeed
a gap in the extant literature.}

Finally, we come to a last and absolutely crucial point involving
sufficiency-thresholds, and its discussion in the literature. Namely,
\textit{how sufficiency-thresholds are determined}. Clearly, that is
definitely a key issue, because any conceivable application of any
sufficientarian rule requires selection of (at least) \textit{one specific
threshold} (or threshold system). But then, how is that selection supposedly
made? Quite remarkably, the extant literature on sufficientarianism is almost
silent on that issue. As previously mentioned in the Introduction, one of the
few exceptions is Crisp (2003) who explicitly invokes the pivotal role of an
impartial observer or spectator in the threshold-selection process (and, more
precisely, the elicitation of her sentiment of compassion as the source of her
decision)\footnote{That is of course a rather transparent reference to the
approach to ethics due to Adam Smith's \textit{Theory of Moral Sentiments }and
partly to his `immediate' precursors, Hutcheson and Hume.}. In a slightly more
practical if similar vein, Hassoun (2021) suggests a \textit{`}%
mechanism\textit{' }to select a sufficiency-threshold regarded as `the
standard necessary for living a minimally good life': such a `mechanism'
consists in fact in a set of recommendations to guide the decisions of any
`reasonable, free, caring person' in charge of the threshold-selection
process. A definitely more operational approach is suggested by Timmer (2022)
who advances the notion of \textit{`political sufficientarianism' }which
amounts to selection of the sufficiency-threshold by means of `fair'
democratic procedures.

As discussed in the Introduction, \emph{the present paper addresses such
important issue of threshold-selection from a mechanism-design perspective,
and establishes the existence of nice and strategy-proof protocols which might
be deployed in order to accomplish that task. Such a result is obtained by
exploiting the structure of the capability-type space as a finite product of
finite linearly ordered sets, which makes it both a finite ordered set and a
finite distributive lattice}\textbf{, }a fact that in turn makes it possible
to rely on a few previous results\footnote{Mainly, Dilworth (1960), Savaglio
and Vannucci (2019), and Vannucci (2019).}\textbf{.} That is so because such a
space, as a finite ordered set, has order filters (upward closed subsets) the
collections of whose \textit{minimal} elements (which may be not unique, but
are of course by construction mutually incomparable or \textit{antichains})
amount in fact to \textit{threshold systems}. And as a distributive lattice,
it also admits nice aggregation rules with a remarkable strategy-proof
property which can be used as protocols to select a unique threshold system.

Preorders on an `opportunity space' with a single threshold defining the
top-ranked indifference class, and possibly several further ranked
indifference classes down below, are introduced and studied in Savaglio and
Vannucci (2007) and Vannucci (2013) under the label `filtral preorders' (since
of course any such threshold induces exactly one preorder filter). However,
the foregoing contributions focus on the special case of simple thresholds
consisting of a \textit{single `point'} in a multidimensional opportunity
space, which indeed correspond to the special subclass of \textit{principal
}order filters. Moreover, as already discussed in some detail above, the only
affordance/achievement space introduced by Chambers and Ye (2024) that is both
more general than our own capability-type space and supports a tight
connection of sufficiency-sets as order filters to thresholds is in fact a
meet-semilattice, and it is only \textit{latticial }(i.e., meet-closed) order
filters that are taken into consideration in the aforementioned paper. But any
latticial order filter having some minimal point is a \textit{principal} order
filter, i.e., it has a \textit{unique }minimal point hence only admits a
trivial threshold system consisting of a \textit{single }threshold.

The very same observation applies to the multidimensional thresholds
considered by Nielsen and Axelsen (2017) which amount indeed to
\textit{single} threshold-points of the (multidimensional) capability space
they propose to work with. General or multi-point threshold preorder or order
filters are also explicitly mentioned in Savaglio and Vannucci (2007) and, in
a slightly different yet obviously related `poverty ranking' setting, in
Peragine, Pittau, Savaglio and Vannucci (2021), but not studied in any detail.
Thus, to the best of the authors' knowledge, sufficientarian rules with
threshold systems consisting of nontrivial antichains of thresholds were
apparently never considered and examined in the previous literature on
sufficientarianism and related topics.

\bigskip

\section{\textbf{Model and results }}

In the present work sufficientarianism is described in terms of a class of
\textit{sufficientarian binary grading rules}, namely basic sufficientarian
\textit{rating} rules which amount to a special class of binary grading
functions of capability-types as defined below.

\subsection{\textbf{Notation and basic definitions: binary grading functions
and (binary) sufficientarian grading rules}}

Let $[n]:=\left \{  1,...,n\right \}  $ be a finite set of agents, and
$(\mathbf{X}_{i},\leqslant_{i})$ with $|\mathbf{X}_{i}|=l_{i}\in \mathbb{Z}%
_{+}\setminus \left \{  0\right \}  $, and $i\in \lbrack m]:=\left \{
1,...,m\right \}  $ the finite family of relevant
`positive'\textit{\ affordances }(and related \textit{achievements}), each one
of them consisting of a finite set $\mathbf{X}_{i}$ of levels ordered by a
linear order $\leqslant_{i}$ (i.e., a reflexive, connected, transitive and
antisymmetric binary relation). Borrowing and adapting the terminology from
Sen (1985, 1999) we denote as \textit{capability} a set of `positive'
affordances/achievements of the capability space to be defined below.
\textit{Capabilities }and their underlying \textit{affordances}%
/\textit{achievements}\footnote{Thus, in our model affordances/achievements
are in fact the counterparts of Sen's \textit{functionings }as constituent
units of capabilities.\textit{\ }} are meant to be \textit{observable }and
\textit{verifiable }attributes of agents. The affordances/achievements we are
going to consider are most typically non-rival, treated as strictly
\textit{individual} attributes\footnote{Capabilities of a collective or even
public nature that are related to access to public goods are thus essentially
ignored. That simplifying move is not meant to imply that public-good-related
capabilities (of which access to scientific knowledge as the output of basic
scientific research is a prominent example) are irrelevant or just not
amenable to treatment within our framework. On the contrary, a simple
adjustment of our model can accommodate capabilities of a public nature (more
on this in the final section of this paper).}, and an individual
capability-assignment to a certain agent is the relevant
\textit{capability-type}, or simply \textit{type}, of that agent. Accordingly,
we define the \textit{capability-type space} as $\mathbf{X}:=%
{\displaystyle \prod \limits_{i=1}^{m}}
\mathbf{X}_{i}$, and denote the finite partially ordered capability-type space
by $(\mathbf{X},\mathbf{\leqslant)}$, where $\mathbf{\leqslant:=}%
{\displaystyle \prod \limits_{i=1}^{m}}
\leqslant_{i}$\footnote{It should be noticed that, by construction, the basic
domain of entities to be assigned to agents essentially amounts to an
arbitrary (finite) partially ordered set. That is so because, though our
capability space is in fact a product of linearly ordered sets, it is well
known that any partially ordered set may be identified with its minimal
decomposition into (disjoint) linear orders.}. Let\ $\mathbf{X}^{n}$ be the
set of all conceivable capability-type-assignments to agents in $[n]$, and
$\left \{  0,1\right \}  $ the two relevant \textit{grades }that may be also
read as no/yes or false/true, respectively.

We refer to any $\mathbf{x}_{[n]}\in$ $\mathbf{X}^{n}$ as an
\textit{assignment of }\ capability-types, and for any such assignment and any
agent $i\in N$, we denote by $\mathbf{x}_{i}\in \mathbf{X}$ the individual
assignment of $i$ at $\mathbf{x}_{[n]}$. Within such a framework, the
information base of sufficientarianism is then a list of judgements on the
individual affordances/achievements asserting for each agent whether her/his
individual assignment is \textit{`sufficient' }or not. Any such judgement can
be expressed through a \emph{unary predicate} `\textit{being
sufficient\textquotedblright \ }that is\textit{\ }defined on capability-types
in $\mathbf{X,}$ and can in fact be identified with its truth-value (either
$1$ if true, or $0$ if false), which in the present context is in fact
precisely a \textit{grade}. Accordingly, a similar $0/1$ judgment can be
extended to a full \textit{assignment} $\mathbf{x}_{[n]}\in$ $\mathbf{X}^{n} $
of such capability-types relying precisely on the given list of judgements on
capability-types (one for each agent $i=1,...,n$). Thus, the judgement `$i $
\textit{has enough }at capability-assignment $\mathbf{x}_{N}$' amounts to the
equivalent judgement `capability-type $\mathbf{x}_{i}$ is sufficient' and its truth-value.

Therefore, the foregoing approach results in the definition of a particular
\emph{binary} \emph{grading function} (BGF) $g:$ $\mathbf{X}^{n}%
\rightarrow \{0,1\}^{n}$ that assigns to any conceivable assignment
$\mathbf{x}_{[n]}\mathbf{\in X}^{n}$ of capability-types to the $n$ agents the
list $g(\mathbf{x}_{[n]})=(g_{1}(\mathbf{x}_{[n]}),...,g_{n}(\mathbf{x}%
_{[n]}))$ of their respective $n$ binary values or \textit{grades}, one such
grade for every agent, that indicate whether or not the corresponding agent
`has enough' according to the capability-type assignment under
consideration\footnote{Sufficientarian grading functions with an arbitrary (or
possibly just bounded or finite) set of grades might also be considered, but
will be not in the present work.}. It should be noticed that such a binary
grading function amount to a special case of a \emph{social grading function},
a convenient tool to describe social states from a normative point of view
that has been recently and successfully advocated by Balinski and Laraki
(2007, 2011, 2014). A social grading function assesses each such social state
by a list of \emph{grades }chosen from a bounded linearly ordered set of
grades, as many grades as there are individual agents, and each grade - which
may be indeed a number as in our case- is intended to assess the social state
as far as the corresponding agent is concerned. The social grading function
returns an aggregate grade for each social state. Thus, in our case a binary
grading functions is precisely a social grading function having
capability-types as states and just two ordered grades. Which is of course
most appropriate since sufficientarianism, or `the doctrine of sufficiency' as
originally formulated by Frankfurt (1987), is precisely about whether or not
each agent `has enough'. Thus, it seems to be natural to restrict the possible
grades of individual assignments of capability-types to the set $\{0,1\}$
(e.g., an agent with such a capability-type `has not enough' or `has enough', respectively).

It should also be emphasized that a discussion of sufficientarian principles
in terms of BGFs is definitely at variance with the bulk of the extant
literature. Indeed, within the recent literature (see, e.g., Alcantud,
Mariotti and Veneziani (2022), Bossert, Cato and Kamaga (2022, 2023) and
Chambers and Ye (2024)), sufficientarianism has been examined through the
properties of a binary relation on the set of social states (namely, entire
affordances/achievements assignments), as opposed to an ordered bipartition of
that set of social states (assumed here to be $\mathbf{X}^{n}$).\ But, as
shown in the next section below, the present approach based on BGFs also makes
it possible to define several natural sufficientarian (total)
\textit{preorders }on $\mathbf{X}^{n}$.

Finally, it must be stressed that it is by no means the case that an arbitrary
binary grading function on $\mathbf{X}^{n}$ qualifies as a sound
\textit{sufficientarian} binary grading rule. To see this, consider the binary
grading function $g^{UD}:\mathbf{X}^{n}\rightarrow \{0,1\}^{n}$ defined as
follows: for any $\mathbf{x}_{[n]}\in \mathbf{X}^{n}$ and $i\in \lbrack n]$,%

\begin{align*}
g_{i}^{UD}(\mathbf{x}_{[n]})  & =0\text{ if there exists }h\in \lbrack n]\text{
such that }\mathbf{x}_{i}\text{ }\leqslant \mathbf{x}_{h}\text{ and }%
\mathbf{x}_{i}\text{ }\neq \mathbf{x}_{h}\\
& =1\text{ otherwise }%
\end{align*}

Clearly, $g^{UD}$ represents a version of the \textit{undominated diversity
(UD) }criterion (an egalitarian criterion discussed at length in Van Parijs
(1995)) which grades `$0$' an achievement/affordance type of an assignment if
there exists another achievement/affordance type of the same assignment that
weakly dominates it, and `$1$' otherwise (see Van Parijs (1995) for a detailed
discussion and advocacy of UD, and Basili and Vannucci (2013) for a
characterization).\ Clearly, in order to classify an achievement/affordance
assignment $g^{UD}$ relies on \textit{comparisons} between types of the
assignment. It follows that it cannot qualify as a \textit{sufficientarian}
binary grading rule, to the extent that on the contrary the latter is supposed
to rely on `absolute' assessments on the `sufficiency' of
achievement/affordance types at any given assignment. Thus, we must introduce
a list of properties of binary grading functions in order to obtain a
characterization of \textit{sufficientarian grading rules }as a proper
subclass of such functions.

\subsection{\textbf{A characterization of sufficientarian grading rules:
endogenizing thresholds}}

To begin with, let us observe that in order to avoid the need for clumsy
clauses or qualifications it is most convenient and quite natural to consider
a capability space that is large enough to include both a capability- type
that is sufficient for any agent and a capability-type that, on the contrary,
is not sufficient for any agent. Moreover, as already mentioned above, the
capabilities we are going to focus on are meant to be observable, verifiable
and \textit{individual} (see, however, the last section of the paper for a few
remarks on \textit{public }capabilities). Accordingly, we shall focus on those
binary grading functions (BGFs) $g:\mathbf{X}^{n}\rightarrow \left \{
0,1\right \}  ^{n}$ (as defined on arbitrary capability-assignments in
$\mathbf{X}^{n}$ where $N:=\left \{  1,...,n\right \}  $ is the set of relevant
agents) that are \textit{nontrivial }in the following sense: there exist
$\mathbf{y}_{[n]}$,$\mathbf{z}_{[n]}\in \mathbf{X}^{n}$ such that
$N_{1}(g(\mathbf{y}_{[n]}))=[n]$ and $N_{0}(g(\mathbf{z}_{[n]}))=[n]$ (where,
for any $\mathbf{x}_{[n]}\in \mathbf{X}^{n},$ $N_{1}(g(\mathbf{x}_{[n]}))$ and
$N_{0}(g(\mathbf{x}_{[n]}))$ denote the subsets of $1 $-graded and $0$-graded
agents at capability-assignment $\mathbf{x}_{[n]}$ according to $g$ , respectively).

Let us now introduce the properties of BGFs to be used in order to provide our
basic characterization of \textit{sufficientarian} \textit{binary grading
rules}.\medskip

\textbf{Definition 1 }(\textbf{Isotony) }A BGF $g:\mathbf{X}^{n}%
\rightarrow \left \{  0,1\right \}  ^{n}$\ is \emph{isotonic} if and only if for
any $\mathbf{x}_{[n]},\mathbf{x}_{[n]}^{\prime}\in \mathbf{X}^{n}$,
$\mathbf{x}_{[n]}\leqslant \mathbf{x}_{[n]}^{\prime}$ entails $g(\mathbf{x}%
_{[n]}\mathbf{)\leqslant}g(\mathbf{x}_{N}^{\prime})$\textbf{.}\medskip

\textbf{Definition 2 }(\textbf{Separability}) A BGF $g:\mathbf{X}%
^{n}\rightarrow \left \{  0,1\right \}  ^{n}$ is \emph{separable}\textit{\ }if
and only if for any $\mathbf{x}_{[n]},\mathbf{x}_{[n]}^{\prime}\in
\mathbf{X}^{n}$ and $i\in \lbrack n]$, if $\mathbf{x}_{i}=\mathbf{x}%
_{i}^{\prime}$ entails $(g(\mathbf{x}_{[n]}\mathbf{))}_{i}=\mathbf{(}%
g(\mathbf{x}_{[n]}^{\prime}))_{i}$\textbf{.}\medskip

\textbf{Definition 3 }(\textbf{Symmetry) }A BGF $g:\mathbf{X}^{n}\rightarrow$
is \emph{symmetric}\textit{\ }if and only if for any $\mathbf{x}_{[n]}%
\in \mathbf{X}^{n}$, and permutation $\sigma:[n]\longrightarrow \lbrack n]$ ,
$g(\mathbf{x}_{\sigma \lbrack n]}\mathbf{)=\sigma(}g(\mathbf{x}_{[n]}%
))$\textbf{.}\medskip

\textit{Isotony} of a BGF is meant to capture the fact that the components of
a capability-type are uncontroversially positively correlated with well-being,
and \textit{Separability} captures the understanding that every component of a
capability type denotes a \textit{strictly individual characteristic} of an
arbitrary agent. \textit{Symmetry }amounts to an \textit{universality}
requirement concerning the assessment standards used in order to establish
what is and what is not `\textit{sufficient}': notice that (as any such
standard amounts to a \textit{system }of several distinct admissible
thresholds, but Symmetry requires that system to be \textit{unique }and apply
to\textit{\ everyone. }

It can be easily shown, and left to the reader to check, that a nontrivial
isotonic, separable and symmetric BGF is also \emph{onto.}\medskip

\textbf{Definition 4. }A BGF $g:\mathbf{X}^{n}\rightarrow \left \{  0,1\right \}
^{n}$ is a (binary)\emph{\ sufficientarian grading rule }if and only if there
exist a positive integer $k$ and $\mathbf{x}_{1}^{\ast},...,\mathbf{x}%
_{k}^{\ast}\in \mathbf{X}$ such that $\mathbf{X}^{\ast}=\left \{  \mathbf{x}%
_{1}^{\ast},...,\mathbf{x}_{k}^{\ast}\right \}  $ is a \emph{threshold system},
i.e., an \emph{antichain}\textit{\ }of $\mathbf{X}$ (namely $\mathbf{x}%
_{j}^{\ast}$ $\nleqslant \mathbf{x}_{h}^{\ast}$ for every $j,h=1,...,k$ with
$j\neq h$) and for every $\mathbf{x}_{[n]}\mathbf{\in X}^{n}$, and
$i\in \lbrack n]$, $g_{i}(\mathbf{x}_{[n]}\mathbf{)=}1$ if and only if
$\mathbf{x}_{j}^{\ast}\leqslant \mathbf{x}_{i}$ for some $j=1,...,k$. We denote
by $\mathcal{S}(\mathbf{X}^{n})$ the class of all sufficientarian grading
rules on $\mathbf{X}^{n}$.\medskip

\textbf{Proposition 1. }Let $g:\mathbf{X}^{n}\rightarrow \left \{  0,1\right \}
^{n}$ be a nontrivial BGF. Then, $g$ is a sufficientarian binary grading rule
if and only if it is isotonic, separable and symmetric.\medskip

\textbf{Remark 1. }The foregoing characterization is tight, since Isotony,
Separability and Symmetry are independent properties (see the Appendix for the
relevant details). Incidentally, observe that the UD binary grading function
$g^{UD}$ introduced in the previous Subsection satisfies both Isotony and
Symmetry, but fails to satisfy Separability.

Notice that any sufficientarian binary grading rule \ $g:\mathbf{X}%
^{n}\rightarrow \left \{  0,1\right \}  ^{n}$ induces by itself a
\textit{sufficientarian judgment} on -or equivalently a
\textit{sufficientarian classification} of- capability profiles in
$\mathbf{X}^{n}$ by the following rule: a capability-profile $\mathbf{x}%
_{[n]}\in \mathbf{X}^{n}$ is $g$-\textit{sufficient }if and only if
$g_{i}(\mathbf{x}_{[n]})=1$ for every $i\in \lbrack n]$. Clearly enough, such a
sufficientarian judgment can be also represented as a \textit{`simple'}
\textit{total preorder} $\widehat{\succcurlyeq}_{g}$ with \textit{precisely
two} indifference classes thanks to nontriviality (indeed, ontoness) of $g$,
namely for any $\mathbf{x}_{[n]},\mathbf{x}_{[n]}^{\prime}\in \mathbf{X}^{n}$
$\mathbf{x}_{[n]}\widehat{\succcurlyeq}_{g}\mathbf{x}_{[n]}^{\prime}$ if and
only if either $\mathbf{x}_{[n]}$ is $g$-\textit{sufficient} or $\mathbf{x}%
_{[n]}^{\prime}$ is \textit{not }$g$-\textit{sufficient}. Arguably, such a
`simple' total preorder is \textit{already }a sound and precise reformulation
of the basic sufficientarian judgement as a ranking criterion.

However, the rest of the literature on sufficientarianism has been focussing
on rankings, namely total preorders of $\mathbf{X}^{n}$ with no explicit
limitations on the number of their admissible indifference classes (except
possibly the size of the population $[n]$ of agents). But then, a further
`\textit{refined' sufficiency-count sufficientarian }total preorder
$\succcurlyeq_{g}$on $\mathbf{X}^{n}$ can be defined through $g$ by the
following most natural rule:\ for any $\mathbf{x}_{[n]},\mathbf{x}%
_{[n]}^{\prime}\in \mathbf{X}^{n}$,%
\[
\mathbf{x}_{[n]}\succcurlyeq_{g}\mathbf{x}_{[n]}^{\prime}\text{ if and only if
}|\left \{  i\in \lbrack n]:g_{i}(\mathbf{x}_{[n]})=1\right \}  |\geq|\left \{
i\in \lbrack n]:g_{i}(\mathbf{x}_{[n]}^{\prime})=1\right \}  |\text{,}%
\]
(or, equivalently, $\frac{|\left \{  i\in \lbrack n]:g_{i}(\mathbf{x}%
_{[n]})=1\right \}  |}{n}\geq \frac{|\left \{  i\in \lbrack n]:g_{i}%
(\mathbf{x}_{[n]}^{\prime})=1\right \}  |}{n}$)\footnote{For any set $Y$, $|Y|$
denotes the cardinality or size of $Y$.}.

Counterparts of the foregoing preorder have been indeed presented and
discussed in the recent literature (notably, Alcantud, Mariotti and Veneziani
(2022), and Nakada and Sakamoto (2024)). A characterization of the basic
sufficientarian preorders $\widehat{\succcurlyeq}_{g}$, $\succcurlyeq_{g}$will
be provided below in the next subsection.

\subsection{\textbf{The sufficientarian total preorders induced by a
sufficientarian grading rule through direct extension: characterizations}}

Let us now proceed from basic sufficientarian \textit{ratings} to the
sufficientarian \textit{rankings} they induce on the space of capability-type assignments.

To begin with, observe that since $\left \{  0,1\right \}  ^{n}$ is the boolean
$n$-hypercube endowed with its own partial order $\mathbf{\geqslant} $, any
sufficientarian grading rule $g:\mathbf{X}^{n}\rightarrow \left \{  0,1\right \}
^{n}$ induces a unique $(\left \{  0,1\right \}  ^{n},\geqslant)$-monotonic and
$g$- consistent \textit{partial order }$\geqslant_{g}$ on $\mathbf{X}^{n}$ by
the following most obvious rule:\ for any $\mathbf{x}_{[n]},\mathbf{x}%
_{[n]}^{\prime}\in \mathbf{X}^{n}$, $\mathbf{x}_{[n]}\geqslant_{g}%
\mathbf{x}_{[n]}^{\prime}$ if and only if $g(\mathbf{x}_{[n]})\geqslant
g(\mathbf{x}_{[n]}^{\prime})$. It should also be stressed that \textit{\ }%
$\geqslant_{g}$ is a `topped partial', namely a partial order with a (unique)
\textit{maximum }whenever $g$ is onto\textit{.} Let us now recall, for the
sake of convenience, the simple and the sufficiency-count sufficientarian
preorders $\widehat{\succcurlyeq}_{g}$ and $\succcurlyeq_{g}$ as introduced
above in the previous section.$\mathbf{\medskip}$

\textbf{Definition 5 }(\emph{Simple sufficientarian preorder})\textbf{.} Let
$g:\mathbf{X}^{n}\rightarrow \left \{  0,1\right \}  ^{n}$ be a sufficientarian
grading rule. Then, the simple sufficientarian preorder $\widehat
{\succcurlyeq}_{g}$ induced by $g$ is defined as follows: for any
$\mathbf{x}_{[n]},\mathbf{x}_{[n]}^{\prime}\in \mathbf{X}^{n}$, $\mathbf{x}%
_{[n]}\widehat{\succcurlyeq}_{g}\mathbf{x}_{[n]}^{\prime}$ if and only if
either $g_{i}(\mathbf{x}_{[n]})=1$ for all $i\in \lbrack n]$, or $g_{i}%
(\mathbf{x}_{[n]}^{\prime})=0$ for some $i\in \lbrack n]$.$\mathbf{\medskip} $

\textbf{Definition 6 }(\emph{Sufficiency-count preorder}). Let $g:\mathbf{X}%
^{n}\rightarrow \left \{  0,1\right \}  ^{n}$ be a sufficientarian grading rule.
Then, the sufficiency-count preorder $\succcurlyeq_{g}$ induced by $g$ is
defined as follows: for any $\mathbf{x}_{[n]},\mathbf{x}_{[n]}^{\prime}%
\in \mathbf{X}^{n}$, $\mathbf{x}_{[n]},\mathbf{x}_{[n]}^{\prime}\in
\mathbf{X}^{N}$, $\mathbf{x}_{[n]}\succcurlyeq_{g}\mathbf{x}_{[n]}^{\prime} $
if and only if $|\left \{  i\in \lbrack n]:g_{i}(\mathbf{x}_{[n]})=1\right \}
|\geq|\left \{  i\in \lbrack n]:g_{i}(\mathbf{x}_{[n]}^{\prime})=1\right \}
|$.$\mathbf{\medskip}$

It can be easily shown (and left to the reader to check) that both
$\widehat{\succcurlyeq}_{g}$ and $\succcurlyeq_{g}$ are in fact extensions of
the partial order $\geqslant_{g}$ to a \textit{total preorder }on
$\mathbf{X}^{N}$. Moreover, $\widehat{\succcurlyeq}_{g}$ $\mathbf{\supseteq
}\succcurlyeq_{g}$ namely $\widehat{\succcurlyeq}_{g}$ is indeed a
\textit{coarser }extension of $\geqslant_{g}$ than $\succcurlyeq_{g}$, and is
in fact a nontrivial extension of $\geqslant_{g}$ to a total preorder on
$\mathbf{X}^{N}$, because $\widehat{\succcurlyeq}_{g}\neq \mathbf{X}^{N}%
\times \mathbf{X}^{N}$ by nontriviality of $g$. The following definitions and
Claim make it precise in what sense it is also a most `natural' extension of
$\geqslant_{g}$.$\mathbf{\medskip}$

\textbf{Definition 7 }(\emph{The top class of a BGF}) Let $\ g:\mathbf{X}%
^{n}\rightarrow \left \{  0,1\right \}  ^{n}$ be a BGF. Then the top class of $g$
is $\ top_{g}(\mathbf{X}^{n}):=g^{-1}(\mathbf{1})$.$\mathbf{\medskip}$

Notice that $\mathbf{X}^{n}\neq top_{g}(\mathbf{X}^{n})\neq \emptyset$ whenever
$g$ is (as in our case) a \textit{nontrivial} BGF.$\mathbf{\medskip}$

\textbf{Definition 8 }(\emph{Top-faithful sufficientarian preorders}\textit{).
}Let \ $g:\mathbf{X}^{n}\rightarrow \left \{  0,1\right \}  ^{n}$ be a
sufficientarian grading rule, and $\succcurlyeq$ a total preorder on
$\mathbf{X}^{n}$ that is an extension of $\geqslant_{g}$, namely
$\  \geqslant_{g}\subseteq \succcurlyeq$. Then, $\succcurlyeq$ is a top-faithful
(sufficientarian) preorder induced by $g$ if $max(\succcurlyeq)=top_{g}%
(\mathbf{X}^{n})$.$\mathbf{\medskip}$

Thus, in plain words, a total preorder over the set $\mathbf{X}^{n}$ of all
possible assignments of capability-types to agents that extends the partial
order induced on $\mathbf{X}^{n}$ by a sufficientarian grading rule is
top-faithful whenever the set of its maximal elements is precisely the top
class of $g$. The largest of such total preorders is indeed the simple
sufficientarian preorder, as made precise by the following (second-order)
characterization.$\mathbf{\medskip}$

\textbf{Claim 1 }Let $g:\mathbf{X}^{n}\rightarrow \left \{  0,1\right \}  ^{n}$
be a sufficientarian grading rule. Then, the simple sufficientarian preorder
$\widehat{\succcurlyeq}_{g}$ is the \textit{coarsest }top-faithful extension
of $\geqslant_{g}$ to a total preorder $\succcurlyeq$ on $\mathbf{X}^{n}%
$.$\mathbf{\medskip}$

Let us now consider the \textit{sufficiency-count} total preorder
$\succcurlyeq_{g}$ induced by a sufficientarian grading rule $g$. To begin
with, observe that $\succcurlyeq_{g}$is also a top-faithful sufficientarian
preorder induced by $g$. Next, in order to proceed to a characterization of
$\succcurlyeq_{g}$ we introduce the following properties for total preorders
$\succcurlyeq$ on $\mathbf{X}^{n}$.\footnote{We denote with $\sim$ and $\succ$
the symmetric and asymmetric components of $\succcurlyeq$, respectively.}%
$\mathbf{\medskip}$

\begin{itemize}
\item ($\mathbf{X}$-\emph{Isotony} ($\mathbf{X}$-\emph{IS})) A total preorder
$\succcurlyeq$ on $\mathbf{X}^{n}$ satisfies $\mathbf{X}$-IS iff for any
$\mathbf{x}_{[n]},\mathbf{x}_{[n]}^{\prime}\in \mathbf{X}^{n}$, $\mathbf{x}%
_{[n]}^{\prime}\leqslant \mathbf{x}_{[n]}$ entails $\mathbf{x}_{[n]}%
\succcurlyeq \mathbf{x}_{[n]}^{\prime}.$
\end{itemize}

$\boldsymbol{X}$-Isotony just express the understanding that we are focusing
here on assignments of access to `achievements' as opposed of, say, `burdens':
thus, more access to achievements cannot results in a lower ranking position.

\begin{itemize}
\item (\emph{Anonymity (AN)}) A total preorder $\succcurlyeq$ on
$\mathbf{X}^{n}$ is \textit{anonymous} if and only if, for any $\mathbf{x}%
_{[n]}\in \mathbf{X}^{n}$, and any permutation $\sigma:[n]\longrightarrow
\lbrack n]$, $\mathbf{x}_{[n]}\sim \mathbf{x}_{\sigma \lbrack n]}.$
\end{itemize}

Anonymity simply requires that the ranking of a capability-type assignment
+does \textit{not }depend on the identity of the agents its capability-types
are assigned to.

It can be easily shown -and left to the reader to check- that Anonymity and
$\mathbf{X}$\textit{-}Isotony\textit{\ }are both satisfied by the simple
sufficientarian preorder $\widehat{\succcurlyeq}_{g}$ \textit{and }the
sufficiency-count preorder $\succcurlyeq_{g}$induced by an arbitrary
sufficientarian binary grading rule on $\mathbf{X}^{n}$.

On the contrary, it can also be easily checked that the following property is
satisfied by sufficiency-count preorders but \textit{not }by simple
sufficientarian preorders induced by sufficientarian grading
rules.$\mathbf{\medskip}$

\textbf{Definition 9 }(\emph{Strict Monotonicity with respect to} $g$
(\emph{SM}($g$))). Let $g:\mathbf{X}^{n}\rightarrow \left \{  0,1\right \}  ^{n}$
be an onto binary grading function and $\succcurlyeq$ a total preorder on
$\mathbf{X}^{n}$ which is an extension of the partial order $\geqslant_{g}$on
$\mathbf{X}^{n}$, and $\mathbf{x}_{[n]},\mathbf{x}_{[n]}^{\prime}\in
\mathbf{X}^{n}$ and $i\in \lbrack n]$ be such that $\ g_{l}(\mathbf{x}%
_{[n]})=g_{l}(\mathbf{x}_{[n]}^{\prime})$ for any $l\in \lbrack n]\backslash
\left \{  i\right \}  $, $g_{i}(\mathbf{x}_{[n]})=1$ and $g_{i}(\mathbf{x}%
_{[n]}^{\prime})=0$, then $\mathbf{x}_{[n]}\succ \mathbf{x}_{[n]}^{\prime}%
$.$\mathbf{\medskip}$

\bigskip

\textbf{Proposition 2 \ }A total preorder $\succcurlyeq$ on $\mathbf{X}^{N}$
is an extension of the partial order $\geqslant_{g}$that satisfies AN and
SM($g$) if and only if $\succcurlyeq=\succcurlyeq_{g}$.$\mathbf{\medskip}$

\bigskip

\textbf{Remark 2.} It should be noticed that the simple sufficientarian
preorder $\widehat{\succcurlyeq}_{g}$ does satisfy AN, but fails to satisfy
SM($g$) whenever $n\geq2$ (indeed, if $n=1$ the simple sufficientarian
preorder and the sufficiency-count preorder do coincide). Thus, for $n\geq2$,
the characterization of the sufficiency-count preorder $\succcurlyeq_{g}$
provided by Proposition 2 is indeed tight (since of course any
\textit{projection} of $g$ induces a total preorder on $\mathbf{X}^{n}$ which
extends partial order $\geqslant_{g}$ and satisfies SM($g$) but violates AN
whenever $n\geq2$).

The sufficiency-count preorder $\succcurlyeq_{g}$ may be regarded as a special
case of the cardinality total preorder of opportunity sets, and could also be
characterized along the same lines of Pattanaik \ and Xu (1990) where three
axioms are used: Indifference between no-choice situations, Independence, and
Strict monotonicity. Essentially, our own characterization replaces their
first two axioms with Anonymity, which is arguably much more `natural' in the
present and more specific setting.

\bigskip

\subsection{Sufficiency-gap total preorders induced by a sufficientarian
grading rule: some characterizations}

Clearly enough, the sufficiency-count total preorder $\succcurlyeq_{g}$ on
capability-type assignments induced by a sufficientarian binary grading rule
$g$ amounts to an analogue of the head-count based poverty ranking. This
simple observation obviously invites the introduction of a sufficiency-gap
ranking counterpart of poverty-gap rankings as an alternative refinement of
the simple sufficientarian total preorder. \newline We introduce a
sufficiency-gap ranking by jointly exploiting the structure of the
capability-type space $\mathbf{X}$ \textit{and} the characteristic threshold
system $\mathcal{X}^{\ast}:=\left \{  \mathbf{x}_{1}^{\ast},...,\mathbf{x}%
_{k}^{\ast}\right \}  $ of the sufficientarian binary grading rule $g$. Indeed,
the capability-type space $(\mathbf{X},\mathbf{\leqslant)}$ with $\mathbf{X}:=%
{\displaystyle \prod \limits_{i=1}^{m}}
\mathbf{X}_{i}$ is by construction a product of bounded linearly ordered sets
and is therefore endowed with a natural \textbf{metric} which can be defined
in several equivalent ways. In what follows, we shall introduce that metric
relying on \emph{shortest paths of the covering graph} of $\mathbf{X}$, as
explained below. In order to proceed, a few new definitions are now
required.$\mathbf{\medskip}$

\textbf{Definition 10. }A \emph{chain}\textit{\ }of poset $(\mathbf{X}%
,\leqslant)$ is a set $\mathbf{Y}\subseteq \mathbf{X}$ such that for any
\textit{distinct} $\mathbf{u,v}\in \mathbf{Y}$ either $\mathbf{u\leqslant v}$
or $\mathbf{v\leqslant u}$ holds, and its \textit{length }$l(\mathbf{Y})$ is
$|\mathbf{Y}|-1$. A chain $\mathbf{Y}$ of $(\mathbf{X},\leqslant)$ having $x
$\textbf{\ }as its $\leqslant$-minimum and $\mathbf{y}$ as its $\leqslant
$-maximum\ is \textit{maximal }if there is no $\mathbf{z\in X\setminus Y}$
such that $\mathbf{x\leqslant z\leqslant y}$.$\mathbf{\medskip}$

\textbf{Definition 11.\ }An \emph{antichain}\textit{\ }of poset $(\mathbf{X}%
,\leqslant)$ is a set $\mathbf{Y\subseteq X}$\textbf{\ }such that for any
\textit{distinct} $\mathbf{u,v\in Y}$ neither $\mathbf{u\leqslant v}$ nor
$\mathbf{v\leqslant u}$ hold.$\mathbf{\medskip}$

\textbf{Definition 12.\ }For any $\mathbf{x,y\in X}$ such that $\mathbf{x<y}$
(i.e. $\mathbf{x\leqslant y}$ and \textit{not} $\mathbf{y\leqslant x}$) the
\emph{length}\textit{\ }of the order-interval $[\mathbf{x,y}]:=\left \{
\mathbf{z\in X:x\leqslant z\leqslant y}\right \}  $\textbf{,} written
$l([\mathbf{x,y}])$, is the length of a (maximal) chain of \textit{maximum
}length having $\mathbf{x}$ as its $\leqslant$-minimum and $\mathbf{y}$ as its
$\leqslant$-maximum. In particular, $\mathbf{x\in X}$ is said to be
\emph{covered}\textit{\ }by $\mathbf{y\in X}$, written $\mathbf{x\ll y}$, iff
$\mathbf{x<y}$ and $[\boldsymbol{x,y}]=\left \{  \mathbf{x,y}\right \}  $,
namely $l(\left[  \mathbf{x,y}\right]  )=1$.$\mathbf{\medskip}$

\textbf{Definition 13.\ }The \emph{covering graph}\textit{\ }$C(\mathbf{X}%
):=(\mathbf{X},E^{\ll})$ of $\mathbf{X}$ is the undirected graph having
$\mathbf{X}$ as vertex-set and $E^{\ll}:=\left \{  \left \{  \mathbf{x,y}%
\right \}  \subseteq \mathbf{X}:\mathbf{x\ll y}\text{ or }\mathbf{y\ll
x}\right \}  $ as edge-set.$\mathbf{\medskip}$

\textbf{Definition 14.\ }A \emph{path}\textit{\ }$\pi_{\mathbf{xy}\text{ }}$of
$C(\mathbf{X})$ connecting two vertices $\mathbf{x}$ and $\mathbf{y}$ is a
maximal chain $\left \{  \mathbf{z}_{0}\mathbf{,....,z}_{k}\right \}  $ of
$\  \mathbf{X}$ such that $\left \{  \mathbf{z}_{0}\mathbf{,z}_{k}\right \}
=\left \{  \mathbf{x,y}\right \}  $ and $\mathbf{z}_{i}\mathbf{\ll z}_{i+1}$,
for any $i=1,...,k-1$, and is of \textit{length }$l(\pi_{\mathbf{xy}}%
)=k$\textit{. }The set of all paths of $C(\mathbf{X})$ connecting $\mathbf{x}%
$\textbf{\ }and $\mathbf{y}$ is denoted by $\Pi_{\mathbf{xy}}$%
.$\mathbf{\medskip}$

\textbf{Definition 15.\ }A \emph{geodesic}\textit{\ }from $\mathbf{x}$ to
$\mathbf{y}$ on $C(\mathbf{X})$ is a path of\textit{\ minimum length} (i.e., a
\textit{shortest path}) connecting $\mathbf{x}$ and $\mathbf{y}$%
.$\mathbf{\medskip}$

It can be easily proved (and left to the reader to check) that the
\textit{shortest length} function $\delta:\mathbf{X\times X}\rightarrow
\mathbb{Z}_{+}$ of $C(\mathbf{X})$ as defined by the rule $\delta
(\mathbf{x,y}):=l(\pi_{\mathbf{xy}})$ for any $\mathbf{x,y\in X}$ (where
$\pi_{\mathbf{xy}}$ is a path of minimum length in $\Pi_{\mathbf{xy}}$) is
indeed a \textit{metric }\footnote{Thus, by definition, $\delta$ has
\textit{non-negative values} and satisfies the following conditions for every
$\mathbf{x,y,z\in X}$:
\par
$(i)$ (\textit{Identity Recognition}) $\delta(\mathbf{x,x)=}0$;
\par
$(ii)$ (\textit{Identity of Indiscernibles}) $\delta(\mathbf{x,y)=}0$ only if
$\mathbf{x=y}$;
\par
$(iii)$ (\textit{Symmetry}) $\delta(\mathbf{x,y)=}\delta(\mathbf{y,x)}$; (iv)
(\textit{Triangular Inequality) }$\delta(\mathbf{x,z)\leq}\delta
(\mathbf{x,y)+}\delta(\mathbf{y,z)}$.
\par
A sketch of the complete argument to establish validity of that statement goes
as follows. By construction $\mathbf{X}$ is a (bounded) distributive lattice
(i.e., it also satisfies for any $\mathbf{x,y,z}$,
\par
(\textit{distributivity}): $\mathbf{x\vee(y\wedge z)=(x\vee y)\wedge(x\vee
z})$ or equivalently $\mathbf{x\wedge(y\vee z)=(x\wedge y)\mathbf{\vee(x\wedge
z)}}$).
\par
But then, $\mathbf{X}$ is also a (bounded) \textit{modular }lattice, i.e., it
satisfies for any $\mathbf{x,y,z}$
\par
(\textit{modularity) if }$\mathbf{x\leqslant z}$ then $\mathbf{x\vee(y\wedge
z)=(x\vee y)\wedge z}$ \ or equivalently, if $\mathbf{z\leqslant x}$ then
$\mathbf{x\wedge(y\vee z)=(x\wedge y)\vee z}$.
\par
And modularity of $\mathbf{X}$, in turn, implies that (i) the length of an
interval of $\mathbf{X}$ is uniquely defined (because modularity implies in
particular validity of the Jordan-Dedekind chain condition for ordered sets
requiring equal length of all maximal chains having the same extrema), and
(ii) $\delta$ as defined above satisfies Triangular Inequality, hence it is
indeed a metric (since the other three conditions are obviously satisfied).
\par
It follows that $(\mathbf{X,\leqslant,}\delta)$ is indeed a \textit{metric
}lattice. See, e.g., Barbut and Monjardet (1970) for more details.}%
.$\mathbf{\medskip}$

\textbf{Definition 16.\ }Let \ $g:\mathbf{X}^{\left[  n\right]  }%
\rightarrow \left \{  0,1\right \}  ^{n}$ be a sufficientarian grading rule,
$\mathbf{x}\in X $ a capability-type and $\mathcal{X}^{\ast}(g):=(\mathbf{x}%
_{1}^{\ast},...,\mathbf{x}_{k}^{\ast})$ the threshold system induced by $g$.
Then, the distance $\delta_{g}(\mathbf{x,}\mathcal{X}^{\ast}(g))$ of
$\mathbf{x}$ from $\mathcal{X}^{\ast}(g)$ is defined as

$\delta_{g}(\mathbf{x},\mathcal{X}^{\ast}(g)):=\left \{
\begin{array}
[c]{c}%
\min_{j\in \lbrack k]}\left \{  \delta(\mathbf{x},\mathbf{x}_{j}^{\ast
})\right \}  \text{ if \ }\mathbf{x}_{j}^{\ast}\nleqslant \mathbf{x}\text{ for
every \ }j\in \lbrack k],\\
0\text{ otherwise}%
\end{array}
\right \}  .\mathbf{\medskip}$

\textbf{Definition 17.}\ Let \ $g:\mathbf{X}^{n}\rightarrow \left \{
0,1\right \}  ^{n}$ be a sufficientarian grading rule, $\mathbf{x}_{[n]}%
\in \mathbf{X}^{n}$ a capability-type assignment and $\mathcal{X}^{\ast
}(g):=(\mathbf{x}_{1}^{\ast},...,\mathbf{x}_{k}^{\ast})$ the threshold system
induced by $g$. The \emph{sufficiency-gap profile} of $\mathbf{x}_{[n]}%
\in \mathbf{X}^{n}$ with respect to $\mathcal{X}^{\ast}(g)$ is $(\delta
_{g}(\mathbf{x}_{i},\mathcal{X}^{\ast}(g)))_{i\in \lbrack n]}\in \mathbb{Z}%
_{+}^{n}$.$\mathbf{\medskip}$

\textbf{Definition 18.\ }\emph{(Gap-Antitony with respect to }$g$
(GA($g$)\emph{)}: Let \ $g:\mathbf{X}^{n}\rightarrow \left \{  0,1\right \}
^{n}$ be a sufficientarian grading rule, $\mathbf{x}_{[n]}\in \mathbf{X}^{n}$ a
capability-type assignment and $\mathcal{X}^{\ast}(g):=(\mathbf{x}_{1}^{\ast
},...,\mathbf{x}_{k}^{\ast})$ the threshold system induced by $g. $ Then, a
preorder $\succcurlyeq$ over $\mathbf{X}^{n\text{ }}$is \emph{gap-antitonic
with respect to }$g$ if, for any $\mathbf{x}_{[n]},\mathbf{x}_{[n]}^{\prime
}\in \mathbf{X}^{n}$, $\  \delta_{g}(\mathbf{x}_{i}\mathbf{,}\mathcal{X}^{\ast
}(g))\leq \delta_{g}(\mathbf{x}_{i}^{\prime}\mathbf{,}\mathcal{X}^{\ast}(g))$
for every $i\in \lbrack n]$ implies $\mathbf{x}_{[n]}\succcurlyeq
\mathbf{x}_{[n]}^{\prime}.\mathbf{\medskip}$

\textbf{Definition 19.\ }(\emph{Sufficiency-gap total preorders induced by a
sufficientarian grading rule }$g$) Let $g:\mathbf{X}^{n}\rightarrow \left \{
0,1\right \}  ^{n}$ be a sufficientarian grading rule. A \emph{sufficiency-gap
preorder induced by }$g$ is a total preorder $\succcurlyeq$ on $\mathbf{X}%
^{n}$ that satisfies Anonymity and Gap-Antitony w.r.t. $g$ as defined
above.$\mathbf{\medskip}$

It should be emphasized that any sufficiency-gap preorder as defined above
also satisfies by construction $\mathbf{X}$-Isotony (precisely as
sufficientarian simple preorders and sufficiency-count preorders do). \ On the
contrary, a sufficiency-gap preorder induced by sufficientarian grading rule
$g$ need \textit{not} be an extension of partial order $\leqslant_{g}$. That
is so because, in general, nothing prevent existence of two capability-type
assignments $\mathbf{x}_{[n]},\mathbf{x}_{[n]}^{\prime}\in \mathbf{X}^{N}$ such that:

$(i)$ $\delta_{g}(\mathbf{x}_{i}\mathbf{,}\mathcal{X}^{\ast}(g))<\delta
_{g}(\mathbf{x}_{i}^{\prime}\mathbf{,}\mathcal{X}^{\ast}(g))$ and $\delta
_{g}(\mathbf{x}_{j}^{\prime}\mathbf{,}\mathcal{X}^{\ast}(g))<\delta
_{g}(\mathbf{x}_{j}\mathbf{,}\mathcal{X}^{\ast}(g))$ for some $i,j\in \lbrack
n]$,

$(ii)$ $|\left \{  h\in \lbrack n]:\delta_{g}(\mathbf{x}_{h}\mathbf{,}%
\mathcal{X}^{\ast}(g))=0\right \}  |<|\left \{  h\in \lbrack n]:\delta
_{g}(\mathbf{x}_{h}^{\prime}\mathbf{,}\mathcal{X}^{\ast}(g))=0\right \}  |$
whence $\mathbf{x}_{[n]}\succ_{g}\mathbf{x}_{[n]}^{\prime}$, and by Symmetry
of $g$, $\mathbf{x}_{[n]}^{\prime}<_{g}\mathbf{x}_{[n]}$. Yet,

$(iii)$ according to some plausible distance-aggregation rule, the aggregate
distance of the sufficiency-gap profile of $\mathbf{x}_{[n]}$ from the
threshold system induced by $g$ on $\mathbf{X}^{n}$ is \textit{also greater}
than the aggregate distance of the sufficiency gap-profile of $\mathbf{x}%
_{[n]}^{\prime}$ from that threshold system.

Of course, whether or not the foregoing conditions $(i)-(ii)-(iii)$ can be
jointly satisfied ultimately depends on the possibility to specify plausible
distance-aggregation rules that validate the claim under consideration. As a
matter of fact, it can be easily shown that there are plenty of them. To
validate that claim, it may be useful to consider a few distinguished examples
of sufficiency-gap preorders, as listed below. It can be easily shown -and
left to the reader to check- that each one of the following four total
preorders over $\mathbf{X}^{n}$ satisfies both Anonymity and Gap-Antitony
w.r.t. $g$.$\mathbf{\medskip}$

\begin{itemize}
\item (\emph{min-average sufficiency-gap preorder}\textit{\ }$\succcurlyeq
_{\delta_{g}}^{\ast av}$). For any $\mathbf{x}_{[n]},\mathbf{x}_{[n]}^{\prime
}\in \mathbf{X}^{n}$,%
\[
\mathbf{x}_{[n]}\succcurlyeq_{\delta_{g}}^{\ast av}\mathbf{x}_{[n]}^{\prime
}\  \text{iff }\frac{\widetilde{\delta}_{g}(\boldsymbol{x}_{[n]},\mathcal{X}%
^{\ast}(g))}{n}\leq \frac{\widetilde{\delta}_{g}(\boldsymbol{x}_{[n]}^{\prime
},\mathcal{X}^{\ast}(g))}{n}\text{ iff }\widetilde{\delta}_{g}(\boldsymbol{x}%
_{[n]},\mathcal{X}^{\ast}(g))\leq \widetilde{\delta}_{g}(\boldsymbol{x}%
_{[n]}^{\prime},\mathcal{X}^{\ast}(g))
\]

\end{itemize}

where, for any $\mathbf{z}_{[n]}\in \mathbf{X}^{n}$,\ $\widetilde{\delta}%
_{g}(\boldsymbol{z}_{[n]},\mathcal{X}^{\ast}(g)):=%
{\displaystyle \sum \limits_{i\in \lbrack n]}}
\min \left \{  \delta_{g}(\mathbf{z}_{i},\boldsymbol{x}_{j}^{\ast}%
):j=1,...,k\right \}  $.\medskip

\begin{itemize}
\item (\emph{min-max sufficiency-gap preorder}\textit{\ }$\succcurlyeq
_{\delta_{g}}^{\ast \max}$). For any $\mathbf{x}_{[n]},\mathbf{x}_{[n]}%
^{\prime}\in \mathbf{X}^{n}$,%
\begin{align*}
\mathbf{x}_{[n]}  & \succcurlyeq_{\delta_{g}}^{\ast \max}\mathbf{x}%
_{[n]}^{\prime}\text{ \  \  \ if and only if}\\
max_{i\in,g[n]_{i}(\mathbf{x}_{N})=0}\min \left \{  \delta_{g}(\mathbf{x}%
_{i},\boldsymbol{x}_{j}^{\ast}):j=1,...,k\right \}   & \leq max_{i\in \lbrack
n],g_{i}(\mathbf{x}_{N}^{\prime})=0}\min \left \{  \delta_{g}(\mathbf{x}%
_{i}^{\prime},\boldsymbol{x}_{j}^{\ast}):j=1,...,k\right \}  \text{.}%
\end{align*}
\medskip

\item (\emph{min-leximax sufficiency-gap preorder}\textit{\ }$\succcurlyeq
_{\delta_{g}}^{\ast l\max}$). For any $\mathbf{x}_{[n]},\mathbf{x}%
_{[n]}^{\prime}\in \mathbf{X}^{n}$, $\mathbf{x}_{[n]}\succcurlyeq_{\delta_{g}%
}^{\ast l\max}\mathbf{x}_{[n]}^{\prime}$ if and only if there exist
permutations $\sigma:[n]\longrightarrow \lbrack n]$, $\tau:[n]\longrightarrow
\lbrack n]$ s.t. for any $i,j\in \lbrack n]$,$\  \mathbf{x}_{\sigma(i)}%
\geq \mathbf{x}_{\sigma(j)}$ if and only if $\sigma(i)\leq \sigma(j)$, and
$\mathbf{x}_{\sigma^{\prime}(i)}^{\prime}\geq \mathbf{x}_{\sigma^{\prime}%
(j)}^{\prime}$ if and only if $\tau(i)\leq \tau(j)$, and there exists $h^{\ast
}\in \lbrack n]$ such that $\mathbf{x}_{\sigma(i)}=\mathbf{x}_{\tau(j)}%
^{\prime}$ for all $i,j\in \lbrack n]$ with $\sigma(i)=\tau(j)\leq h^{\ast}$,
and either $h^{\ast}<n$ and $\min \left \{  \delta_{g}(\mathbf{x}_{\tau
^{-1}(h^{\ast}+1)},\boldsymbol{x}_{j}^{\ast}):j=1,...,k\right \}  >\min \left \{
\delta_{g}(\mathbf{x}_{\sigma^{-1}(h^{\ast}+1)},\boldsymbol{x}_{j}^{\ast
}):j=1,...,k\right \}  $, or $h^{\ast}=n$, hence $\mathbf{x}_{\sigma \lbrack
n]}=\mathbf{x}_{\tau \lbrack n]}^{\prime}$.\medskip

\item (\emph{min-upper-middlemost sufficiency-gap preorder}\textit{\ }%
$\succcurlyeq_{\delta_{g}}^{\ast m^{+}}$). For any $\mathbf{x}_{[n]}%
,\mathbf{x}_{[n]}^{\prime}\in \mathbf{X}^{n}$, $\mathbf{x}_{[n]}\succcurlyeq
_{\delta_{g}}^{\ast m^{+}}\mathbf{x}_{[n]}^{\prime}$ if and only if
\end{itemize}

$m^{+}(\min \left \{  \delta_{g}(\mathbf{x}_{i},\boldsymbol{x}_{j}^{\ast
}):j=1,...,k\right \}  :i\in \lbrack n],g_{i}(\mathbf{x}_{[n]})=0)\leq$

$\leq m^{+}(\min \left \{  \delta_{g}(\mathbf{x}_{i}^{\prime},\boldsymbol{x}%
_{j}^{\ast}):j=1,...,k\right \}  :i\in \lbrack n],g_{i}(\mathbf{x}_{[n]}%
^{\prime})=0)$

where for any $\mathbf{z\in}\mathbb{Z}^{l}$, $m^{+}(\mathbf{z})$ is the
upper-middlemost value of $\left[  \mathbf{z}\right]  :=\left \{
z_{1},...,z_{l}\right \}  $, namely $m^{+}(\mathbf{z}):=\max \left \{
z_{i^{\ast}},z_{j^{\ast}}\right \}  $ with $i^{\ast},j^{\ast}\in \left \{
1,...,l\right \}  $ such that

$||\left \{  z_{i}\in \left[  \mathbf{z}\right]  \text{: }z_{i}\leq z_{i^{\ast}%
}\right \}  |-|\left \{  z_{i}\in \left[  \mathbf{z}\right]  \text{: }z_{i^{\ast
}}\leq z_{i}\right \}  ||=||\left \{  z_{i}\in \left[  \mathbf{z}\right]  \text{:
}z_{i}\leq z_{j^{\ast}}\right \}  |-|\left \{  z_{i}\in \left[  \mathbf{z}%
\right]  \text{: }z_{j^{\ast}}\leq z_{i}\right \}  ||\in \left \{  0,1\right \}
.\medskip$

Clearly, if $l$ is odd, then $z_{i^{\ast}}=z_{j^{\ast}}$ and $m^{+}%
(\mathbf{z})$ is the \emph{median}\textit{\ }of $\left[  \mathbf{z}\right]  $:
in that case $m^{+}(\mathbf{z})$ is the nonnegative integer $z^{\ast}$ that
minimizes the sum $\sum_{i=1}^{l}|z^{\ast}-z_{i}|$ and that property might be
used in order to characterize $\succcurlyeq_{\delta_{g}}^{\ast m^{+}}$itself
as the total preorder induced by minimization of the value that minimizes the
sum of modules of its differences from the individual sufficientarian-gaps:
see e.g. Bandelt and Barth\'{e}lemy (1984).

In what follows, we shall provide characterizations of both the min-average
and the min-leximax sufficiency-gap preorders that can also be regarded as
examples of sufficientarian rankings that embody generalized utilitarian and
egalitarian principles, respectively.\medskip

\bigskip

\textbf{Remark 3. }As an alternative approach to identification of suitable
sufficiency-gap rankings, one may start from the subclass of sufficiency-gap
total preorders that also satisfy the Weak Majorization principle, as defined
below. Let $g:\mathbf{X}^{n}\rightarrow \left \{  0,1\right \}  ^{n}$ be a
sufficientarian grading rule. A weak-majorization sufficientarian-gap ranking
$\succcurlyeq_{\delta_{g}}$induced by $g$ is a total preorder on
$\mathbf{X}^{n}$ which satisfies the following condition: for any
$\mathbf{x}_{[n]},\mathbf{x}_{[n]}^{\prime}\in \mathbf{X}^{n}$, and any pair of
permutations $\sigma:N\rightarrow N$, $\sigma^{\prime}:N\longrightarrow N$
such that for all $i,j\in \lbrack n]$ with $i<j$, $\delta_{g}(\mathbf{x}%
_{\sigma(j)},\mathcal{X}^{\ast}(g)(g))\leq$ $\delta_{g}(\mathbf{x}_{\sigma
(i)},\mathcal{X}^{\ast}(g))$ and $\delta_{g}(\mathbf{x}_{\sigma^{\prime}%
(j)}^{\prime},\mathcal{X}^{\ast}(g))\leq$ $\delta_{g}(\mathbf{x}%
_{\sigma^{\prime}(i)}^{\prime},\mathcal{X}^{\ast}(g))$, if for every
$k\in \lbrack n]$:%
\[
\  \sum_{i\in \lbrack k],g(\sigma(i))=0}(\delta_{g}(\mathbf{x}_{\sigma
(i)}\mathbf{,}\mathcal{X}^{\ast}(g))\leq \sum_{i\in \lbrack k],g(\sigma
(i))=0}(\delta_{g}(\mathbf{x}_{\sigma(i)}^{\prime}\mathbf{,}\mathcal{X}^{\ast
}(g))
\]

then $\mathbf{x}_{[n]}\succcurlyeq \mathbf{x}_{[n]}^{\prime}$.\medskip

Of course, the Weak Majorization principle by itself defines a \emph{non-total
partial preorder }on $\mathbf{X}^{n}$ which satisfies, by construction, both
Anonymity and Gap-Antitony w.r.t. $g$. But then, one may obtain extensions of
that partial preorder to a total preorder by using as completion criteria any
one of the min-average, min-max, min-leximax, or min-upper middlemost
sufficiency-gap rankings as defined above. It should also be mentioned that
the Weak Majorization principle boils down to a finite family of partial sums'
inequalities (indexed by $[n]$) that a partial preorder on $\mathbf{X}^{n}$ is
required to `respect'. In that connection, the min-average and the min-max
sufficiency-gap total preorders can also be identified as those total
preorders that `respect' just \emph{one }of the inequalities of such a family,
as indexed by values $k=n$ and $k=1$, respectively.\medskip

In order to provide characterizations of the min-average and min-leximax
sufficiency-gap total preorders, we only need to observe that for every
sufficientarian grading rule $g$ on $\mathbf{X}^{n}$ the distances of
capability-types from the threshold system $\mathcal{X}^{\ast}(g)$ of $g$ are
non-negative \textit{integer vectors }induced by $g$ over $\mathbf{X}^{n} $,
and amount to a \textit{bounded }subset $\mathcal{Z}$ of the ordered set
$(\mathbb{Z}_{+}^{n},\leqslant)$ (where $\leqslant$ denotes the natural
component-wise partial order and an arbitrary element of $\mathcal{Z} $ is
denoted by $\mathbf{d}$). Thus, we can reformulate the two conditions defining
sufficiency-gap total preorders with no explicit reference whatsoever to
sufficiency-gap profiles or sufficientarian grading rules. In particular,
while the Anonymity (AN) condition stays unaltered, the family of Gap-Antitony
(GA($g$)) conditions are reformulated as Antitony. Then we start from the
following two axioms for sufficiency-gap total preorders (we also replicate
here the definition of Anonymity just for the sake of convenience):\medskip

\emph{Anonymity (AN) For any }$\mathbf{d}\in \mathcal{Z}$\emph{, and every
permutation }$\pi$\emph{\ :}$[n]\longrightarrow \lbrack n]$\emph{, }%
$\mathbf{d}\sim \mathbf{d}_{\pi}$\emph{, where }$\mathbf{d}_{\pi}:=\left(
d_{\pi \left(  1\right)  },...,d_{\pi \left(  n\right)  }\right)  $%
\emph{.\medskip}

\emph{Antitony (ANT)}\textbf{\ }For any $\mathbf{d}$,$\mathbf{d}^{\prime}%
\in \mathcal{Z}$ if $d_{i}\leq d_{i}^{\prime}$ for all $i\in \lbrack n]$ then
$\mathbf{d}\succeq \mathbf{d}^{\prime}$.\medskip

A few further axioms for our characterizations are now to be
introduced.\medskip

\emph{Strong Antitony (S-ANT)}\textbf{\ }For any $\mathbf{d},\mathbf{d}%
^{\prime}\in \mathcal{Z}$ if $d_{i}\leq d_{i}^{\prime}$ for all $i\in \lbrack
n]$ then $\mathbf{d\succ d}^{\prime}$ if $\mathbf{d\neq d}^{\prime} $ and
$\mathbf{d}\succeq \mathbf{d}^{\prime}$ otherwise.\medskip

\emph{Restricted Translation Invariance (RTI)}\textbf{\ }For all
$\mathbf{d,d}^{\prime}\in \mathcal{Z}$ and $\mathbf{z}\in \mathbb{Z}_{+}^{n}$
such that both $\mathbf{d}\boldsymbol{+}\mathbf{z}\in \mathcal{Z}$ and
$\mathbf{d}^{\prime}\mathbf{+z}\in \mathcal{Z}$, if $\mathbf{d\succeq
d}^{\prime}$ then $\mathbf{d+z}\succeq \mathbf{d}^{\prime}\mathbf{+z.}$\medskip

\emph{Restricted Hammond Equity (RHE)}\textbf{\ }For any $\mathbf{d,d}%
^{\prime}\in \mathcal{Z}$, if $d_{i}=d_{i}^{\prime}$ for all $i\in \lbrack
n]\setminus \left \{  h,k\right \}  $, $d_{h}=\max_{i\in \lbrack n]}d_{i}$,
$d_{k}^{\prime}=\max_{i\in \lbrack n]}d_{i}^{\prime}$, and $d_{h}^{\prime}\leq
d_{k}<d_{h}\leq d_{k}^{\prime}$, then $\mathbf{d\succeq d}^{\prime}%
.$\footnote{The label of that axiom is meant to recall that it amounts to a
version of the `Equity Axiom' first introduced by Hammond (1976, 1979) as a
generalization and strengthening of the `Weak Equity Axiom' previously
suggested by Sen in the first, 1973 edition of Sen (1997), and further
discussed and used in Sen (1977).}\medskip

In plain words, RTI\ requires invariance of the preorder with respect to
changes of origin of the underlying space whose points' distances are being
considered. RHE requires instead that if two assignments only differ both in
the points of maximal distance and in the respective value of such maximal
distances, then the assignment with a smaller difference between the distances
of such two points of maximum distance is either strictly better or
indifferent to the other.\medskip

\bigskip

\textbf{Proposition 3 }Let $\succeq$ be a total preorder over a bounded set
$\mathcal{Z\subseteq}\mathbb{Z}_{+}^{n}$, and $\succeq^{\ast av}$the total
preorder over $\mathcal{Z}$ defined as follows: for every
$\mathbf{d\mathbf{:=(}}d_{1}...,d_{n}\mathbf{),d}^{\prime}:=(d_{1}^{\prime
},...,d_{n}^{\prime})\in \mathcal{Z}$, $\mathbf{d}\succeq^{\ast av}%
\mathbf{d}^{\prime}$ if and only if \ $\sum_{i=1}^{n}d_{i}\geq \sum_{i=1}%
^{n}d_{i}^{\prime}$. Then, $\succeq$ satisfies AN, S-ANT and RTI if and only
if $\succeq=\succeq^{\ast av}$.\medskip

\bigskip

\textbf{Remark 4. }The proof of the previous characterization (see the
Appendix) is an adaptation and application of the `inductive' proof technique
first used in a similar setting by Sen (1977) to prove a different but related
proposition, and subsequently also deployed in Hammond (1979) and d'Aspremont
(1985).\medskip

\bigskip

\textbf{Corollary 1. }Let $\succeq$ be a total preorder over the bounded set
$L_{\mathbf{X}}^{n}$ $\subseteq \mathbb{Z}_{+}^{n}$, where $L_{\mathbf{X}%
}:=\prod_{i=1}^{m}\left \{  0,1,...,l_{i}\right \}  $ and $L_{\mathbf{X}}^{n}$
is ordered according to the natural component-wise partial order of
$\mathbb{Z}_{+}^{n}$. Then, $\succeq=\succcurlyeq_{\delta_{g}}^{\ast av}$ if
and only if $\succeq$ satisfies AN, S-ANT and RTI.\medskip

\textbf{Proof. }It is easily checked that $L_{\mathbf{X}}$ is indeed the set
of possible distances between points of our capability-type space $\mathbf{X}
$. Then, the Corollary follows immediately from Proposition 3.\qquad
\qquad \qquad \qquad \qquad \qquad \qquad \qquad \qquad \qquad \qquad \qquad \qquad
\qquad \qquad$\square$\medskip

Let us now proceed in a similar way in order to produce a characterization of
the min-max aggregation rule $\succcurlyeq_{\delta_{g}}^{\ast mM}$as a special
case of the characterization of min-max aggregation rules on a bounded set
$\mathcal{Z\subseteq}\mathbb{Z}_{+}^{n}$.

To begin with, let us establish the validity of the following claim.\medskip

\textbf{Claim 2. }Let $\succeq$ be a total preorder over a bounded set
$\mathcal{Z\subseteq}$ $Z^{n}$ that satisfies $RHE$. Then $\succeq$ also
satisfies $\mathnf{AN}$.\medskip

The present proof of Proposition 4 below relies again (as the proof of
previous Proposition 3) on the definition of a family of auxiliary total
preorders $\widehat{\succeq}^{m}$ over $\mathcal{Z}$ indexed by $m\in$
$[n]\setminus \left \{  1\right \}  $, and amounts to a `restricted' induction
argument on $[n]\setminus \left \{  1\right \}  $. We also need, for the sake of
completeness and convenience, an explicit general definition of the
min-leximax total preorder $\succeq^{\ast l\max}$over $\mathcal{Z}.$\medskip

\bigskip

\textbf{Definition 20\ }(\emph{min-leximax preorder}\textit{\ }$\succeq^{\ast
l\max}$). For any $\mathbf{d},\mathbf{d}^{\prime}\in \mathcal{Z}$,
$\mathbf{d}\succcurlyeq^{\ast l\max}\mathbf{d}^{\prime}$ if and only if there
exist permutations $\sigma:[n]\longrightarrow \lbrack n]$, $\tau
:[n]\longrightarrow \lbrack n]$ such that, for any $i,j\in \lbrack
n]$,\ $\mathbf{d}_{\sigma(i)}\geq \mathbf{d}_{\sigma(j)}$ if and only if
$\sigma(i)\leq \sigma(j)$, and $\mathbf{x}_{\tau(i)}^{\prime}\geq
\mathbf{x}_{\tau(j)}^{\prime}$ if and only if $\tau(i)\leq \tau(j)$, and
$h^{\ast}\in \lbrack n]$ such that $\mathbf{d}_{\sigma(i)}=\mathbf{d}_{\tau
(j)}^{\prime}$ for all $i,j\in \lbrack n]$ with $\sigma(i)=\tau(j)\leq h^{\ast
}$, and either $[h^{\ast}<n$ \ and \ $d_{\tau^{-1}(h^{\ast}+1)}^{\prime
}>d_{\sigma^{-1}(h^{\ast}+1)}]$ or $h^{\ast}=n$ hence $\mathbf{d}_{\sigma
}=\mathbf{d}_{\tau}^{\prime}$.\medskip

\bigskip

\textbf{Proposition 4. \ }Let $\widehat{\succeq}$ be a total preorder over a
bounded set $\mathcal{Z\subseteq}\mathbb{Z}_{+}^{n}$, and $\succeq^{\ast
l\max}$the min-leximax total preorder over $\mathcal{Z}$ . Then,
$\widehat{\succeq}$ satisfies S-ANT and RHE if and only if $\widehat{\succeq
}=\succeq^{\ast l\max}$.\textbf{\medskip}

\bigskip

\textbf{Corollary 2. }Let $\succeq$ be a total preorder over the bounded set
$L_{\mathbf{X}}^{n}$ $\subseteq \mathbb{Z}_{+}^{n}$, where $L_{\mathbf{X}%
}:=\prod_{i=1}^{m}\left \{  0,1,...,l_{i}\right \}  $ and $L_{\mathbf{X}}^{n}$
is ordered according to the natural component-wise partial order of
$\mathbb{Z}_{+}^{n}$. Then, $\succeq=\succcurlyeq_{\delta_{g}}^{\ast l\max}$
if and only if $\succeq$ satisfies S-ANT and RHE.\textbf{\medskip}

\textbf{Proof. }It is easily checked that $L_{\mathbf{X}}$ is indeed the set
of possible distances between points of our capability-type space $\mathbf{X}
$. Then, the Corollary follows immediately from Proposition 4.\qquad
\qquad \qquad \qquad \qquad \qquad \qquad \qquad \qquad \qquad \qquad \qquad \qquad
\qquad \qquad \qquad$\square$\textbf{\medskip}

\bigskip

Let us now try and summarize the main findings of the present section. To
begin with, it has been shown that a sufficientarian assessment of capability
profiles as specified by its simple or sufficiency-count total preorders is a
valuable and self-standing basic benchmarking criterion of distributive
justice. Different considerations apply when it comes to the design or
assessment of policies to improve on profiles that fail (perhaps grossly) the
sufficientarian criterion\ itself.\ As mentioned above, such an assessment
typically requires criteria to identify priorities concerning the agents who
should be allotted more resources among those who fail to achieve the
sufficiency threshold (and perhaps those agents who achieve such threshold and
could afford larger transfers to the former), precisely as in the case of
poverty abatement policies which typically rely on poverty-gap criteria.

Concerning such policy-related issues, we have just shown that a
sufficientarian perspective can also provide a distinctive contribution to
defining priority criteria to be used as a guidance or assessment of policies
aimed at improving on allocations that definitely fail to meet sufficientarian
criteria. But we should emphasize that, arguably, relying on sufficientarian
criteria in order to assess social progress in achievement allocation is in
principle consistent with the endorsement of \textit{virtually any other
criterion} when it comes to guiding or assessing remedial policies to be
applied to `non-sufficientarian' affordance/achievement allocations (e.g.
utilitarian, leximin, undominated diversity, inequality reduction or more
generally any suitable \textit{prioritarian }criterion either egalitarian or
not: see, among others, Atkinson and Bourguignon (1982), Moulin (1988), Sen
(1997), Parfit (1997), Roemer (2004), Arneson (2006), Cohen (2011)).

\bigskip

\section{\textbf{Strategy--proof identification and selection of a
sufficientarian grading rule}}

Let us now address the issue of designing protocols to actually implement
sufficientarian grading rules (i.e. sufficientarian \textit{BGFs). }As
mentioned above, we take achievement vectors of the capabilitiy space to be
observable and verifiable. However, a sufficientarian BGF $g$ relies in fact
on a unique \textit{threshold system, }that must be somehow defined, and
conversely. Thus we have to address the problem of choosing a
\textit{specific} sufficientarian BGF $g$, or equivalently a specific
threshold system (i.e., antichain). We claim that such a threshold system can
and should be determined in a consensual manner by \emph{aggregating the
opinions} on that matter of the agents involved (or perhaps of their
representatives, or appointed experts). Thus, we have to supplement our model
with the (true) judgements of the relevant agents concerning the appropriate
`sufficientarian' system of thresholds, by asking them to disclose such a
(private) true judgement. Now, for any agent such a judgement can be regarded
as her/his most preferred `sufficientarian' $g$. But then, since the actual,
true judgments of the relevant agents are private information of the latter
since they are typically not observable and verifiable, it follows that the
aggregation rules to be used should be \textit{strategy-proof} in order to
prevent strategic manipulation of outcomes (namely, the chosen threshold
systems) by submission of incorrect/false judgements.

Thus, in a more formal language, the situation can be described as follows.
Let $\mathcal{S}(\mathbf{X}^{n})$ denote the class of all sufficientarian BGFs
over $\mathbf{X}^{n}$ and $\mathcal{A}_{\mathbf{X}}$ the set of all antichains
(or threshold systems) of $\mathbf{X}$. Our aim is to define a well-behaved
protocol enabling a given set of agents/stakeholders $[n^{\prime}%
]\subseteq \lbrack n]$ to select some specific sufficientarian BGF
$g:\mathbf{X}^{n}\rightarrow \left \{  0,1\right \}  ^{n}$ , which is of course
induced by one specific antichain $\mathcal{X}^{\ast}:=(\mathbf{x}_{1}^{\ast
},...,\mathbf{x}_{k}^{\ast})\in \mathcal{A}_{\mathbf{X}}$, and $\mathcal{A}%
_{\mathbf{X}}$. Suppose also, for the sake of simplicity, that all agents in
$[n]$ are involved, i.e., that $[n^{\prime}]=[n]$. Each agent is required to
propose the most appropriate sufficientarian grading rule over $\mathbf{X}%
^{n}$, or equivalently and more conveniently, the most appropriate antichain
$\mathcal{X}$ of $\mathbf{X}$ (namely, $\mathcal{X\in} $ $\mathcal{A}%
_{\mathbf{X}}$).

The first key point to notice is that in the present framework based upon
finite capability-type space $\mathbf{X}$, the set $\mathcal{S}(\mathbf{X}%
^{n})$ of sufficientarian grading rules over $\mathbf{X}^{n}$ ordered by the
natural point-wise partial order $\leqslant$ \footnote{More formally, for any
pair of sufficientarian BGFs $\ g,g^{\prime}\in \mathcal{S}(\mathbf{X}^{n})$,
$g\leqslant g^{\prime}$ if and only if $g(\mathbf{x}_{[n]})\leq g^{\prime
}(\mathbf{x}_{[n]})$ for every $\mathbf{x}_{[n]}\in \mathbf{X}^{n}$.
\par
{}} and the set $\mathcal{A}_{\mathbf{X}}$ of antichains of $\mathbf{X}$
endowed with weak dominance $\succapprox$ are essentially the same thing. This
is made precise by the following definitions and lemma.\textbf{\medskip}

\textbf{Definition 21 (}\emph{The partially ordered set \ }$(S(X^{n}%
),\leqslant)$\emph{\ of sufficientarian grading rules}). For any pair of
sufficientarian BGFs $\ g,g^{\prime}\in \mathcal{S}(\mathbf{X}^{n})$,
$g\leqslant g^{\prime}$ if and only if $g(\mathbf{x}_{[n]})\leq g^{\prime
}(\mathbf{x}_{[n]})$ for every $\mathbf{x}_{[n]}\in \mathbf{X}^{n}$.
Furthermore, for every $\mathbf{x}_{[n]}\in \mathbf{X}^{n}$, $(g\mathbf{\vee
}g^{\prime})(\mathbf{x}_{[n]}):=g(\mathbf{x}_{[n]})\vee g^{\prime}%
(\mathbf{x}_{[n]})$ and $(g\mathbf{\wedge}g^{\prime})(\mathbf{x}%
_{[n]}):=g(\mathbf{x}_{[n]})\wedge g^{\prime}(\mathbf{x}_{[n]})$
.\textbf{\medskip}

\textbf{Definition 22 (}\emph{The relational system }$(A_{\mathbf{X}%
},\succapprox)$\emph{\ of antichains endowed with the weak dominance
relation)}. For any pair of\ threshold systems or antichains $\mathcal{X}%
,\mathcal{X}^{\prime}\in \mathcal{A}_{\mathbf{X}}$, $\mathcal{X}\succapprox
\mathcal{X}^{\prime}$ if and only if for every $\mathbf{x}_{i}\in \mathcal{X}$
there exists $\mathbf{x}_{j}^{\prime}\in \mathcal{X}^{\prime}$ such that
$\mathbf{x}_{j}^{\prime}\leqslant \mathbf{x}_{i}$.\textbf{\medskip}

We also recall here, for the sake of completeness, that a lattice
$(X,\vee,\wedge)$ is \emph{distributive }if and only if $x\vee(y\wedge
z)\mathbf{=}(x\vee y)\wedge(x\vee z)$, or equivalently $x\wedge(y\vee
z)=(x\wedge y)\vee(x\wedge z)$ for all $x,y,z$ $\in$ $X.$

\bigskip

\textbf{Lemma 1}$.$The weak dominance relation $\succapprox$ is a partial
order, hence both $(\mathcal{S}(\mathbf{X}^{n}),\leqslant)$ and $(\mathcal{A}%
_{\mathbf{X}},\succapprox)$ are partially ordered sets. Moreover,
$(\mathcal{S}(\mathbf{X}^{n}),\leqslant)$ and $(\mathcal{A}_{\mathbf{X}%
},\succapprox)$ are in fact two \emph{isomorphic distributive lattices.}%
\textbf{\medskip}

\bigskip

Such a Lemma relies heavily on a theorem due to Dilworth which establishes
that the antichains (subsets of mutually incomparable elements) of a finite
partially ordered set are indeed a (finite) \textit{distributive lattice
}(Dilworth (1960), see also Anderson (1987)). Then, the Lemma asserts that
such a lattice is indeed, by construction, isomorphic to the lattice of
sufficientarian BGFs induced by the natural point-wise partial order which is
of course also a distributive lattice. But then, a natural \textit{metric} can
be defined on the \textit{set }$\mathcal{A}_{\mathbf{X}}$ \textit{of
antichains of} $\mathbf{X}$ (or equivalently on the set of all sufficientarian
BGF over $\mathbf{X}^{n}$ along the same lines of the geodesic-based metric on
$\mathbf{X}$ defined in the previous section). To put it in simple words, such
a natural metric is defined as follows: the distance between two
sufficientarian BGFs $g,g^{\prime}\in \mathcal{S}(\mathbf{X}^{n})$ is precisely
the distance between their respective threshold systems or antichains
$\mathcal{X}^{\ast}(g),\mathcal{X}^{\ast}(g^{\prime}).$ And the distance
between antichains $\mathcal{X}^{\ast}(g)$ and $\mathcal{X}^{\ast}(g^{\prime
})$ is the minimum distance among the distances between pairs $\mathbf{x,x}%
^{\prime}\in \mathbf{X}$ such that $\mathbf{x\in}\mathcal{X}\mathbf{^{\ast}%
(}g\mathbf{)}$ and $\mathbf{x}^{\prime}\in \mathcal{X}\mathbf{^{\ast}%
(}g^{\prime}\mathbf{)}$. It should be noticed that such a distance on
antichains of $\boldsymbol{X}$ is in fact the obvious extension to
$\mathcal{A}_{\mathbf{X}}$ of the `natural' metric $d$ on $\mathbf{X}$ defined
above \footnote{We are indeed denoting by $d$ both the metrics of $\mathbf{X}$
and its `extension' to $\mathcal{A}_{\mathbf{X}}$, which is strictly speaking
a slight abuse of language, but a quite innocuous one. That is so because, by
definition, any singleton $\left \{  \mathbf{x}\right \}  $, with $\mathbf{x\in
X}$, is a (degenerate) antichain of $\mathbf{X}$. It follows that one might as
well start by first defining $d$ over $\mathcal{A}_{\mathbf{X}}$ and then
identifying the distance on $\mathbf{X}$ with the restriction of $d$ to the
subset of `degenerate' singleton antichains of $\mathbf{X}$.}, and makes it
also possible to introduce a \textit{metric betweenness} ternary relation
$B_{d}$ on $\mathcal{A}_{\mathbf{X}}$ as defined below.\textbf{\medskip}

\bigskip

\textbf{Definition 23 }\emph{(Metric Betweenness over Antichains of
}$\mathbf{X}$\emph{)}. The $d$\emph{-metric betweenness }relation $B_{d}$ is
the ternary relation on $\mathcal{A}_{\mathbf{X}}$ defined by the following
rule: for any $\mathcal{X}$, $\mathcal{Y}$, $\mathcal{V}\in \mathcal{A}%
_{\mathbf{X}}$, $(\mathcal{X}$, $\mathcal{V}$,$\mathcal{Y)\in}B_{d}$ (namely,
$\mathcal{V}$ is between $\mathcal{X}$ and $\mathcal{Y}$ according to the
$d$-metric) if and only if $d(\mathcal{X},\mathcal{Y)=}d(\mathcal{X}$,
$\mathcal{V})+d(\mathcal{V},\mathcal{Y)}$.\footnote{It is well-known but worth
recalling that in any bounded distributive lattice the metric betweenness is
identical to both the \textit{median betweenness} and the
\textit{interval-length betweenness} (see, e.g., Barbut and Monjardet (1970)
for the relevant definitions and details).}\textbf{\medskip}

\bigskip

It follows that, under the natural assumption that any agent regards her/his
own judgement concerning the appropriate sufficiency-threshold system as the
best one and considers any other judgement on that matter to be the better the
\textit{closer} it is to her/his own according to \textit{that} `natural'
metric. Such a notion of an agent's preferences between threshold systems
makes perfect sense, but requires a precise notion of \textit{one }threshold
system \textit{being closer} than \textit{another} to a \textit{third }one.
And the metric betweenness $B_{d}$ over $\mathcal{A}_{\mathbf{X}} $\ provides
\textit{exactly} that kind of notion with the resulting \textit{ternary space
}$(\mathcal{A}_{\mathbf{X}},B_{d})$\footnote{It should also be noticed that
such a ternary space is in particular a \textit{median space }since (as
mentioned in the previous note) the metric betwenness $B_{d}$ is also the
\textit{median} \textit{betweenness }of $\mathcal{A}_{\mathbf{X}}$ (see also
Nehring and Puppe (2007)).}.

Once this further `intrinsic' structure of the set $\mathcal{A}_{\mathbf{X}}$
of threshold systems/antichains of $\mathbf{X}$ as a (median) ternary space is
made explicit and put in place, one can immediately appreciate the
`naturality' and generality of single-peaked preferences over $(\mathcal{A}%
_{\mathbf{X}},B_{d})$, as defined below.\textbf{\medskip}

\bigskip

\textbf{Definition 25 }(\emph{Single-Peaked Preferences over Antichains of}
$\mathbf{X}$). Let $(\mathcal{A}_{\mathbf{X}},B_{d})$ be the ternary space of
antichains of $\mathbf{X}$ induced by metric betweenness $B_{d}$, and
$\succcurlyeq$ a preorder i.e. a reflexive and transitive binary relation over
$\mathcal{A}_{\mathbf{X}}$ (we shall denote by $\succ$ and $\sim$ its
asymmetric and symmetric components, respectively, by $Top(\succcurlyeq)$ the
possibly empty set of its maxima, and by $||$ the set of its
\textit{incomparable }ordered pairs i.e. $x||y$ if and only if neither
$x\succcurlyeq y$ nor $y\succcurlyeq x$ hold).

Then, $\succcurlyeq$ is said to be \emph{single-peaked}\textbf{\  \ }in
$(\mathcal{A}_{\mathbf{X}},B_{d})$ if and only if $SP$-$(i)$ there is a
\textit{unique maximum} of $\succcurlyeq$ in $\mathcal{A}_{\mathbf{X}}$ , its
\textit{top }antichain -denoted $top(\succcurlyeq)$- and $SP$-$(ii)$ for all
$\mathcal{X},\mathcal{Y},\mathcal{V}\in \mathcal{A}_{\mathbf{X}}$, if
$\mathcal{(X},\mathcal{V},\mathcal{Y})\in$ $B_{d}$ then \emph{not}
$\mathcal{Y}$ $\succ \mathcal{V}$.

The set of all single-peaked preference preorders in $(\mathcal{A}%
_{\mathbf{X}},B_{d})$ is denoted $\mathcal{D}_{B_{d}}.$\textbf{\medskip}

\bigskip

Observe that preferences that are\textit{\ preorders with a unique maximum
that are single-peaked }with respect to metric betweenness $B_{d}$ are a most
appropriate representation of preferences over threshold-systems/antichains of
$\mathbf{X}$ (or, equivalently, sufficientarian grading rules on
$\mathbf{X}^{n}$) such that each agent's opinion is her unique top antichain,
while concerning any other antichain is regarded to be the better the closer
it is to the top antichain, as dictated by $B_{d}$). Thus, for any profile of
(true) judgements on the appropriate threshold-system/antichain $\mathcal{X}$
(or equivalently sufficientarian BGF $g$ on $\mathbf{X}^{n}$) we end up with a
`natural' profile of single-peaked preferences on sufficientarian BGFs on
$\mathbf{X}^{n}$).

But then, we are now in a position to consider aggregation rules for
threshold-systems/antichains of $\mathbf{X}$ (or equivalently sufficientarian
BGFs on and their properties, including strategy-proofness
properties\textbf{\medskip}

\bigskip

\textbf{Definition 26. }An \emph{aggregation rule} for $([n],\mathcal{A}%
_{\mathbf{X}})$ is a function $f:\mathcal{A}_{\mathbf{X}}^{n}\rightarrow
\mathcal{A}_{\mathbf{X}}$.\textbf{\medskip}

\textbf{Definition 27 }(\emph{Strategy-Proofness on} $\mathcal{D}_{B_{d}}^{n}
$\emph{of an aggregation rule \ for} $([n],\mathcal{A}_{\mathbf{X}})$). An
aggregation rule $f$ \ for $([n],\mathcal{A}_{\mathbf{X}})$ is
\emph{strategy-proof} on $\mathcal{D}_{B_{d}}^{n}$ iff for all single--peaked
$[n]$-profiles $(\succcurlyeq_{i})_{i\in N}$ in $(\mathcal{A}_{\mathbf{X}%
},B_{d}) $, and for all $i\in \lbrack n]$, $\mathbf{y}_{i}\in \mathbf{X}$, and
$(\mathbf{x}_{j})_{j\in \lbrack n]}\in \mathbf{X}^{n}$ such that $\mathbf{x}%
_{j}=top(\succcurlyeq_{j})$ for each $j\in \lbrack n]$, \emph{not}%
\textit{\ }$f((\mathbf{y}_{i},(\mathbf{x}_{j})_{j\in([n]\smallsetminus \left \{
i\right \}  )}))\succ_{i}f((\mathbf{x}_{j})_{j\in \lbrack n]})$.\textbf{\medskip
}

Non-trivial strategy-proof aggregation rules should be -at least to some
extent- \emph{input-responsive}\textit{\ }and \emph{output-unbiased}\textit{.
}A few requirements can be deployed to present several versions and degrees of
input-responsiveness and output-unbiasedness of aggregation rules,
namely\textbf{\medskip}

\emph{Inclusiveness}: an aggregation rule for $([n],\mathcal{A}_{\mathbf{X}})
$ is \emph{inclusive}\textbf{\ }if and only if for each voter $i\in \lbrack n]$
there exist $\mathbf{x}_{[n]}\in \mathbf{X}^{n}$ and $\mathbf{y}_{i}%
\in \mathbf{X}$ such that $f(\mathbf{x}_{^{[n]\smallsetminus \left \{  i\right \}
}},\mathbf{y}_{i})\neq f(\mathbf{x}).$\textbf{\medskip}

\emph{Anonymity}\textbf{: }an aggregation rule $f$ for $([n],\mathcal{A}%
_{\mathbf{X}})$ is \emph{anonymous}\textbf{\ }if for each $\mathbf{x}_{[n]}%
\in \mathbf{X}^{N}$ and each permutation $\sigma:[n]\longrightarrow \lbrack n]$
, $f(\mathbf{x}_{[n]})=f(\mathbf{x}_{\sigma \lbrack n]})$ (where $\mathbf{x}%
_{\sigma \lbrack n]}=(\mathbf{x}_{\sigma(1)},...,\mathbf{x}_{\sigma(n)}%
)$).\textbf{\medskip}

\emph{Idempotence}\textbf{: }an aggregation rule $f$ for $([n],\mathcal{A}%
_{\mathbf{X}})$ is \emph{idempotent}\textbf{\ }(or \emph{unanimity-respecting}%
\textit{) }if\ $f(\mathbf{x},...,\mathbf{x})=\mathbf{x}$ for each
$\mathbf{x}\in \mathbf{X}.$\textbf{\medskip}

\emph{Sovereignty}: an aggregation rule $f$ for $([n],\mathcal{A}_{\mathbf{X}%
})$ is \emph{sovereign}\textbf{\ }if for each $\mathbf{y}\in \mathbf{X}$ there
exists $\mathbf{x}_{[n]}\in \mathbf{X}^{n}$ such that $f(\mathbf{x}%
_{[n]})=\mathbf{y}$ i.e. $f$ \ is an \emph{onto}\textit{\ }%
function.\textbf{\medskip}

\emph{Neutrality}\textbf{: }an aggregation rule $f$ \ for $([n],\mathcal{A}%
_{\mathbf{X}})$ is \emph{neutral}\textbf{\ }if for each $\boldsymbol{x}%
_{[n]}\in \mathbf{X}^{n}$ and each permutation $\pi$: $\mathbf{X\longrightarrow
X},$ $f(\pi(\mathbf{x}_{[n_{]}}))=\pi(f(\mathbf{x}_{[n]})) $ (where
$\pi(\mathbf{x}_{[n]})=(\pi(\mathbf{x}_{1}),...,\pi(\mathbf{x}_{n}%
))).$\textbf{\medskip}

Notice that both \emph{Idempotence }and \emph{Neutrality}\textbf{\ }imply
\emph{Sovereignty}\textbf{\ (}but not conversely\textbf{), }while
\emph{Anonymity}\textbf{\ }and \emph{Sovereignty}\textbf{\ }jointly imply
\emph{Inclusiveness}\textbf{\ }(but not conversely). However, it is easily
checked that \textit{if} \textit{Strategy-Proofness holds, }%
Sovereignty\textbf{\ }and Idempotence are in fact equivalent.

Now, we are looking for some aggregation rule $G:(\mathcal{S}(\mathbf{X}^{n})
$ $)^{n}\longrightarrow \mathcal{S}(\mathbf{X}^{n})$ that for any profile of
proposed sufficientarian BGFs returns a single sufficientarian BGF and
respects three key properties of any \textit{reliable }`democratic'
aggregation protocol: \textit{anonymity} (it is the list of proposals that
counts not the identity of the proponent of each proposal),
\textit{idempotence} or respect for unanimity (if everyone makes the same
proposal, that proposal must the chosen one), and \textit{strategy-proofness
}(nobody should be ever in the position to obtain the choice of a proposal
that is better, i.e., closer to his/her truly preferred proposal by making a
proposal \textit{other than }the latter) with respect to domain $D$ of
single-peaked preferences over $\mathcal{A}_{\mathbf{X}}$ as defined above.

Let us now proceed to consider the address the task of designing and
implementing an aggregation rule $G:(\mathcal{S}(\mathbf{X}^{n})$
$)^{n}\longrightarrow \mathcal{S}(\mathbf{X}^{n})$ that satisfy
\textit{anonymity, idempotence and strategy-proofness }by means of a protocol
that relies on the isomorphic bijection $\varphi:\mathcal{S}(\mathbf{X}%
^{n})\longrightarrow \mathcal{A}_{\mathbf{X}}$ between sufficientarian BGFs
over $\mathbf{X}^{n}$ and antichains of $\mathbf{X}$.

$(i)$ To begin with, every agent $i\in$ $[n]$ submits a sufficientarian BGF
$g^{i}\in \mathcal{S}(\mathbf{X}^{n})$ so that a profile $\mathbf{g}%
_{[n]}=(g^{1},...,g^{n})$ $\in(\mathcal{S}(\mathbf{X}^{n})$ $)^{n}$.

$(ii)$ For any $g^{i}$ we consider $\varphi(g^{i})$ the antichain (or order
filter) of $\mathbf{X}$ also denoted $\mathcal{X}_{i}(g^{i})\in \mathcal{A}%
_{\mathbf{X}}$, and obtain a profile $\varphi \left(  \mathbf{g}_{[n]}\right)
:=(\varphi(g^{1}),...,\varphi(g^{n}))$ as an input to an aggregation rule
$F:(\mathcal{A}_{\mathbf{X}})^{n}\longrightarrow \mathcal{A}_{\mathbf{X}}$ that
returns an antichain $\mathcal{X}(\mathbf{g}_{[n]}):=F((\mathcal{X}_{1}%
(g^{1}),...,\mathcal{X}_{n}(g^{n}))$ as its output.

$(iii)$ Finally, we define $G(\mathbf{g}_{[n]}):=\varphi^{-1}(\mathbf{X}%
(\mathbf{g}_{[n]}))$. Moreover, we say that $G$ is truthfully implementable or
strategy-proof on a certain domain or sufficientarian BGFs if $F$ is
strategy-proof on the corresponding domain $\mathcal{D}_{B_{d}}^{n}$of
(single-peaked) preorders over $\mathcal{A}_{\mathbf{X}}$.

Clearly enough, each one of the three properties Anonymity, Idempotence, and
Strategy-Proofness holds for $G$ if and only if it also holds for $F$.

Here, we can also take avail of the following Lemma which summarizes some
previous results on inclusive, anonymous and idempotent aggregation rules over
bounded distributive lattices that are also strategy-proof on suitably defined
single-peaked domains (such a Lemma is directly implied by Theorem 1 of
Savaglio and Vannucci (2019), but see also Monjardet (1990), Nehring and Puppe
(2007), and Vannucci (2019) for strictly related results).\textbf{\medskip}

\bigskip

\textbf{Lemma 2. }Let $(X,\leqslant)$ be a bounded distributive lattice and
$B_{d}$ its metric betweenness relation. Then there is a class of anonymous
and idempotent aggregation rules for $([n],X)$ that are also single-peaked in
$(X,B_{d})$. Moreover, such a class includes the simple majority rule if $n$
is odd.\textbf{\medskip}

\bigskip

The following proposition shows the existence of aggregation rules $G$ for
sufficientarian BGF that are anonymous, idempotent and strategy-proof by
showing the existence a class of aggregation rules $F:(\mathcal{A}%
_{\mathbf{X}})^{n}\longrightarrow \mathcal{A}_{\mathbf{X}}$ that are indeed
anonymous, idempotent and strategy-proof on the `natural' and large domain of
single-peaked preorders on $\mathcal{A}_{\mathbf{X}}$ defined above, and also
enable implementation of the corresponding $G$.\textbf{\medskip}

\bigskip

\textbf{Proposition 5. }Suppose $n$ is an odd number. Then, the simple
majority aggregation rule $G:(\mathcal{S}(\mathbf{X}^{n}$ $))^{n}%
\longrightarrow \mathcal{S}(\mathbf{X}^{n})$ for sufficientarian BGFs on
capability profiles in $\mathbf{X}^{n}$ is strategy-proof, namely it is
implementable by a protocol $\varphi_{\lbrack n]}^{-1}\circ F\circ \varphi
^{-1}$ (where $F:(\mathcal{A}_{\mathbf{X}})^{n}\longrightarrow \mathcal{A}%
_{\mathbf{X}}$ is an aggregation rule which is \textit{strategy-proof} on the
domain $\mathcal{D}_{B_{d}}^{n}$ of \ preference profiles over $\mathcal{A}%
_{\mathbf{X}}$ that are single-peaked in $(\mathcal{A}_{\mathbf{X}},B_{d}),$
and $\varphi$ is an isomorphism of $\mathcal{S}(\mathbf{X}^{n})$ and
$\mathcal{A}_{\mathbf{X}}$).\textbf{\medskip}

\bigskip

\textbf{Remark 6.} The foregoing protocol can be further refined by adjoining
to it a subprotocol consisting of a further strategy-proof aggregation rule to
endogenously select a committee of representative agents (or expert agents)
out of $[n]$ by the agents in $[n]$ themselves (e.g. by repeatedly sampling
$[n]$ without replacement, via \textit{uniform random dictatorship} as
implemented through a suitable modular arithmetic component). Notice that such
a subprotocol might also be used in order to select \textit{`endogenously' }on
every single occasion a \textit{president }endowed with a double vote when $n$
is even, to the effect of making the practical import of the \textit{oddness
}requirement for $n$ of Proposition 5 a quite minor one. Furthermore, the same
protocol can be easily extended to cover any case under which more than one
threshold is to be elicited (say, the sufficiency threshold system and a
typically higher `limitarian'\ one to establish preferential eligibility as
targets of redistributive taxation, and/or a typically lower poverty
threshold). To address such a double-threshold elicitation problem just
consider the product antichain space $\mathcal{A}_{\mathbf{X}}\times
\mathcal{A}_{\mathbf{X}}$ (which is also finite distributive lattice by
construction) as the new individual strategy space. Each agent is required to
select an ordered pair of antichains in $\mathcal{A}_{\mathbf{X}}$: the two
required threshold systems may be obtained by aggregating, respectively, the
meets and joins of each one of the submitted individual pairs. A similar
approach can be taken to address the task of selecting any \textit{finite
}number of threshold systems.

\bigskip

\section{Concluding remarks}

In this paper, under the general label of `sufficientarian grading rules',
basic sufficientarian rating rules have been introduced and characterized,
relying on binary grading functions (BGFs) as defined on a finite
capability-type space which consists in a finite product of finite linear
orders. And, remarkably, the characterization of basic sufficientarian rating
rules so provided only requires a version of three independent conditions
(i.e., Symmetry, Separability, and Isotony) that are largely used in the
extant literature, eschewing any reference to thresholds. Moreover, such a
characterization highlights a one-to-one correspondence between basic
sufficientarian rules and \textit{threshold systems }given by sets of mutually
incomparable (or antichains of) threshold-points of the capability-type space.
Sufficientarian \textit{ranking} rules induced by basic sufficientarian rating
rules are also defined and characterized, including \textit{sufficiency-count}
ranking rules and a pair of \textit{sufficiency-gap} ranking rules.

Furthermore, and last but not least, the largely neglected but crucial issue
concerning the identification of practical ways to select one specific
threshold system has been squarely addressed from a mechanism-design
perspective. In particular, it has been shown that nicely democratic and
strategy-proof opinion aggregation rules on the set of threshold-systems do
exist, and can be easily deployed to design the required protocols for
threshold-system selection. When it comes to implementation of sufficientarian
rules, It should be fully appreciated the expediency of such protocols. Of
course, selection of any specific sufficientarian rating or ranking rule as
required for any practical application amounts to choosing precisely one
threshold system. However, as previously noticed and discussed, adoption of a
sufficientarian stance may well suggest consideration and possibly
specification of \textit{further} threshold systems (perhaps a `higher' and/or
a `lower' one in addition to the sufficiency-threshold system itself). But
then, the very same protocol may well be used \textit{repeatedly} to select
\textit{all} the required threshold systems. Moreover, it should be noticed
that, as previously mentioned, BGFs and the rating rules they induce may be
used in other contexts including \textit{poverty} analysis,
\textit{exploitation }assessment, and \textit{expertise }evaluation, where
identification of thresholds and possibly caps could have a pivotal role. In
all of those cases, the class of protocols introduced in the present work
might provide a useful blueprint of sorts.

Finally, we should also like to mention a couple of possible significant
extensions of sufficientarian grading rules as defined in this paper.

To begin with, let us get back to the issue of \textit{burden }distribution,
which is perceived as a major challenge to sufficientarian principles by some
authors, including both advocates and critics of sufficientarianism (see,
e.g., Nielsen (2019) and Knight (2022)). We surmise that such an issue can be
successfully addressed relying on a relatively minor extension of the
sufficientarian grading rules as defined in this paper. Namely, it suffices to
enlarge the capability-type space adjoining a finite set of \textit{costly
}affordances/qualifications to denote burdens. Indeed, such
affordances/qualifications can also be represented by a finite number of
linearly ordered ranks denoted by nonnegative numbers in decreasing order
(with $0$ as the top rank, the most costly one). Accordingly, new extended
sufficientarian grading rules can be defined on such enlarged capability-type
space. But then, the rest of the analysis provided in the previous sections
can be replicated on such an extended space.

There is also a further extension of the BGF-based approach to
sufficientarianism we should mention. In the present paper individual
affordances/achievements related to \textit{public goods }have been explicitly
ignored. However our model can be easily extended to cover such public-good
related capabilities, including \textit{global} public goods of which basic
scientific research activities and outputs are a prominent example. In a
well-known and most influential paper, Merton described the rules underlying
proper and successful scientific research activities in terms of four
principles summarized by the acronym CUDOS (namely, `communism',
`universalism', `disinterestedness', `organized skepticism': see Merton
(1942)). Arguably, an augmented version of sufficientarianism as discussed
above with a specific public-good related component of the capability-type
space (possibly enriched with a `burden'-subspace as discussed above) might be
fruitfully used to contribute a precise version of distribution rules emboding
at least some of the four Merton's principles, including at least the first
two of his well-known and highly regarded list. However, further details
concerning those possible extensions of the model presented in this paper are
best left as a possible topic for future research.

\section{Appendix\textbf{\medskip}}

\textbf{Proof of Proposition 1}

$\left(  \Longrightarrow \right)  $ Suppose $g$ is sufficientarian, and
consider any $\mathbf{x}_{[n]}$, $\mathbf{x}_{[n]}^{\prime}\in \mathbf{X}^{n}$
with $\mathbf{x}_{i}=\mathbf{x}_{i}^{\prime}$. By definition, there exists an
antichain $(\mathbf{x}_{1}^{\ast},...,\mathbf{x}_{k}^{\ast})$ of $\mathbf{X}$
such that for all $i\in \lbrack n]$, there is an $\mathbf{x}_{h}^{\ast}$,
$h=1,...,k$, with $g_{i}(\mathbf{x}_{[n]}\mathbf{)=}1$ if $\mathbf{x}%
_{h}^{\ast}\leqslant \mathbf{x}_{i}$ iff $\mathbf{x}_{h}^{\ast}\leqslant
\mathbf{x}_{i}^{\prime}$ iff $g_{i}(\mathbf{x}_{[n]}^{\prime}\mathbf{)=}1$.
Hence, $g$ is of course separable. Moreover, consider any $\mathbf{x}%
_{[n]},\mathbf{x}_{[n]}^{\prime}\in \mathbf{X}^{n}$ with $\mathbf{x}%
_{[n]}\leqslant \mathbf{x}_{[n]}^{\prime}$, and any $i\in N$. If $g_{i}%
(\mathbf{x}_{[n]}\mathbf{)=}0$ there is nothing to prove. So, suppose that
$g_{i}(\mathbf{x}_{[n]}\mathbf{)=}1$. Hence, by definition, there is an
$\mathbf{x}_{j}^{\ast}\leqslant \mathbf{x}_{i}\mathbf{\leqslant}$
$\mathbf{x}_{i}^{\prime}$, thus, $g_{i}(\mathbf{x}_{[n]}^{\prime}\mathbf{)=}%
1$. It follows that $g$ is also isotonic. Finally, consider any $\mathbf{x}%
\in \mathbf{X}^{[n]}$, any permutation $\sigma:[n]\longrightarrow \lbrack n]$
and $i,j\in \lbrack n]$ with $j:=\sigma(i)$. By definition, $g_{i}%
(\mathbf{x}_{[n]}\mathbf{)=}1$ if and only if $g_{j}(\mathbf{x}_{[n]}%
)=g_{\sigma(i)}(\mathbf{x}_{[n]})=g_{i}(\mathbf{x}_{\sigma \lbrack n]})=1$.
Thus, $g$ is indeed symmetric.

$\left(  \Longleftarrow \right)  $\ Suppose $g$ is isotonic, separable and
symmetric, and consider any $\mathbf{x}_{[n]}\in \mathbf{X}^{n}$, $i\in \lbrack
n]$ and $g_{i}^{-1}(1):=$ $\left \{  \mathbf{x}_{[n]}\in \mathbf{X}^{n}%
:g_{i}(\mathbf{x}_{[n]}\mathbf{)}=1\right \}  $.

Now, for any $i\in \lbrack n]$, take $\mathbf{X(}i,1,g):=\left \{  \mathbf{x\in
X:x=x}_{i}\text{ for some }\mathbf{x}_{[n]}\in g_{i}^{-1}(1)\right \}  $, and
posit $\mathbf{X}^{u}\mathbf{(}i,1,g):=\left \{  \mathbf{x\in X:x}_{i}%
\leqslant \mathbf{x}\text{ for some }\mathbf{x}\in \mathbf{X(}i,1,g)\right \}  $,
and observe that by \textit{isotony} and \textit{separability} $\mathbf{X}%
^{u}\mathbf{(}i,1,g)\subseteq \mathbf{X(}i,1,g)$, whence, in fact,
$\mathbf{X(}i,1,g)=\mathbf{X}^{u}\mathbf{(}i,1,g)$. It follows that any
$\mathbf{X(}i,1,g)$ is actually an \emph{order filter} of the partially
ordered set $\mathbf{X}$ (which is a finite product of finite linearly ordered
sets, hence, in\ particular, a finite distributive lattice). Notice however
that any such order filter $\mathbf{X}^{u}\mathbf{(}i,1,g)$ is uniquely
determined by its \textit{basis }namely the set $\mathbf{X}_{\min}%
^{u}\mathbf{(}i,1,g)$ of its \textit{minimal }elements, and as it is easily
checked it must be the case for every $\mathbf{x,x}^{\prime}\in \mathbf{X}%
_{\min}^{u}\mathbf{(}i,1,g)$ if $\mathbf{x\neq}$ $\mathbf{x}^{\prime}$ then
$\mathbf{x\nleqslant x}^{\prime}$ namely $\mathbf{X}_{\min}^{u}\mathbf{(}%
i,1,g)$ is by construction a (finite) \textit{antichain }$\left(
\mathbf{x}_{i1},...,\mathbf{x}_{ik}\right)  $ of $\mathbf{X}$\textbf{.
}Moreover, symmetry implies that $\mathbf{X}_{\min}^{u}\mathbf{(}%
i,1,g)=\mathbf{X}_{\min}^{u}\mathbf{(}j,1,g)=\mathcal{X}^{\ast}:=\left(
\mathbf{x}_{1}^{\ast},...,\mathbf{x}_{k}^{\ast}\right)  $ for any
$i,j\in \lbrack n]$ (otherwise there is no guarantee that for every $i,j$ and
$\mathbf{x}_{[n]}$ both $g_{i}(\mathbf{x}_{[n]}\mathbf{)=}1$ and
$g_{i}(\mathbf{x}_{\sigma \lbrack n]})=1$ hold). But then, it follows that for
any $\mathbf{x}_{[n]}\in \mathbf{X}^{n}$ and any $i\in \lbrack n]$,
$g_{i}(\mathbf{x}_{[n]})=1$ if and only if there exists $\mathbf{x}_{h}^{\ast
}\in \left \{  \mathbf{x}_{1}^{\ast},...,\mathbf{x}_{k}^{\ast}\right \}  $ such
that $\mathbf{x}_{h}^{\ast}\leqslant \mathbf{x}_{i}$, i.e. $g$ is indeed a
\textit{sufficientarian BGF }as required.\medskip

The foregoing characterization of sufficientarian BGFs is tight. To check
validity of that statement consider the following three examples:\smallskip

$(i)$ $g^{\prime}:\mathbf{X}^{n}\rightarrow \left \{  0,1\right \}  ^{n}$ is such
that there exist a positive integer $k$ and $\mathbf{x}_{1}^{\ast
},...,\mathbf{x}_{k}^{\ast}\in \mathbf{X}$ with $\mathcal{X}^{\ast}%
=(\mathbf{x}_{1}^{\ast},...,\mathbf{x}_{k}^{\ast})$ being an \textit{antichain
}of $\mathbf{X}$ (namely $\mathbf{x}_{j}^{\ast}$ $\nleqslant \mathbf{x}%
_{h}^{\ast}$ for every $j,h=1,...,k$ with $j\neq h$) and for every
$\mathbf{x}_{N}\mathbf{\in X}^{N}$, $g_{i}^{\prime}(\mathbf{x}_{N}%
\mathbf{)=}1$ if and only if $\mathbf{x}_{i}\leqslant \mathbf{x}_{h}^{\ast}$
for some $h=1,...,k|$. It can be checked that $g^{\prime}$ is separable and
symmetric but not isotonic (indeed, it is antitonic);\smallskip

$(ii)$ $g^{\prime \prime}:\mathbf{X}^{n}\rightarrow \left \{  0,1\right \}  ^{n} $
is such that there exist a positive integer $k$ and $\mathbf{x}_{1}^{\ast
},...,\mathbf{x}_{k}^{\ast}\in \mathbf{X}$ with $\mathcal{X}^{\ast}=$
$(\mathbf{x}_{1}^{\ast},...,\mathbf{x}_{k}^{\ast})$ being an \textit{antichain
}of $\mathbf{X}$ and for every $\mathbf{x}_{[n]}\mathbf{\in X}^{n}$,
$g_{i}^{\prime \prime}(\mathbf{x}_{[n]}\mathbf{)=}1$ if and only if
$\mathbf{x}_{i}\leqslant \mathbf{x}_{h}^{\ast}$ and $\mathbf{x}_{j}%
\leqslant \mathbf{x}_{h}^{\ast}$ for some $h=1,...,k$, and $j\in \lbrack
n]\setminus \left \{  i\right \}  $. Clearly, $g^{\prime \prime}$ is isotonic and
symmetric but not separable.\smallskip

$(iii)$\ $g^{\prime \prime \prime}:\mathbf{X}^{n}\rightarrow \left \{
0,1\right \}  ^{n}$ is such that for every $i$ there exist a positive integer
$k_{i}$ and $\mathbf{x}_{1}^{\ast},...,\mathbf{x}_{k_{i}}^{\ast}\in \mathbf{X}$
with $(\mathbf{x}_{1}^{\ast},...,\mathbf{x}_{k_{i}}^{\ast})$ being an
\textit{antichain }of $\mathbf{X}$ and $k_{i}\neq k_{j}$ for some
$i,j\in \lbrack n]$, and for every $\mathbf{x}_{[n]}\mathbf{\in X}^{n}$,
$g_{i}^{\prime \prime \prime}(\mathbf{x}_{[n]}\mathbf{)=}1$ if and only if
$\mathbf{x}_{i}\leqslant \mathbf{x}_{h}^{\ast}$ for some $h=1,...,k_{i}$. It
can be checked that $g^{\prime \prime \prime}$ is separable and isotonic but not
symmetric. $\  \qquad \qquad \qquad \qquad \qquad \square$

\bigskip

\textbf{Proof of Claim 1}

Let $\mathbf{x}_{[n]},\mathbf{x}_{[n]}^{\prime}\in \mathbf{X}^{n}$ and
$\mathbf{x}_{[n]}\geqslant_{g}\mathbf{x}_{[n]}^{\prime}$ , i.e. by definition
$g(\mathbf{x}_{[n]})\geqslant g(\mathbf{x}_{[n]}^{\prime}).$ Two cases are to
be distinguished : $(i)$ $g_{i}(\mathbf{x}_{[n]})=1$ for all $i\in \lbrack n]$
and $(ii)$ $g_{i}(\mathbf{x}_{[n]})=0$ for some $i\in \lbrack n]$. In the first
case, $\mathbf{x}_{[n]}\widehat{\succcurlyeq}_{g}\mathbf{x}_{[n]}^{\prime}$ by
definition. In the second case $0=g_{i}(\mathbf{x}_{[n]})\geq g_{i}%
(\mathbf{x}_{[n]}^{\prime})\geq0$ whence $g_{i}(\mathbf{x}_{[n]}^{\prime})=0$
which in turn implies $\mathbf{x}_{[n]}\widehat{\succcurlyeq}_{g}%
\mathbf{x}_{[n]}^{\prime}$, by definition. It follows that $\widehat
{\succcurlyeq}_{g}$ is indeed an \textit{extension }of $\geqslant_{g}$. It is
also easily checked that $\widehat{\succcurlyeq}_{g}$ is a total preorder.
That is so because both \textit{reflexivity} and \textit{connectedness} of
$\widehat{\succcurlyeq}_{g}$ follow immediately by its definition. Moreover,
$\widehat{\succcurlyeq}_{g}$ is also transitive: indeed, suppose that
$\mathbf{x}_{[n]}\widehat{\succcurlyeq}_{g}\mathbf{x}_{[n]}^{\prime}$ and
$\mathbf{x}_{[n]}^{\prime}\widehat{\succcurlyeq}_{g}\mathbf{x}_{[n]}%
^{\prime \prime}$ . Then, again, cases $(i)$ and $(ii)$ are to be distinguished
concerning $\mathbf{x}_{[n]}$. If $(i)$ holds, then $\mathbf{x}_{[n]}%
\widehat{\succcurlyeq}_{g}\mathbf{x}_{[n]}^{\prime \prime}$ by definition. If
on the contrary $(ii)$ holds, $\mathbf{x}_{[n]}\widehat{\succcurlyeq}%
_{g}\mathbf{x}_{[n]}^{\prime}$ implies that $g_{i}(\mathbf{x}_{[n]}^{\prime
})=0$ for some $i\in \lbrack n]$ . But then, $\mathbf{x}_{[n]}^{\prime}%
\widehat{\succcurlyeq}_{g}\mathbf{x}_{[n]}^{\prime \prime}$ implies in turn
that $g_{i}(\mathbf{x}_{[n]}^{\prime \prime})=0$ for some $i\in \lbrack n]$. It
follows again that $\mathbf{x}_{[n]}\widehat{\succcurlyeq}_{g}\mathbf{x}%
_{[n]}^{\prime \prime}$ by definition, \textit{transitivity} of $\widehat
{\succcurlyeq}_{g}$ is thus confirmed, and $\widehat{\succcurlyeq}_{g}$ is a
well-defined total preorder.

To check that $\widehat{\succcurlyeq}_{g}$ is also \textit{top-faithful},
observe that by definition if $\mathbf{x}_{[n]}\widehat{\succcurlyeq}%
_{g}\mathbf{x}_{[n]}^{\prime}$ for every $\mathbf{x}_{[n]}^{\prime}\in$
$\mathbf{X}^{n}$ then, since $g$ is onto, it must be the case that
$g_{i}(\mathbf{x}_{[n]})=1$ for all $i\in \lbrack n]$, which in turn implies
that $\mathbf{x}_{[n]}\in top_{g}(\mathbf{X}^{[n]})$. Conversely, if
$\mathbf{x}_{[n]}\in top_{g}(\mathbf{X}^{n})$ then by definition
$g(\mathbf{x}_{[n]})=\mathbf{1}$, which in turn implies that $\mathbf{x}%
_{[n]}\widehat{\succcurlyeq}_{g}\mathbf{x}_{[n]}^{\prime}$ for every
$\mathbf{x}_{[n]}^{\prime}\in$ $\mathbf{X}^{n}$, by definition of
$\widehat{\succcurlyeq}_{g}$ itself.

Finally, suppose that $\succcurlyeq$ is a total preorder on $\mathbf{X}^{n}$
which is also a top-faithful extension of $\geqslant_{g}$, and consider any
pair $\mathbf{x}_{[n]},\mathbf{x}_{[n]}^{\prime}\in$ $\mathbf{X}^{n}$ such
that $\mathbf{x}_{[n]}\succcurlyeq \mathbf{x}_{[n]}^{\prime}$. Two cases are to
be distinguished: $(i)$ $\left \{  \mathbf{x}_{[n]},\mathbf{x}_{[n]}^{\prime
}\right \}  \cap$ $top_{g}(\mathbf{X}^{n})\neq \emptyset$. In this case, since
by assumption $top_{g}(\mathbf{X}^{n})=\max(\succcurlyeq)=\max(\widehat
{\succcurlyeq}_{g})$ it must be the case that $\mathbf{x}_{[n]}\widehat
{\succcurlyeq}_{g}\mathbf{x}_{[n]}^{\prime}$ as well, by construction; (ii)
$\left \{  \mathbf{x}_{[n]},\mathbf{x}_{[n]}^{\prime}\right \}  \cap$
$top_{g}(\mathbf{X}^{n})=\emptyset$. But then, both $\mathbf{x}_{[n]}%
\widehat{\succcurlyeq}_{g}\mathbf{x}_{[n]}^{\prime}$ and $\mathbf{x}%
_{[n]}^{\prime}\widehat{\succcurlyeq}_{g}\mathbf{x}_{[n]}$ do hold, by
construction of $\widehat{\succcurlyeq}_{g}$. It follows that $\succcurlyeq
\subseteq \widehat{\succcurlyeq}_{g}$ as required, and the proof of our Claim
is now complete.$\qquad \qquad \qquad \qquad \qquad \qquad \qquad \qquad \square$

\bigskip

\textbf{Proof of Proposition 2}

To begin with, notice that $\succcurlyeq_{g}$ is by construction both a total
preorder on $X^{n}$ and an extension of $\geqslant_{g}$. That $\succcurlyeq
_{g}$ satisfies AN and SM is also clearly the case. Indeed, for any
$\mathbf{x}_{[n]}\in \mathbf{X}^{n}$ and any permutation $\sigma
:[n]\longrightarrow \lbrack n]$, $\ |\left \{  i\in \lbrack n]:g_{i}%
(\mathbf{x}_{[n]})=1\right \}  |=|\left \{  i\in \lbrack n]:g_{i}(\mathbf{x}%
_{\sigma \lbrack n]})=1\right \}  |$ \ by construction. Thus $\mathbf{x}%
_{[n]}\sim_{g}\mathbf{x}_{\sigma \lbrack n]}$ (where $\sim_{g}$ is of course
the symmetric component of $\succcurlyeq_{g}$) and AN holds. Moreover, let
$\mathbf{x}_{[n]},\mathbf{x}_{[n]}^{\prime}\in \mathbf{X}^{n}$ and $i\in \lbrack
n]$ be such that $\ g_{l}(\mathbf{x}_{[n]})=g_{l}(\mathbf{x}_{[n]}^{\prime})$
for any $l\in \lbrack n]\backslash \left \{  i\right \}  $, $g_{i}(\mathbf{x}%
_{[n]})=1$ and $g_{i}(\mathbf{x}_{[n]}^{\prime})=0$. Then, by definition
$|\left \{  i\in \lbrack n]:g_{i}(\mathbf{x}_{[n]})=1\right \}  |=|\left \{
i\in \lbrack n]:g_{i}(\mathbf{x}_{[n]}^{\prime})=1\right \}  |+1>|\left \{
i\in \lbrack n]:g_{i}(\mathbf{x}_{[n]}^{\prime})=1\right \}  |$. Therefore,
$\mathbf{x}_{[n]}\succ_{g}\mathbf{x}_{[n]}^{\prime}$ and SM also holds.

Conversely, let $\succcurlyeq$ be a total preorder on $\mathbf{X}^{n}$ that is
an extension of the partial order $\geqslant_{g}$ and satisfies both AN and
SM, and consider any $\mathbf{x}_{[n]},\mathbf{x}_{[n]}^{\prime}\in
\mathbf{X}^{n}$. Next, suppose w.l.o.g. that $|\left \{  i\in \lbrack
n]:g_{i}(\mathbf{x}_{[n]})=1\right \}  |\geq|\left \{  i\in \lbrack
n]:g_{i}(\mathbf{x}_{[n]}^{\prime})=1\right \}  |$ .

We may distinguish two cases:

(i) $|\left \{  i\in \lbrack n]:g_{i}(\mathbf{x}_{[n]})=1\right \}  |=|\left \{
i\in \lbrack n]:g_{i}(\mathbf{x}_{[n]}^{\prime})=1\right \}  |$ . If that is the
case, then there exists a permutation $\sigma:[n]\rightarrow \lbrack n]$ such
that, for any $i\in \lbrack n]$, $g_{i}(\mathbf{x}_{[n]})=1$ if and only if
$g_{\sigma(i)}(\mathbf{x}_{\sigma \lbrack n]}^{\prime})=1$. Now, $\mathbf{x}%
_{[n]}^{\prime}\sim \mathbf{x}_{\sigma \lbrack n]}^{\prime}$ by Anonymity of
$\succcurlyeq$, and $g(\mathbf{x}_{[n]})=g(\mathbf{x}_{\sigma \lbrack
n]}^{\prime}$ $).$ Therefore, $\mathbf{x}_{[n]}\sim \mathbf{x}_{\sigma \lbrack
n]}^{\prime}$ since $\succcurlyeq$ is an extension of $\geqslant_{g}$. It
follows\ that $\mathbf{x}_{[n]}\sim \mathbf{x}_{[n]}^{\prime}$\ by transitivity
of $\succcurlyeq$. \  \  \  \  \  \  \  \  \ 

(ii) $|\left \{  i\in \lbrack n]:g_{i}(\mathbf{x}_{[n]})=1\right \}  |>|\left \{
i\in \lbrack n]:g_{i}(\mathbf{x}_{[n]}^{\prime})=1\right \}  |$,

with $|\left \{  i\in \lbrack n]:g_{i}(\mathbf{x}_{[n]})=1\right \}  |=k$, and
$|\left \{  i\in \lbrack n]:g_{i}(\mathbf{x}_{[n]}^{\prime})=1\right \}  |=h$. If
that is the case, there exist a permutation

$\sigma^{\ast}:[n]\rightarrow \lbrack n]$ and $i_{1}^{\ast},...,i_{k--h}^{\ast
}$ $\in \left \{  i\in \lbrack n]:g_{i}(\mathbf{x}_{[n]})=1\right \}
\setminus \left \{  i\in \lbrack n]:g_{i}(\mathbf{x}_{[n]}^{\prime})=1\right \}  $
such that, for any $i\in \lbrack n]$,

$g_{i}(\mathbf{x}_{\sigma^{\ast}[n]})=1$ iff $i\in \left \{  i_{1}^{\ast
},...,i_{k--h}^{\ast}\right \}  $ $\cup$\ $\left \{  i\in \lbrack n]:g_{i}%
(\mathbf{x}_{[n]}^{\prime})=1\right \}  $.

Therefore, by construction of $\mathbf{x}_{\sigma^{\ast}[n]}$ and Anonymity of
$\succcurlyeq$, it follows that $\mathbf{x}_{[n]}\sim \mathbf{x}_{\sigma^{\ast
}[n]}$\textbf{. }Next, consider a sequence $\mathbf{x}_{\sigma^{\ast}[n]}%
^{j}\in \mathbf{X}^{n}$, $j=0,1,...,k-h$ defined as follows: $\mathbf{x}%
_{\sigma^{\ast}[n]}^{j}=(\mathbf{x}_{\sigma^{\ast}[n]|[n]^{\ast j}}%
,\mathbf{x}_{[n]|[n]\setminus \lbrack n]^{\ast j}}^{\prime})$ where $[n]^{\ast
j}:=\left \{  i\in \lbrack n]:g_{i}(\mathbf{x}_{[n]}^{\prime})=1\right \}
\cup \left \{  i_{1}^{\ast},...,i_{j}^{\ast}\right \}  $ for $j=1,...,k-h$, while
$\mathbf{x}_{\sigma^{\ast}[n]}^{j}=\mathbf{x}_{[n]}^{\prime}$ for $j=0.$
Moreover, observe that, by construction of that sequence and Strict
Monotonicity w.r.t. $g$ of $\succcurlyeq$, \ $\mathbf{x}_{\sigma^{\ast}%
[n]}^{j+1}\succ \mathbf{x}_{\sigma^{\ast}[n]}^{j}$, for any $j=0,1,...,k-(h+1)$%
. Hence, in particular, $\mathbf{x}_{\sigma^{\ast}[n]}^{1}\succ \mathbf{x}%
_{[n]}^{\prime}$, while $\mathbf{x}_{\sigma^{\ast}[n]}^{k-h}$ $=$
$\mathbf{x}_{\sigma^{\ast}[n]}\sim \mathbf{x}_{[n]}$. As a consequence,
$\mathbf{x}_{[n]}\succ \mathbf{x}_{[n]}^{\prime}$ holds, by transitivity of
$\succcurlyeq$. It follows that $|\left \{  i\in \lbrack n]:g_{i}(\mathbf{x}%
_{[n]})=1\right \}  |\geq|\left \{  i\in \lbrack n]:g_{i}(\mathbf{x}%
_{[n]}^{\prime})=1\right \}  |$ implies $\mathbf{x}_{[n]}\succcurlyeq
\mathbf{x}_{[n]}^{\prime}$. Hence $\succcurlyeq_{g}\subseteq \succcurlyeq$.
Now, suppose that for some $\mathbf{x}_{[n]},\mathbf{x}_{[n]}^{\prime}%
\in \mathbf{X}^{n}$, both $\mathbf{x}_{[n]}\succcurlyeq \mathbf{x}_{[n]}%
^{\prime}$ and \textit{not }$\mathbf{x}_{[n]}\succcurlyeq_{g}\mathbf{x}%
_{[n]}^{\prime}$ hold. Since $\succcurlyeq_{g}$ is a total preorder, it must
be the case that $\mathbf{x}_{[n]}^{\prime}\succ_{g}$ $\mathbf{x}_{[n]}$ and
thus $\mathbf{x}_{[n]}^{\prime}\succcurlyeq$ $\mathbf{x}_{[n]}$ . Therefore,
by definition $|\left \{  i\in \lbrack n]:g_{i}(\mathbf{x}_{[n]}^{\prime
})=1\right \}  |>|\left \{  i\in \lbrack n]:g_{i}(\mathbf{x}_{[n]})=1\right \}
|.$But then, there exist $\mathbf{x}_{[n]}^{\prime \prime},\mathbf{x}%
_{[n]}^{\prime \prime \prime}$ $\in \mathbf{X}^{n}$ such that $\mathbf{x}%
_{[n]}^{\prime \prime}\sim_{g}\mathbf{x}_{[n]}$, $\mathbf{x}_{[n]}%
^{\prime \prime \prime}$ $\sim_{g}\mathbf{x}_{[n]}^{\prime}$ and $\mathbf{x}%
_{[n]}^{\prime \prime \prime}>_{g}$ $\mathbf{x}_{[n]}^{\prime \prime}$. Thus,
$\mathbf{x}_{[n]}^{\prime \prime \prime}\succ$ $\mathbf{x}_{[n]}^{\prime \prime}$
as well since $\succcurlyeq$ is also an extension of $\geqslant_{g}.$ On the
other hand $\succcurlyeq_{g}\subseteq \succcurlyeq$,$\  \mathbf{x}_{[n]}%
^{\prime \prime}\sim_{g}\mathbf{x}_{[n]}$ and $\mathbf{x}_{[n]}^{\prime
\prime \prime}$ $\sim_{g}\mathbf{x}_{[n]}^{\prime}$ imply $\mathbf{x}%
_{[n]}^{\prime \prime}\sim \mathbf{x}_{[n]}$ and $\mathbf{x}_{[n]}^{\prime
\prime \prime}$ $\sim \mathbf{x}_{[n]}^{\prime}$ hence in particular
$\mathbf{x}_{[n]}^{\prime}\succ \mathbf{x}_{[n]}$, a contradiction. It follows
that $\succcurlyeq \subseteq \succcurlyeq_{g}$as well. Hence $\succcurlyeq
_{g}=\succcurlyeq$ and the proof is complete.$\qquad \qquad \qquad \qquad
\qquad \qquad \square$

\bigskip

\textbf{Proof of Proposition 3}

$\Longleftarrow$ Since we are dealing here exclusively with the additive
fragment of elementary arithmetic, and addition (which the `natural' order
$\leq$ over $\mathbb{Z}_{+}$ relies on for its very definition) is defined by
means of the successor function $\mathrm{S:}\mathbb{Z}_{+}\longrightarrow
\mathbb{Z}_{+}$\ a few basic points are worth recalling here. Namely,
$\mathrm{S}$ is a one-to-one function such that every \textit{positive}
integer $x$ is the successor of a nonnegative integer, while $0$ is not, and
of course \textrm{S}$(0)=1$. Moreover, addition $+$ is defined by the rule:
$x+\mathrm{S}(y)=\mathrm{S}(y)+x:=\mathrm{S}(x+y)$, $x+0=0+x:=x$, and the
`natural' order $\leq$ is defined by the rule: for any $x,y\in \mathbb{Z}_{+}$,
$x\leq y$ if there exists $z\in \mathbb{Z}_{+}$ such that $y=x+z$. Clearly, it
follows that addition is \textit{commutative} by definition, and it can be
easily checked that it is also \textit{associative}.

From all of the above it follows that $\succcurlyeq_{\delta_{g}}^{\ast av}%
$satisfies AN as a plain consequence of associativity and commutativity of
addition, and both S-ANT and RTI as a consequence of the definition of $\leq$
and of associativity and commutativity of addition.

$\Longrightarrow$Now, suppose $\succeq$ is a total preorder over $\mathcal{Z}$
that satisfies AN, S-ANT, RTI. The present proof relies on the definition of a
family of auxiliary total preorders over $\mathcal{Z}$ indexed by
$[n]\setminus \left \{  1\right \}  $, namely $\left \{  \succeq^{m}%
:m=2,...,n\right \}  $ defined by the following rule: $\succeq^{m}$

is such that for any $\boldsymbol{d,d}^{\prime}\in$ $\mathcal{Z}$ $\ $and any
subset $M\subseteq \lbrack n]$ of cardinality $m$ with $d_{h}=d_{h}^{\prime}$
for every $h\in \lbrack n]\setminus M$, $\mathbf{d\succeq d}^{\prime}$ if and
only if $\sum_{i\in M}d_{i}^{\prime}\geq \sum_{i\in M}^{n}d_{i}$.

Then, the bulk of the proof amounts to a sort of restricted induction argument
on $[n]\setminus \left \{  1\right \}  $ which consists of two steps, namely: (I)
$\succeq=\succeq^{2}$, (II) for any $m$, $2\leq m<n$ \ if $\succeq=\succeq
^{m}$then $\succeq=\succeq^{m+1}$. That is so, because once both (I) and (II)
are established one only has to observe that they jointly imply that in
particular $\succeq=\succeq^{n}$. Moreover, as it is easily checked
$\succeq^{n}=\succcurlyeq_{\delta_{g}}^{\ast av}$, by definition. Thus, it
follows that $\succeq=\succcurlyeq_{\delta_{g}}^{\ast av}$as required.

Therefore, we only need to prove that both (I) and (II) hold true.

Step (I). Let us first consider any $i,j\in \lbrack n]$, $i\neq j$ and
$\boldsymbol{d,d}^{\prime}\in$ $\mathcal{Z}$ such that $d_{i}^{\prime}%
+d_{j}^{\prime}\geq d_{i}+d_{j}$, and $d_{h}=d_{h}^{\prime}$ for every
$h\in \lbrack n],i\neq h\neq j$ (and suppose without loss of generality that
$i=1$ and $j=2$).

If $\mathbf{d=d}^{\prime}$ then of course $\mathbf{d\sim d}^{\prime}$ by
reflexivity of $\succeq$, so we assume without loss of generality that
$\boldsymbol{d\neq d}^{\prime}$. Five cases are to be considered:

(i) $d_{1}^{\prime}>d_{1}$ and $d_{2}^{\prime}\geq d_{2}$ or $d_{1}^{\prime
}\geq d_{1}$ and $d_{2}^{\prime}>d_{2}$ : in this case $\boldsymbol{d\succ
d}^{\prime}$ follows immediately from S-ANT;

(ii) $d_{1}^{\prime}>d_{1}$ , $d_{2}>d_{2}^{\prime}$ and $d_{1}^{\prime}%
+d_{2}^{\prime}>d_{1}+d_{2}$: let us first denote by $\mathbf{e}_{h}%
:=(e_{h1},...,e_{hn})\in \mathbb{Z}^{n}$, for any $h\in \lbrack n]$, the
\textit{unit vector} with $e_{hh}=1$ and $e_{hh^{\prime}}=0$ for any
$h^{\prime}\in \lbrack n]$, $h^{\prime}\neq h.$ Moreover, let $k_{1}:=$
$(d_{1}^{\prime}-d_{1})>0$, and $k_{2}:=(d_{2}-d_{2}^{\prime})>0$, whence
$k:=(k_{1}-k_{2})>0$. Furthermore, AN implies that $\mathbf{e}_{h}%
\sim \mathbf{e}_{h^{\prime}}$ for any $h,h^{\prime}\in \lbrack n]$: hence, in
particular, $\mathbf{e}_{1}\sim \mathbf{e}_{2}$. Next, observe that

$\mathbf{d}^{\prime}\mathbf{=}d_{1}\mathbf{e}_{1}+k_{1}\mathbf{e}_{1}%
+d_{2}^{\prime}\mathbf{e}_{2}$ and $\mathbf{d=}d_{1}\mathbf{e}_{1}%
+d_{2}^{\prime}\mathbf{e}_{2}+k_{2}\mathbf{e}_{2}$.

Now, RTI implies that $\mathbf{e}_{1}+\mathbf{e}_{2}\sim \mathbf{e}%
_{2}+\mathbf{e}_{2}$ and $\mathbf{e}_{2}+\mathbf{e}_{1}\sim \mathbf{e}%
_{1}+\mathbf{e}_{1}$ whence, by transitivity of $\succeq$ and commutativity of
addition, $\mathbf{e}_{1}+\mathbf{e}_{2}\sim \mathbf{e}_{1}+\mathbf{e}_{1}%
\sim \mathbf{e}_{2}+\mathbf{e}_{2}$. But then, it also follows from a repeated
application of RTI and commutativity plus associativity of addition that
$\mathbf{d}^{\prime}\mathbf{\sim}d_{1}\mathbf{e}_{1}+d_{2}^{\prime}%
\mathbf{e}_{1}+k_{2}\mathbf{e}_{1}+k\mathbf{e}_{1}$, and $\mathbf{d\sim}%
d_{1}\mathbf{e}_{1}+d_{2}^{\prime}\mathbf{e}_{1}+k_{2}\mathbf{e}_{1}$. Thus,
$\mathbf{d}^{\prime}\sim \mathbf{d+}k\mathbf{e}_{1}$, hence $\mathbf{d\succ
d}^{\prime}$ by transitivity of $\succeq$ since $\mathbf{d\succ d+}%
k\mathbf{e}_{1}$\textbf{\ }by S-ANT.

(iii) $d_{i}^{\prime}>d_{i}$ , $d_{j}>d_{j}^{\prime}$ and $d_{i}^{\prime
}+d_{j}^{\prime}=d_{i}+d_{j}$; by replicating the argument previously used for
case (ii) and using the same notation, we obtain the same identities with
$k_{1}=k_{2}$ and consequently $k=0$. It follows that $\mathbf{d\sim
d}^{\prime}$ .

(iv) $d_{i}>d_{i}^{\prime}$ , $d_{j}^{\prime}>d_{j}$ and $d_{i}^{\prime}%
+d_{j}^{\prime}>d_{i}+d_{j}$; \ by replicating the same argument used for case
(ii) and using a similar notation, with $k_{1}:=(d_{1}-d_{1}^{\prime})>0$,
$k_{2}:=(d_{2}^{\prime}-d_{2})>0$ and $k:=(k_{2}-k_{1})>0$, we also obtain
$\mathbf{d\succ d}^{\prime}$.

(v) $d_{i}>d_{i}^{\prime}$ , $d_{j}^{\prime}>d_{j}$ and $d_{i}^{\prime}%
+d_{j}^{\prime}=d_{i}+d_{j}$; \ by replicating the same argument used for case
(iii) and using a similar notation, with $k_{1}:=(d_{1}-d_{1}^{\prime})>0$,
$k_{2}:=(d_{2}^{\prime}-d_{2})>0$ and $k:=(k_{2}-k_{1})=0$, we obtain again
$\mathbf{d\sim d}^{\prime}$ .

Thus, we have shown that for any $\boldsymbol{d,d}^{\prime}\in$ $\mathcal{Z}$
$\ $and any $i,j\in \lbrack n]$, $i\neq j$ such that $d_{i}^{\prime}%
+d_{j}^{\prime}\geq d_{i}+d_{j}$, and $d_{h}=d_{h}^{\prime}$ for every
$h\in \lbrack n],i\neq h\neq j$, it must be the case that $\mathbf{d\succeq
d}^{\prime}$, and $\mathbf{d\succ d}^{\prime}$ if in particular $d_{i}%
^{\prime}+d_{j}^{\prime}>d_{i}+d_{j}.$

Conversely, suppose that $\mathbf{d\succeq d}^{\prime}$ for some
$\boldsymbol{d,d}^{\prime}\in$ $\mathcal{Z}$ such that for some $i,j\in \lbrack
n]$, $i\neq j$, both $d_{h}=d_{h}^{\prime}$ for every $h\in \lbrack n],i\neq
h\neq j$ yet $d_{i}+d_{j}>d_{i}^{\prime}+d_{j}^{\prime}.$ Then, it follows
from the previous argument that $\mathbf{d}^{\prime}\succ \mathbf{d}$, a
contradiction. Thus, we have in fact shown that for any $\boldsymbol{d,d}%
^{\prime}\in$ $\mathcal{Z}$ $\ $and any $i,j\in \lbrack n]$, $i\neq j$ such
that $d_{h}=d_{h}^{\prime}$ for every $h\in \lbrack n],i\neq h\neq j$,
$\mathbf{d\succeq d}^{\prime}$ if and only if $d_{i}^{\prime}+d_{j}^{\prime
}\geq d_{i}+d_{j}$ or, equivalently, that $\succeq=\succeq^{2}$.

Step (II). Suppose that $\succeq=\succeq^{m}$for some $m$, $2\leq m<n$, and
consider any pair of vectors $\boldsymbol{d,d}^{\prime}\in$ $\mathcal{Z}$ such
that $d_{h}=d_{h}^{\prime}$ for every $h\in M^{\prime}:=[n]\setminus \left \{
1,m+1\right \}  $. Then, consider a third vector $\mathbf{d}^{\ast}%
\in \mathcal{Z}$ such that : $d_{1}^{\ast}+d_{m+1}^{\ast}=d_{1}+d_{m+1}$,
$d_{m+1}^{\ast}=d_{m+1}^{\prime}$, and $d_{h}^{\ast}=d_{h\text{ }}$for all
$h\in M^{\prime}.$ Therefore, $\mathbf{d}\sim \mathbf{d}^{\ast}$ since
$\succeq=\succeq^{2}$. Moreover, $\succeq=\succeq^{m}$implies, by definition,
that $\mathbf{d}^{\ast}$ $\succeq \mathbf{d}^{\prime}$ if and only if
$\sum_{i=1}^{m}d_{i}^{\prime}\geq \sum_{i=1}^{m}d_{i}^{\ast}$. It follows that,
since by construction $d_{m+1}^{\ast}=d_{m+1}^{\prime}$, $\mathbf{d}$
$\succeq \mathbf{d}^{\prime}$ if and only if $\sum_{i=1}^{m+1}d_{i}^{\prime
}\geq \sum_{i=1}^{m+1}d_{i}$ or, equivalently, $\succeq=\succeq^{m+1}$as required.

Notice that such a characterization is \textit{tight}, as established by the
following three counterexamples:

(i) Let $\succeq^{i^{\ast}}$be a $i^{\ast}$-weakly dictatorial preorder for
some $i^{\ast}\in \lbrack n]$, defined as follows: for any $\mathbf{d:=(}%
d_{1}...,d_{n}),\mathbf{d}^{\prime}:=(d_{1}^{\prime},...,d_{n}^{\prime}%
)\in \mathcal{Z}$, $\mathbf{d\succeq^{i^{\ast}}d}^{\prime}$ if and only if
either $d_{i^{\ast}}<d_{i^{\ast}}^{\prime}$ or $\ [d_{i^{\ast}}=d_{i^{\ast}%
}^{\prime}$ and$\  \sum_{i=1}^{n}d_{i}\leq \sum_{i=1}^{n}d_{i}^{\prime}]$. It
can be easily checked that $\succeq^{i^{\ast}}$satisfies both S-ANT and RTI,
but violates AN.

(ii) Let $\succeq:=$ $\mathcal{Z}^{2}$ the trivial total preorder over
$\mathcal{Z}$ that consists of a unique indifference class. It satisfies AN
and RTI, but violates S-ANT.

(iii) Let us now consider a total preorder $\succeq$ over a bounded subset
$\mathcal{Z}$ $\subseteq \mathbb{Z}_{+}^{2}$ that is defined as follows: for
any $\mathbf{d},\mathbf{d}^{\prime}\in \mathcal{Z}$, $\mathbf{d}\succeq
\mathbf{d}^{\prime}$ if and only if $\quad f(\mathbf{d})\; \leq \;f(\mathbf{d}%
^{\prime}),$ where $f:\mathbb{Z}_{+}^{2}\rightarrow \mathbb{Z}_{+}$ is in turn
defined by the rule

$\  \  \  \ f(\mathbf{d})\;=\;%
\begin{cases}
2(d_{1}+d_{2})+1 & \text{if }d_{1}=d_{2},\\
2(d_{1}+d_{2}) & \text{if }d_{1}\neq d_{2}.
\end{cases}
$.

Intuitively, $\succeq$ ranks vectors primarily by sum of their components
(smaller is better, reflecting a smaller sufficiency-gap), but breaks ties
among vectors with equal sums of their components by penalising vectors whose
two components coincide. It is immediately checked that $\succeq$ satisfies AN
(due to commutativity of addition, and symmetry of equality). Moreover, it
also satisfies S-ANT:\ that is so, because if $\mathbf{d=d}^{\prime}$ then
$\mathbf{d\sim d}^{\prime}$ by definition of $\succcurlyeq$ and there is
nothing to prove. If instead $d_{i}\leq d_{i}^{\prime}$ for every
$i\in \{1,2\}$ with $\mathbf{d}\neq \mathbf{d}^{\prime}$ then $\ d_{1}%
+d_{2}<d_{1}^{\prime}+d_{2}^{\prime}$ hence $(d_{1}+d_{2})+1\leq d_{1}%
^{\prime}+d_{2}^{\prime}$. Therefore,
\[
f(d)\; \leq \;2(d_{1}+d_{2})+1\;=\;2(d_{1}+d_{2}+1)-1\; \leq \;2(d_{1}^{\prime
}+d_{2}^{\prime})-1\;<\; \;2(d_{1}^{\prime}+d_{2}^{\prime})\leq \;f(d^{\prime})
\]

which in turn $\mathbf{d}\succ \mathbf{d}^{\prime}$ as required.

However, $\succeq$ fails to satisfy RTI. To see this, consider $\mathbf{d}%
=(1,1)$ and $\mathbf{d}^{\prime}=(0,2)$. Then
\[
f(1,1)=2\cdot2+1=5\quad \text{and}\quad f(0,2)=2\cdot2+0=4,
\]
so $\mathbf{d}^{\prime}\succ \mathbf{d}$ by definition of $\succeq$, whence a
fortiori $\mathbf{d}^{\prime}\succeq \mathbf{d}$ . Now add $\mathbf{z}=(2,0)$
to both profiles: $\mathbf{d+z}=(3,1)$ and $\mathbf{d}^{\prime}\mathbf{+z}%
=(2,2)$. Then,
\[
f(3,1)=2\cdot4+0=8\quad \text{and}\quad f(2,2)=2\cdot4+1=9,
\]
so $\mathbf{d+z\succ d}^{\prime}\mathbf{+z}$ \ hence \textit{not }%
$\mathbf{d}^{\prime}+\mathbf{z\succeq d+z}$ and RTI is indeed violated.
$\qquad \qquad \qquad \qquad \square$

\bigskip

\textbf{Proof of Claim 2. }

Let $\succeq$ be a total preorder over $\mathcal{Z}$. Since every permutation
$\sigma:[n]\longrightarrow \lbrack n]$ decomposes into a finite composition of
transpositions $\pi_{i}:[n]\longrightarrow \lbrack n]$, $i=1,...,t$ and
$\succcurlyeq$ is transitive, it suffices to show that $\mathbf{d}%
\sim \mathbf{d}_{\pi}$ for every $\mathbf{d}\in \mathcal{Z}$ and every
transposition $\pi$ of two indices $h\neq k$. So, fix $\mathbf{d}%
\in \mathcal{Z}$ and a transposition $\pi$ of $h\neq k$. If $d_{h}=d_{k}$ then
$\mathbf{d}_{\pi}=\mathbf{d}$ and the conclusion is immediate. Assume
therefore, without loss of generality, that
\begin{equation}
d_{h}>d_{k}.\label{eq:strict}%
\end{equation}
Set $\mathbf{d}^{\prime}:=\mathbf{d}_{\pi}$, so that $d_{h}^{\prime}=d_{k}
$,\ $d_{k}^{\prime}=d_{h}$, and $d_{i}^{\prime}=d_{i}$ for all $i\notin
\{h,k\}$.

\noindent \textit{Step~1}\textbf{: $d\succeq d^{\prime}$.}\ We verify that the
pair $(\mathbf{d},\mathbf{d}^{\prime})$ satisfies the hypothesis of RHE with
$h$ playing the role of the maximiser in $\mathbf{d}$ and $k$ playing the role
of the maximiser in $\mathbf{d}^{\prime}$.

\begin{itemize}
\item All components outside $\{h,k\}$ coincide: $d_{i}=d_{i}^{\prime}$ for
all $i\notin \{h,k\}$.

\item Position $h$ attains the larger value in $\mathbf{d}$: by assumption,
$d_{h}>d_{k}$.

\item Position $k$ attains the larger value in $\mathbf{d}^{\prime}$: since
$d_{k}^{\prime}=d_{h}>d_{k}=d_{h}^{\prime}$.

\item The chain of inequalities required by RHE reads $d_{k}^{\prime}\geq
d_{h}>d_{k}\geq d_{h}^{\prime}$, i.e.\ $d_{h}\geq d_{h}>d_{k}\geq d_{k}$,
which holds by~construction.
\end{itemize}

Therefore, RHE yields $\mathbf{d\succeq d}^{\prime}$.

\noindent \textit{Step~2}\textbf{: $d^{\prime}\succeq d$.}\ Apply the same
argument to the pair $(\mathbf{d}^{\prime},\mathbf{d})$, swapping the roles of
$h$ and $k$. Now $d_{k}^{\prime}=d_{h}>d_{h}^{\prime}=d_{k}$, so position $k$
attains the largest value in $\mathbf{d}^{\prime}$ and position $h$ attains
the largest value in $\mathbf{d}$. The hypothesis of RHE is satisfied
symmetrically, and RHE yields $\mathbf{d}^{\prime}\succeq \mathbf{d}$.

\textit{\noindent Conclusion.}\ From Steps~1 and~2, $\mathbf{d\succeq
d}^{\prime}$ and $\mathbf{d}^{\prime}\succeq \mathbf{d}$, hence $\mathbf{d}%
\sim \mathbf{d}^{\prime}=\mathbf{d}_{\pi}$. Since both $\mathbf{d}$ and
transposition $\pi$ of $h\neq k$ \ were arbitrary and, as previously
mentioned, every permutation is a finite composition of transpositions,
transitivity of $\succcurlyeq$ extends the result to all permutations. Hence
AN holds: notice that, as a result, any total preorder on the capability-type
space $\mathbf{X}$ that satisfies ANT and RHE with respect to `sufficientarian
distances' does qualify as a sufficiency-gap preorder as previously defined.
$\square$

\bigskip

\textbf{Proof of Proposition 4}

$\Longleftarrow$ It is easily checked that $\succeq^{\ast l\max}$ satisfies
both S-ANT and RHE by definition.

$\Longrightarrow$ We have to prove that if $\widehat{\succeq}$ satisfies S-ANT
and RHE then, for any $\mathbf{d,d}^{\prime}\in \mathcal{Z}$, $\mathbf{d}%
\widehat{\succeq}\mathbf{d}^{\prime}$ if and only if $\mathbf{d}\succeq^{\ast
l\max}\mathbf{d}^{\prime}$. As mentioned above in the text, the proof relies
again on a family of auxiliary total preorders $\widehat{\succeq}^{m}$ over
$\mathcal{Z}$ indexed by $m\in$ $[n]\setminus \left \{  1\right \}  $, namely
$\left \{  \widehat{\succeq}^{m}:m=2,...,n\right \}  $ as defined by the rule
$\mathbf{d\widehat{\succeq}}^{m}\mathbf{d}^{\prime}$ if and only if
$\mathbf{d}\succeq^{\ast l\max}\mathbf{d}^{\prime}$ for any $M\subseteq \lbrack
n]$ with $|M|=m$, and any $\boldsymbol{d,d}^{\prime}\in$ $\mathcal{Z}$ $\ $
with $d_{h}=d_{h}^{\prime}$ for every $h\in \lbrack n]\setminus M$. And, again,
the proof itself amounts to a `restricted' induction argument on
$[n]\setminus \left \{  1\right \}  $ which consists of two steps, namely: (I)
$\widehat{\succeq}=\widehat{\succeq}^{2} $, (II) for any $m$, $2\leq m<n$ \ if
$\widehat{\succeq}=\widehat{\succeq}^{m}$then $\succeq=\widehat{\succeq}%
^{m+1}.$

Step (I). Let us first consider any $i,j\in \lbrack n]$, $i\neq j$ and
$\mathbf{d,d}^{\prime}\in \mathcal{Z}$ such that $d_{i}\geq d_{j}$ ,
$d_{i}^{\prime}\geq d_{j}^{\prime}$ and $d_{h}=d_{h}^{\prime}$ for every
$h\in \lbrack n],i\neq h\neq j$, and $\mathbf{d}\widehat{\succeq}^{2}%
\mathbf{d}^{\prime}$ , i.e. $\mathbf{d}\succeq^{\ast l\max}\mathbf{d}^{\prime
}$. Suppose also (without any loss of generality since $\widehat{\succeq}$
satisfies AN by Claim 2), that $i=1$ and $j=2$. Thus, actually, $d_{1}\geq
d_{2}$ and $d_{1}^{\prime}\geq d_{2}^{\prime}$. If $\mathbf{d=d}^{\prime}$
then of course $\mathbf{d\sim d}^{\prime}$ by reflexivity of $\widehat
{\succeq}$, so we assume without loss of generality that $\boldsymbol{d\neq
d}^{\prime}$.

If $d_{1}=d_{1}^{\prime}$, then $d_{2}\neq d_{2}^{\prime}$ hence
$\mathbf{d}\succeq^{\ast l\max}\mathbf{d}^{\prime}$ implies $d_{2}%
<d_{2}^{\prime}$ which in turn implies $\mathbf{d}\widehat{\succ}%
\mathbf{d}^{\prime}$ by S-ANT (hence in particular $\mathbf{d}\widehat
{\succeq}\mathbf{d}$) .

If $d_{1}<d_{1}^{\prime}$ we can have the following three cases:

$\left(  1:\right)  $ $d_{1}^{\prime}>d_{1}\geq d_{2}\geq d_{2}^{\prime}$ that
by RHE entails that $\mathbf{d}\widehat{\succ}\mathbf{d}^{\prime}$;

$\left(  2:\right)  $ $d_{1}^{\prime}>d_{2}^{\prime}>d_{1}\geq d_{2}$ that by
S-ANT entails again that $\mathbf{d}\widehat{\succ}\mathbf{d}^{\prime}$;

$\left(  3:\right)  $ $d_{1}^{\prime}>d_{1}\geq d_{2}^{\prime}\geq d_{2}$,
then consider a vector $\mathbf{c}\in \mathcal{Z}$ arranged in nonincreasing
order w.l.o.g. by AN (which holds true because it follows from RHE as
established by Claim 2 above) differs from $d^{\prime}$ only in the first two
components $\left \{  1,2\right \}  $, i.e. $\mathbf{c}=\left(  d_{1}%
,d_{2}^{\prime},d_{3}^{\prime},...,d_{n}^{\prime}\right)  =\left(  d_{1}%
,d_{2}^{\prime},d_{3},...,d_{n}\right)  $. By the previous argument, since the
first components of $\mathbf{d}$ and $\mathbf{c}$ are the same while
$d_{2}^{\prime}\geq d_{2}$, it follows by S-ANT (actually, by ANT) that
$\mathbf{d}\widehat{\succeq}\mathbf{c}$. We further observe that
$d_{1}^{\prime}>d_{1}\geq d_{2}^{\prime}\geq d_{2}^{\prime}$, the first
inequality by assumption, the second by construction and the third by
reflexivity. Hence, by RHE, we have that $\mathbf{c}\widehat{\succeq
}\mathbf{d}^{\prime}$ that together with $\mathbf{d}\widehat{\succeq
}\mathbf{c}$ entails by transitivity $\mathbf{d}\widehat{\succeq}%
\mathbf{d}^{\prime}$ as required. It follows that $\widehat{\succeq}%
^{2}\subseteq \widehat{\succeq}$. \ 

Conversely, suppose that for some $i,j\in \lbrack n]$, $i\neq j$ and
$\mathbf{d,d}^{\prime}\in \mathcal{Z}$ such that $d_{i}\geq d_{j}$ ,
$d_{i}^{\prime}\geq d_{j}^{\prime}$ and $d_{h}=d_{h}^{\prime}$ for every
$h\in \lbrack n],i\neq h\neq j$, $\mathbf{d}\widehat{\succeq}\mathbf{d}%
^{\prime}$ yet \textit{not }$\mathbf{d}\widehat{\succeq}^{2}\mathbf{d}%
^{\prime}$, namely \textit{not }$\mathbf{d}\succeq^{\ast l\max}\mathbf{d}%
^{\prime}$ which in turn implies by definition that $\mathbf{d}^{\prime
}\mathbf{\succ}^{\ast l\max}\mathbf{d}$. It follows that either $d_{1}%
^{\prime}<d_{1}$, or $d_{1}^{\prime}=d_{1}$ and $d_{2}^{\prime}<d_{2} $. In
any case, it follows by S-ANT that $\mathbf{d}^{\prime}\widehat{\succ
}\mathbf{d}$, a contradiction. As a result, $\widehat{\succeq}\subseteq
\widehat{\succeq}^{2}$ also holds, hence $\widehat{\succeq}=\widehat{\succeq
}^{2}$ as required.

Step (II) Suppose that $\widehat{\succeq}=\widehat{\succeq}^{m}$ for some $m$,
$2\leq m<n$. We have to prove that\ $\widehat{\succeq}=\widehat{\succeq}%
^{m+1}$.

Suppose that $\widehat{\succeq}=\widehat{\succeq}^{m}$for some $m$, $2\leq
m<n$, and consider any pair of vectors $\boldsymbol{d,d}^{\prime}\in$
$\mathcal{Z}$ such that $d_{h}=d_{h}^{\prime}$ for every $h\in M^{\prime
}:=[n]\setminus \left \{  1,m+1\right \}  $. Then, consider a third vector
$\mathbf{d}^{\ast}\in \mathcal{Z}$ such that : $\max \left \{  d_{1}^{\ast
},d_{m+1}^{\ast}\right \}  =\max \left \{  d_{1},d_{m+1}\right \}  $, and
$d_{h}^{\ast}=d_{h\text{ }}$for all $h\in M^{\prime}.$ Therefore,
$\mathbf{d}\widehat{\sim}\mathbf{d}^{\ast}$ since $\widehat{\succeq}%
=\widehat{\succeq}^{2}$. Moreover, $\widehat{\succeq}=\widehat{\succeq}^{m}%
$implies, by definition, that $\mathbf{d}^{\ast}$ $\widehat{\succeq}%
\mathbf{d}^{\prime}$ if and only if $\mathbf{d}^{\ast}\succeq^{\ast l\max
}\mathbf{d}$. It follows that $\mathbf{d}$ $\widehat{\succeq}\mathbf{d}%
^{\prime}$ by transitivity or equivalently, since by construction $\mathbf{d}$
$\widehat{\succeq}\mathbf{d}^{\prime}$ if and only if $\mathbf{d}\succeq^{\ast
l\max}\mathbf{d}^{\prime}$ that $\succeq=\succeq^{m+1}$as required.

The characterization is also tight. Indeed, it is easily checked that the
trivial preorder $\succeq$ $:=\mathcal{Z}^{2}$ satisfies RHE but violates
S-ANT. Conversely the min-average preorder $\succeq^{\ast av}$does satisfy
S-ANT but violates RHE (to see this, consider $\mathbf{d}=(10,9),\mathbf{d}%
^{\prime}=(2,11)$: clearly $d_{2}^{\prime}>d_{1}>d_{2}>d_{1}^{\prime}$, hence
RHE would require $\mathbf{d\succ d}^{\prime}$. Yet $\mathbf{d}^{\prime}%
\succ^{\ast av}\mathbf{d}$). $\  \  \  \  \  \  \  \  \  \  \  \  \qquad \qquad \  \square$

\bigskip

\textbf{Proof of Lemma 1}

The proof of Lemma 1 consists of two parts.

(a) The first and most important one is the proof that $(\mathcal{A}%
_{\mathbf{X}},\succapprox)$ is a distributive lattice, which is indeed a
classic result due to Dilworth (1960), namely

\begin{itemize}
\item Let $\mathbf{Y}=(Y,\leqslant)$ be a finite partially ordered set,
$\mathcal{A}_{\mathbf{Y}}$ the set of all \textit{antichains} of $\mathbf{Y}$
(i.e., sets of mutually $\leqslant$-incomparable elements of $Y$), and
$\succapprox$ the binary relation on $\mathcal{A}_{\mathbf{Y}}$ defined by the
following rule: for any $\mathcal{X}\mathbf{=}\left \{  x_{1},...,x_{k}%
\right \}  ,$ $\mathcal{Y=}\left \{  y_{1},...,y_{h}\right \}  \mathbf{\in
}\mathcal{A}_{\mathbf{Y}}$, $\mathcal{X}\succapprox \mathcal{Y}$ if and only if
for every $x_{i}\in \mathcal{X}$ there exists $y_{j}\in \mathcal{Y}$ such that
$\ y_{j}\leqslant x_{i}$. Then, $(\mathcal{A}_{\mathbf{Y}},\succapprox)$ is a
distributive lattice.
\end{itemize}

(b) $(\mathcal{A}_{\mathbf{X}},\succapprox)$ is isomorphic to the lattice
$(\mathcal{S(}\mathbf{X}^{n}\mathbf{),\leqslant)}$ of sufficientarian BGFs on
$\mathbf{X.}$

The proof of part (a) to be presented below is \textit{not} the original one
due to Dilworth (1960), but rather a \textit{dualized }version of the proof
proposed by Anderson (1987) focussing on the correspondence between antichains
and \textit{order ideals} (or downward closed sets) as opposed to the
correspondence between antichains and \textit{order filters }(or upward closed
sets) of $\mathbf{Y}$, which is the relevant one in our case. Thus, for the
sake of both convenience and completeness, we report our `dualized' version of
Anderson's proof in some detail below. To begin with, recall that a set
$F\subseteq Y$ is an \textit{order filter }of $\mathbf{Y}=(Y,\leqslant)$ -also
written $F\in \mathcal{F}_{\mathbf{Y}}$- if and only if [ for every $x,y\in Y$,
if $x\in F$ and $x\leqslant y$ \ then $y\in F$].

Part (a) Then, the proof consists in\ establishing the validity of the claims
attached to the following five steps.

Step (i) $(\mathcal{A}_{\mathbf{Y}},\succapprox)$ is a partially ordered set
(or poset). It must be checked that $\succapprox$ is reflexive, antisymmetric
and transitive. Clearly, for any $\mathcal{X}\in \mathcal{A}_{\mathbf{Y}}$,
$\mathcal{X}\succapprox \mathcal{X}$ \ follows immediately, by definition, from
reflexivity of $\leqslant$: thus, \textit{reflexivity} of $\succapprox$ holds.
Moreover, consider any pair of antichains $\mathcal{X},\mathcal{Y\in
A}_{\mathbf{Y}}$ such that $\mathcal{X\succapprox Y}$ and
$\mathcal{Y\succapprox X}$: by definition,\ for every $x\in \mathcal{X}$
\ there exists $y\in \mathcal{Y}$ \ such that $y\leqslant x$, and for every
$y^{\prime}\in \mathcal{Y}$\ there exists $x^{\prime}\in \mathcal{X}$ \ such
that $x^{\prime}\leqslant y^{\prime}$. Now, consider the first pair with
$x\in \mathcal{X}$ arbitrarily chosen and $y\in \mathcal{Y}$ with $y\leqslant
x$. Since $\mathcal{Y\succapprox X}$ , there exists $x^{\prime}\in$
$\mathcal{X}$ such that $\ x^{\prime}\leqslant y$. \ But then, both
$y\leqslant x$ and $x^{\prime}\leqslant y$ . Thus, by transitivity of
$\leqslant$, $x^{\prime}\leqslant x$ while both of them belong to antichain
$\mathcal{X}$. It follows that $x=x^{\prime}$, that in turn implies that
$y=x$. Therefore, $\mathcal{X\subseteq Y}$. But then, a similar argument
applied to the second pair $x^{\prime},y^{\prime}$ mentioned above, with
$y^{\prime}\in \mathcal{Y}$ arbitrarily chosen, and $x^{\prime}\in \mathcal{X}$
such that $x^{\prime}\leqslant y^{\prime}$, establishes that
$\mathcal{Y\subseteq X}$. Thus, $\mathcal{X=Y}$ and \textit{antisymmetry} of
$\succapprox$ holds. Finally, consider any $\mathcal{X}$,$\mathcal{Y}%
$,$\mathcal{Z\in A}_{\mathbf{Y}}$ such that $\mathcal{X\succapprox Y}$ and
$\mathcal{Y\succapprox Z}$. Next, consider an arbitrary $x\in \mathcal{X}$.
Since $\mathcal{X\succapprox Y}$, there exists $y\in \mathcal{Y}$ such that
$y\leqslant x$: Also, since $\mathcal{Y\succapprox Z}$, there exists
$z\in \mathcal{Z}$ such that $z\leqslant y$. But then, $z\leqslant x$ holds by
transitivity of $\leqslant$: it follows that \textit{transitivity }of
$\succapprox$ also holds. Hence $(\mathcal{A}_{\mathbf{Y}},\succapprox)$ is
indeed a poset, as required.

Step (ii) For any antichain $\mathcal{Y\in}(\mathcal{A}_{\mathbf{Y}%
},\succapprox)$ the set

\ $\widehat{\mathcal{Y}}:=\left \{  y\in Y:\text{there exists an }%
x\in \mathcal{Y}\text{ such that }x\leqslant y\right \}  \in \mathcal{F}%
_{\mathbf{Y}}$, i.e., is an order filter of $\mathbf{Y}$. That is immediate,
by definition. Indeed, let $\mathcal{Y\in A}_{\mathbf{Y}}$ and $x,y\in Y$ such
that $\ x\in \widehat{\mathcal{Y}}$ and $x\leqslant y$. Then, by definition,
there exists $x^{\prime}\in \mathcal{Y}$ such that $x^{\prime}\leqslant x$.
Therefore by transitivity $x^{\prime}\leqslant y$ as well, and by definition
$y\in$ $\widehat{\mathcal{Y}}$. \ Hence $\widehat{\mathcal{Y}}$ is indeed an
order filter of $\mathbf{Y}$.

Step (iii) For any order filter $F\in \mathcal{F}_{\mathbf{Y}}$ the set

$F_{\min}:=\left \{  y\in F:\text{for any }x\in Y\text{, }x\leqslant y\text{
only if }x=y\right \}  \in \mathcal{A}_{\mathbf{Y}}$, i.e., is an antichain of
$\mathbf{Y}$. Hence, in particular, for any order filter $F\in \mathcal{F}%
_{\mathbf{Y}}$ there exists an antichain $\mathcal{Y}:\mathcal{=}F_{\min
}\mathcal{\in A}_{\mathbf{Y}}$ such that $F=\widehat{\mathcal{Y}}$, and
conversely if $F=\widehat{\mathcal{Y}}$ \ for some $\mathcal{Y\in
A}_{\mathbf{Y}}$ then $\mathcal{Y=}F_{\min}$. That is also immediate, by
definition. To see this, consider any two \textit{distinct} $x,y\in F_{\min}%
.$Clearly, neither $x\leqslant y$ nor $y\leqslant x$ is the case, because each
one of those inequalities implies $x=y$, a contradiction. Hence, $F_{\min}$ is
in fact an antichain of $\mathbf{Y}$. Moreover, for any $F\in \mathcal{F}%
_{\mathbf{Y}}$, $F=\widehat{F}_{\min}$ and for any $\mathcal{Y\in
A}_{\mathbf{Y}}$, $F=\widehat{\mathcal{Y}}$ only if $\mathcal{Y}=F_{\min}$ by construction.

Step (iv) For any pair of antichains $\mathcal{X},\mathcal{Y\in A}%
_{\mathbf{Y}}$, [$\mathcal{X\succapprox Y}$ \ if and only if \ $\widehat
{\mathcal{X}}\subseteq \widehat{\mathcal{Y}}$ ] and [$\mathcal{X\approx Y}$
\ if and only if \ $\widehat{\mathcal{X}}=\widehat{\mathcal{Y}}$ ]. Suppose
that \ $\mathcal{X\succapprox Y}$ $\ $and $z\in \widehat{\mathcal{X}}$ (with
$\mathcal{Y=}\widehat{\mathcal{Y}}_{\min}$, and $\mathcal{X=}\widehat
{\mathcal{X}}_{\min}$, by construction, as shown under step (iii)). Then, by
definition of $\widehat{\mathcal{X}}$, there exists $x\in \mathcal{X}$ such
that $x\leqslant z.$ Now, $\mathcal{X\succapprox Y}$ implies that for any
$x\in \mathcal{X}$ \ there exists $y^{\prime}\in \mathcal{Y}$ such that
$y^{\prime}\leqslant x.$ Hence $y^{\prime}\leqslant z$ by transitivity of
$\leqslant$, and $z\in \widehat{\mathcal{Y}}$ which in turn implies that
$\widehat{\mathcal{X}}\subseteq \widehat{\mathcal{Y}}$. Conversely, suppose
that $\widehat{\mathcal{X}}\subseteq \widehat{\mathcal{Y}}.$ Then, for every
$x\in \widehat{\mathcal{X}}$ \ there exists $y\in \mathcal{Y}$ such that
$x\leqslant y$ (since $x\in \widehat{\mathcal{Y}}$ as well). Thus, in
particular, for every $x\in \widehat{\mathcal{X}}_{\min}=\mathcal{X}$ \ there
exists $y\in \mathcal{Y}$ such that $x\leqslant y.$ It follows that, by
definition, $\mathcal{X\succapprox Y}$. Finally, suppose that
$\mathcal{X\approx Y}$, i.e., both $\mathcal{X\succapprox Y}$ and
$\mathcal{Y\succapprox X}$ . Then, it follows from the previous part of the
present step that both $\widehat{\mathcal{X}}\subseteq \widehat{\mathcal{Y}}$
and $\widehat{\mathcal{Y}}\subseteq \widehat{\mathcal{X}}$ hold. Therefore,
$\widehat{\mathcal{X}}=\widehat{\mathcal{Y}}.$ Conversely, by the very same
argument, if $\widehat{\mathcal{X}}=\widehat{\mathcal{Y}}$ then
$\mathcal{X\approx Y}$ as required.

Step (v): $(\mathcal{F}_{\mathbf{Y}},\subseteq$) is a distributive lattice and
is isomorphic to $(\mathcal{A}_{\mathbf{Y}},\succapprox)$. Hence
$(\mathcal{A}_{\mathbf{Y}},\succapprox)$ is also a distributive lattice. The
previous steps (ii),(iii),(iv) jointly imply that $(\mathcal{F}_{\mathbf{Y}%
},\subseteq)$ is isomorphic to $(\mathcal{A}_{\mathbf{Y}},\succapprox).$ Now,
consider an arbitrary pair of order filters $F,F^{\prime}\in \mathcal{F}%
_{\mathbf{Y}}$ and their set-theoretic intersection $F\cap F^{\prime}$, and
union $F\cup F^{\prime}$. Let $x,y\in Y$ be such that $x\leqslant y$ and $x\in
F\cap F^{\prime}$. Then, by definition, there exist $z\in F$ and $z^{\prime
}\in F^{\prime}$ such that $z\leqslant x$ and $z^{\prime}\leqslant x.$ It
follows that both $z\leqslant y$ and $z^{\prime}\leqslant y$, by transitivity
of $\leqslant$. Thus, $y$ $\in F\cap F^{\prime}$. It follows that $F\cap
F^{\prime}\in \mathcal{F}_{\mathbf{Y}}$, i.e., is an order filter of
$\mathbf{Y.}$

Moreover, let $x,y\in Y$ be such that $x\leqslant y$ and $x\in F\cup
F^{\prime}$. Then, by definition, there exists either a $z\in F$ such that
$z\leqslant x$ or a $z^{\prime}\in F^{\prime}$ such that $z^{\prime}\leqslant
x$ (or both of them). Suppose then without any loss of generality that there
exists a $z\in F$ such that $z\leqslant x.$ It follows again that $z\leqslant
y$, by transitivity of $\leqslant$. Thus, $y$ $\in F\cup F^{\prime}$. It
follows that $F\cup F^{\prime}\in \mathcal{F}_{\mathbf{Y}}$, i.e., it is also
an order filter of $\mathbf{Y.}$ But then, $(\mathcal{F}_{\mathbf{Y}%
},\subseteq)$ is a \textit{lattice} with a well-defined g.l.b or \textit{meet}
operation $\wedge:=\cap$, and a well-defined l.u.b. or \textit{join }operation
$\vee:=\cup$, and the (mutual) distributivity laws satisfied by set-theoretic
intersection $\cap$ and union $\cup$ imply that $(\mathcal{F}_{\mathbf{Y}%
},\subseteq)$ is a \textit{distributive} lattice (and the same holds for
$(\mathcal{A}_{\mathbf{Y}},\succapprox)$).

Part (b): Let us now consider our case where $\mathbf{Y}=\mathbf{X.}$ To prove
that $(\mathcal{A}_{\mathbf{X}},\succapprox)$ is isomorphic to $\mathcal{S}%
(\mathbf{X}^{n})$, just consider the function $\varphi:\mathcal{S}%
(\mathbf{X}^{n})\longrightarrow \mathcal{A}_{\mathbf{X}}$ defined as follows:
for any $g\in \mathcal{S}(\mathbf{X}^{n})$, $\varphi(g):=\mathcal{X(}g)$. Then,
it is easily checked that $\varphi$ is a latticial isomorphism.$\qquad
\qquad \qquad \square$

\bigskip

\textbf{Proof of Proposition 5}

By Lemma 1 $(\mathcal{A}_{\mathbf{X}},\succapprox)$ is a distributive lattice
that is isomorphic to $(\mathcal{S}(\mathbf{X}^{n}),\leqslant)$. Since
$(\mathcal{A}_{\mathbf{X}},\succapprox)$ is a distributive lattice, two
important corollaries follow, namely

(I) A \textit{metric} betweenness $B_{d}$ (which as observed previously in the
text is also a median betweenness) can be defined on $(\mathcal{A}%
_{\mathbf{X}},\succapprox)$ , and a natural domain of preference preorders
$\succeq$ on $\mathcal{A}_{\mathbf{X}}$ with a unique top antichain that are
single-peaked with respect to $B_{d}$ (namely, for any $(\mathcal{X}%
,\mathcal{Y},\mathcal{Z)\in}B$ if $\mathcal{X=}top(\succeq)\neq \mathcal{Z}$
then $\mathcal{Y\succeq Z}$) (see Savaglio and Vannucci (2019)).

(II) By Lemma 2, there exists a class of (lattice-polynomial) anonymous and
idempotent aggregation rules $F:(\mathcal{A}_{\mathbf{X}})^{n}\longrightarrow
\mathcal{A}_{\mathbf{X}}$ on the preference domain $\mathcal{D}_{B_{d}}$ as
defined above (see Savaglio and Vannucci (2019) and Vannucci (2019)). If $n$
is odd, that class includes the simple majority aggregation rule $F^{maj}$
that is defined as follows: for any profile of antichains $\mathcal{X}%
_{[n]}=(\mathcal{X}_{1},...,\mathcal{X}_{n})\in$ $(\mathcal{A}_{\mathbf{X}%
})^{n}$, $F^{maj}(\mathcal{X}_{[n]})=%
{\displaystyle \bigvee \limits_{S\in \mathcal{W}^{maj}}}
{\displaystyle \bigwedge \limits_{i\in S}}
\mathcal{X}_{i}$, where $\mathcal{W}^{maj}:=\left \{  S\subseteq N:|S|\geq
\left \lfloor \frac{|N|+2}{2}\right \rfloor \right \}  $. Thus, the protocol
$\varphi_{\lbrack n]}^{-1}\circ F^{maj}\circ \varphi^{-1}$ is a strategy-proof
implementation of the aggregation rule $G:(\mathcal{S}(\mathbf{X}^{n})$
$)^{n}\longrightarrow \mathcal{S}(\mathbf{X}^{n})$ \ where $\varphi_{\lbrack
n]}:=(\varphi,...,\varphi)$. $\qquad \qquad \qquad \qquad \qquad \qquad \qquad
\qquad \square$

\end{document}